\documentclass[structabstract]{aa}  
\pdfoutput=1

\usepackage{graphicx}
\usepackage{txfonts}
\usepackage[utf8]{inputenc}
\usepackage{natbib}
\usepackage[draft]{hyperref}
\usepackage{multirow}
\bibpunct{(}{)}{;}{a}{}{,} 
\usepackage{longtable}
\def\aap{A\&A\ }

\def\apjl{ApJL\ }

\def\apjs{ApJS\ }

\begin{document}
    \title{A library of near-infrared integral field spectra of young M-L dwarfs\thanks{Based on observations made with ESO telescopes at the La Silla Paranal Observatory under programs 076.C-0379, 078.C-0510, 078.C-0800, 080.C-0590, 279.C-5010, and 083.C-0595 collected at the European Organization for Astronomical Research in the Southern Hemisphere, Chile.}$^{,}$ \thanks{The library can be downloaded at this adress:  \texttt{http://ipag.\-o\-sug.fr/\-$\sim$gchau\-vin/add\-ma\-te\-rial.html}}}


   \author{M. Bonnefoy   \inst{1}
	\and
	 G.~Chauvin \inst{2,}  \inst{1}
	\and
  	 A.-M.~Lagrange \inst{2}
  	\and 
	P. Rojo \inst{3}
   \and 
	 F.~Allard \inst{4}
 	\and
	C.~Pinte \inst{2}
	\and
	C. Dumas \inst{5} 
	\and 
	D. Homeier \inst{4}
       }

   \institute{Max-Planck-Institut für Astronomie, Königstuhl 17, D-69117 Heidelberg, Germany\\
	\email{bonnefoy@mpia.de}
	\and
	UJF-Grenoble 1 / CNRS-INSU, Institut de Planétologie et d’Astrophysique de Grenoble (IPAG) UMR 5274, Grenoble 38041, France
	\and
	Departamento de Astronomia, Universidad de Chile, Casilla 36-D, Santiago, Chile
 	\and
    CRAL-ENS, 46, All\'ee d'Italie , 69364 Lyon Cedex 07, France
    \and
    European Southern Observatory, Casilla 19001, Santiago 19, Chile}

   \date{Received 13/10/2011; accepted 14/06/2013}

 
  \abstract
  {At young ages, low surface gravity affects the atmospheric properties of ultracool dwarfs. The  impact on medium-resolution near-infrared  (NIR) spectra has only been slightly investigated at the M-L transition sofar.}
   {We present a library of near-infrared (1.1--2.45 $\mathrm{\mu}$m) medium-resolution (R$\sim$1500-2000) integral field spectra of 15 young  M6-L0 dwarfs. We aim at deriving updated NIR spectral type, luminosity and physical parameters ($\mathrm{T_{eff}}$, log g , M, L/L$_{\odot}$) of each source. This work also aims at testing the latest generation of  BT-SETTL atmospheric models.}
   {We estimated spectral types using spectral indices and  spectra of reference young objects classified in the optical. We used the 2010 and 2012 releases of the BT-SETTL synthetic spectral grid and cross-checked the results with the DRIFT-PHOENIX models to derive the atmospheric properties of  the sources.}
   {We do not evidence significant differences between the spectra of young companions and of  reference young isolated brown-dwarfs in the same spectral type range. We derive infrared spectral types L0$\pm$1, L0$\pm$1, M9.5$\pm$0.5, M9.5$\pm$0.5, M9.25$\pm$0.25,  M8$^{+0.5}_{-0.75}$, and M8.5$\pm$0.5 for  AB Pic b, Cha J110913-773444, USco CTIO 108B, GSC 08047-00232 B, DH Tau B, CT Cha b, and HR7329B respectively. BT-SETTL and DRIFT-PHOENIX models  yield close T$\mathrm{_{eff}}$ and log g  estimates for each sources. The models seem to evidence a  $600_{-300}^{+600}$ K drop of the effective temperature at the M-L transition. Assuming the former temperatures are correct, we derive new mass estimates which confirm that DH Tau B, USco CTIO 108B, AB Pic b, KPNO Tau 4, OTS 44, and Cha1109 lies inside or at the boundary of the planetary mass range. We combine the empirical luminosities  of the M9.5--L0 sources to the $\mathrm{T_{eff}}$ to derive semi-empirical radii estimates that do not match ``hot-start" evolutionary models predictions at 1--3 Myr.  We use complementary data to demonstrate that atmospheric models are able to reproduce  the combined optical and infrared spectral energy distribution, together with the near-infrared spectra of these sources simultaneously. But the models still fail to represent the dominant features in the optical. This issue rises  doubts on the ability of  these models to  predict effective temperatures from near-infrared spectra alone.}
   {The library provides templates for the characterization of  other young and late type objects. The study advocates the use of photometric and spectroscopic informations over a broad range of wavelengths to study the properties of very low mass young companions to be detected with the planet imagers (Subaru/SCExAO, LBT/LMIRCam, Gemini/GPI, VLT/SPHERE).}


   \maketitle

\section{Introduction}


Since the discovery of the first bound substellar objects GD\,165~B \citep{1988Natur.336..656B} and Gl\,229~B \citep{1995Natur.378..463N}, the development of large infrared surveys led to an explosion of discoveries of very low-mass stars and mature brown dwarfs in the field. Many of these "ultracool dwarfs"  failed to entered the MK spectroscopic classification scheme and required the creation of two new classes ‘‘L’’ \citep{1999ApJ...519..802K, 1999AJ....118.2466M} and ‘‘T’’ \citep{2002ApJ...564..466G, 2002ApJ...564..421B}.  In the near-infrared (NIR), spectra of late-M and L field dwarfs are dominated by a mix of atomic lines and molecular bands \citep{1994MNRAS.267..413J, 1995AJ....110.2415A, 1996ApJS..104..117L, 2001ApJ...548..908L, 2001AJ....121.1710R}.  At low resolution, the variations in strengths of these features were followed to define a classification scheme coherent with that found at optical wavelengths \citep{2001AJ....121.1710R, 2001ApJ...552L.147T, 2002ApJ...564..466G}. More recently,  \cite{2003ApJ...596..561M} (ML03) and \cite{2005ApJ...623.1115C} (C05) built medium-resolution (R$\sim$2000) NIR libraries and quantified the variations of narrow features that can give additional constraints on spectral types. Models aiming at reproducing the emergent flux of ultracool atmospheres showed that these changes are best explained by a decreasing of the effective temperature that leads in turns to the formation of a dusty cloud deck bellow $\mathrm{\lesssim}$ 2700 K \citep{1996A&A...308L..29T, 2001ApJ...556..357A, 2008MNRAS.391.1854H}. 

Evolutionary models \citep{2000ApJ...542..464C, 2001RvMP...73..719B} predict that substellar objects form and contract with age, being hotter, larger, and more luminous at very young ($\mathrm{\lesssim}$ 100 Myrs) ages. As a consequence, a population of young very low mass objects that would be too faint to study once they have aged were rapidly discovered in young clusters and in young nearby ($<$ 100 pc) associations \citep[see the review of][]{2008hsf2.book..757T} during deep infrared surveys \citep[e.g.][for instance]{2010AJ....139..950R}, observations campaigns with adaptive optics devices \citep[e.g.][]{2003A&A...404..157C, 2010A&A...509A..52C, 2011ApJ...726..113I}, and with space observatories \citep[\textit{Hubble}, \textit{Spitzer}: e.g ][]{1999ApJ...512L..69L, 2000ApJ...541..390L, 2005AJ....130.1845L, 2007ApJ...654..570L, 2010ApJ...714L..84T}. Among these discoveries, planetary mass objects ($\mathrm{\leqslant 13.6 M_{Jup}}$, see the definition of the \textit{International Astronomical Union}) were found free-floating \citep[e.g.][]{1999ApJ...526..336O, 2000MNRAS.314..858L, 2000Sci...290..103Z},  orbiting brown dwarfs (2MASSW J1207334-393254, \cite{2004A&A...425L..29C}; USCO CTIO 108, \cite{2008ApJ...673L.185B}; 2MASS J04414489+2301513 \cite{2010ApJ...714L..84T}) or as wide ($>$ 5 AU) companions to stars (AB Pic, \cite{2005A&A...438L..29C}; DH Tau,  \cite{2005ApJ...620..984I}; CHXR 73, \cite{2006ApJ...649..894L}; GQ Lup, \cite{2005A&A...435L..13N}; CT Cha, \cite{2008A&A...491..311S}; Fomalhaut, \cite{2008Sci...322.1345K}; HR8799, \cite{2008Sci...322.1348M, 2010Natur.468.1080M}; $\beta$ Pictoris, \cite{2010Sci...329...57L}; GSC 06214-00210 \cite{2011ApJ...726..113I}; 1RXS J235133.3+312720 \cite{2012ApJ...753..142B}; $\kappa$ Andromedae, \cite{2013ApJ...763L..32C}), or even binaries (SR 12 AB, \cite{2011AJ....141..119K}; 2MASS J01033563-5515561 AB, \cite{2013arXiv1303.4525D}). This variety of configurations gives precious benchmarks to planet and brown-dwarf formation models \citep[e.g.]{2009ApJ...695L..53B, 2010ApJ...710.1375K, 2012A&A...547A.111M, 2012A&A...544A..32L, 2012MNRAS.419.3115B}.

Together with the temperature, the reduced surface gravity of these objects modifies the chemical and physical properties of their atmospheric layers. This translates into peculiar spectroscopic features that have been identified in the optical spectra of late-M dwarfs \citep{1996ApJ...469..706M, 1997ApJ...489L.165L, 1999ApJ...525..440L, 2002ApJ...575..484G, 2003ApJ...590..348L} and early-L (L0--L5) dwarfs \citep{2006ApJ...639.1120K, 2009AJ....137.3345C, 2009ApJ...703..399L}. In the near-infrared, the most noticeable effects are the triangular shape of the H band \citep{2001MNRAS.326..695L, 2004ApJ...600.1020M, 2006ApJ...639.1120K,2007ApJ...657..511A, 2008MNRAS.383.1385L, 2009MNRAS.392..817W, 2010A&A...512A..52B} and the reduced strength of alkali  (Na I, K I) lines \citep{2003ApJ...593.1074G, 2004ApJ...600.1020M, 2006ApJ...639.1120K}. These differences can be used as precious age proxy \citep{2003ApJ...593.1074G, 2004ApJ...610.1045S, 2007ApJ...657..511A, 2010A&A...519A..93B} to find cluster candidates and/or to confirm their membership. However, using these changes to accurately access the intrinsic physical parameters of these objects is still limited by the lack of a classification scheme that accounts for surface gravity variations \citep{2005ARA&A..43..195K}, and associated T$\mathrm{_{eff}}$/spectral type conversion scales for late-type objects. In order to break the temperature-age-surface gravity degeneracy, grids of synthetic spectra generated for a given set of atmospheric parameters can be compared to empirical spectra \citep{2005ESASP.576..565B, 2008ApJ...675L.105H, 2008ApJ...683.1104F, 2013arXiv1302.6559A}. However, recent studies \citep{2001ApJ...548..908L, 2001MNRAS.326..695L,2010A&A...512A..52B, 2010ApJS..186...63R} show that models must be used with care. Obtention of high quality NIR spectral libraries of well-classified objects are then necessary to define anchor spectra for the classification and to constrain the models. Only recently, \cite{2008MNRAS.383.1385L} built a NIR spectral sequence of M8-L2 candidates Upper Sco members ($\sim$ 5 Myrs) classified in the NIR. But \cite{2009ApJ...696.1589H} found that several of the latest sources of their sample are M dwarfs in the optical.

In this paper, we present the results of an observational campaign to build a homogeneous library of near-infrared (1.1--2.5 $\mu$m) integral field spectra of 15 young M and early-L dwarfs at medium resolution (R$\sim$1500-2000). In contrast with other studies, our spectra are less sensitive to chromatic slit losses that occur on AO-fed spectrographs with narrow entrance slits  \citep{2002ApJ...567L..59G, 2005A&A...430.1027C}. Our original source sample (see Table~\ref{table:1}) is composed of 15 objects close to the planetary mass range found isolated (Cha J110913-773444, hereafter Cha 1109;  OTS 44; KPNO Tau 4), or as companions to star (AB Pic b; DH Tau B, GSC 08047-00232 B, hereafter GSC08047 B, TWA 5B, Gl 417B, USco CTIO 108B, and HR7329B). We observed in addition  2MASS J03454316+2540233 (2M0345), a member of the Ursa Major moving group \citep[300 to 600 Myr][]{2002MNRAS.334..193C, 2003AJ....125.1980K, 2007MNRAS.378L..24B}, to get the reference spectrum of an mature L0 dwarf. Apart Gl 417 B \citep[80-300 Myr][]{2000AJ....120..447K}, our young objects are all members of young nearby associations (TW Hydrae, age$\sim$8 Myr; Tucana-Horologium, age$\sim$30 Myr; Colomba, age$\sim$30 Myr; Carina, age$\sim$30 Myr) and young clusters/star-forming regions (Chameleon I, age$\sim$1-3 Myr; Taurus, age $\sim$1 Myr; Upper-Sco, age=5-11 Myr).  We also reduced and re-analyzed spectra of the tight binary TWA 22AB \citep[see ][]{2009A&A...506..799B}, of the young companion CT Cha b \citep{2008A&A...491..311S}, and of the young field dwarf 2MASS J01415823-4633574 \citep[2M0141;][]{2006ApJ...639.1120K}, and 2MASSW J1207334-393254 A \citep[2M1207 A;][]{2002ApJ...575..484G}.  In section 2, we describe the observations, and the associated data reduction. We detail our spectral analysis and derive the physical parameters of each object in section 3. Finally, in section 4, we discuss our results and try to identify and quantify different biases that could have affected the analysis.   

\begin{table*}[ht]
\begin{minipage}[ht]{\linewidth}
\caption{Source properties taken from the literature.}
\label{table:1}
\centering
\begin{tabular}{ l l l l l l l l l l l}     
\hline\hline
Name											&		D			& Sep. 		&		Memb.	&	Sp. type		&		A$\mathrm{_{v}}$		&	T$\mathrm{_{eff}}$\tablefootmark{b}	&		log g \tablefootmark{b}  &	L$\mathrm{_{bol}}$				&		Mass			&		References \\
													&		(pc)		&('')	 		&							&							&		(mag)							&		(K)								&		(dex)	  							&		(dex	)				&	(M$\mathrm{_{Jup}}$)		&	   				     				 \\
\hline
Cha1109										&	$\sim$165 	&	--	 		&		Cha I				&	M9.5$\pm$0.5\tablefootmark{a}						&	1$\pm$1		&		--									&		--		 		 					&		-3.33				&8$^{+7}_{-3}$&		1,2,41		\\
AB Pic b											&45.5	&	5.5	& Tuc-Hor		&		L0--L1			& 0.27$\pm$0.02				&	2000$^{+100}_{-300}$		&	4.0$\pm$0.5					&		-3.6$\pm$0.2		&	10--14			&	3,4,6,7 \\
							&	&	& Carina		&						& 				&	&					&				&		&	5 \\
DH Tau B					&  $\sim$140						&			2.3		&		Taurus		&		--		&	2.7--2.8					&		2700-2800	&	4.0--4.5				&	$\sim$ -2.44		&	30--50				   &	8	\\
							&	142	&						&						&				&	1.1$\pm$1.1				&							&								&	-2.71$\pm$0.12	&	11$^{+10}_{-3}$ & 9 \\
GSC08047  B 	&	85-92	&	 3.3	&	Tuc-Hor		&		M9.5$\pm$1	  & --	&		--	 &		--		&  -3.3$\pm$0.1		&		25$\pm$10		&		5,10,11,12, 46		\\
TWA	22A		&		17.5	&	0.09	&	$\beta$ Pic 	&		M6$\pm$1	&	--	&		2900$\pm$200	&	4.0--5.5	&	-2.11$\pm$0.13	&	220$\pm$21\tablefootmark{c} &  13,14, 46	\\
TWA	22B 		&		17.5	&	0.09	&	$\beta$ Pic		&		M6$\pm$1 &	--		&		2900$^{+200}_{-100}$	&	4.0--5.5	&	-2.30$\pm$0.16	&	220$\pm$21\tablefootmark{c} & 13,14, 46 	\\
2M1207 A		&	52.4	 &	-- 	&	TWA	&	M8.25$\pm$0.50\tablefootmark{a} &	$\sim$0.0 	&	2550$\pm$150	&	4.0$\pm$0.5 &	-2.72		&		24$\pm$6	& 15,16,17,18 \\		
OTS44			&	$\sim$165		&		-- 	&	Cha I	&	M9.5$\pm$1\tablefootmark{a}		&	0.0 &	--	&	--	&	-3.11	&	$\sim$15 & 	2,19,20\\
KPNO 4		&	$\sim$140	&	--  & Taurus	&	M9.5$^{+0.5}_{-0.25}$\tablefootmark{a}  &	0.0 &  --  &  --  &  -2.64  & $\sim$15  & 20, 21, 22  \\
						&						&		 &				&					& 4		  &		&		&	-2.0		&	10			&	23			\\			
					&						&		 &				&				&	2.45			  &		&		&				&					&	24			\\	
2M0141			&	$\sim$41	&	--	 &	Field	 &		L0\tablefootmark{a}	&	--	&	2000$\pm$100	&	4.0$\pm$0.5	&	-3.5 & 12$^{+13}_{-6}$ & 25,26 \\	
TWA 5B				&	50		&	1.96 &	TWA	&		M8.5--M9\tablefootmark{a}  & --	&			2750$\pm$200			&			3.9$\pm$0.6		&	-2.68$\pm$0.07	&	20--25 	& 27,28,29, 47\\
Gl 417 B\tablefootmark{d}			&		21.7		&	$\sim$92	&	 Field		&	L4.5\tablefootmark{a} 	&	-- 	&			1800--1900								&				5.5--6.0						&						--		&		35$\pm$15			&	30,31,32,33 \\
2M0345			&	26.9 	&		-- 	&		UMa 	&	L0\tablefootmark{a}	 &	 --	 &	 2000	 & 	6 	&	 -3.60	 & 	--	 & 	34,35,36,38,39 \\	
						&				&				&					&									 &			 &	 1900	 &		5.5&				&			&		37 \\
						&				&				&					&									 &			 &	 2200	 &		5.0&				&			&		40	\\										
CT Cha b          &   $\sim$165  & 2.67  &  Cha I  &  M8-L0  &  $5.2\pm0.8$   &  $2600\pm250$  &   3.5   &   $-2.68\pm0.21$   &   8-23   &   42  \\   
USco 108 B   &   145   &  4.6 &  USco  &  M9.5\tablefootmark{a}   &   --    &  --   &   --  &  $-3.14\pm0.20$   &   $14^{+2}_{-8}$   &   43 \\  
HR7329 B   &   $48.2$   &  4.2  &  $\beta$ Pic  &  M7-M8\tablefootmark{a}   &   --    &   --   &   --  &  $-2.63\pm0.09$   &   20-50   &   44, 45 \\  
\hline
\end{tabular}
\end{minipage}
\textbf{References:} (1) \cite{2005ApJ...635L..93L}, (2) \cite{2007ApJS..173..104L}, (3) \cite{2003ApJ...599..342S}, (4)  \cite{2009MNRAS.394.1925V}, (5) \cite{2008hsf2.book..757T}, (6) \cite{2010A&A...512A..52B}, (7) \cite{2005A&A...438L..29C}, (8) \cite{2005ApJ...620..984I}, (9) \cite{2006ApJ...649..894L}, (10) \cite{2003A&A...404..157C}, (11) \cite{2005A&A...430.1027C}, (12) \cite{2004A&A...420..647N}, (13) \cite{2009A&A...506..799B}, (14) \cite{2009A&A...503..281T}, (15) \cite{2002ApJ...575..484G}, (16) \cite{2009ApJ...696.1589H}, (17) \cite{2008A&A...477L...1D},  (18) \cite{2007ApJ...657.1064M}, (19) \cite{1998Sci...282.1095T}, (20) \cite{2004ApJ...617..565L}, (21) \cite{2002ApJ...580..317B}, (22) \cite{2004ApJ...600.1020M}, (23) \cite{2006ApJ...649..306K}, (24) \cite{2007A&A...465..855G}, (25) \cite{2006ApJ...639.1120K}, (26) \cite{2009AJ....137.3345C}, (27) \cite{1999ApJ...512L..69L}, (28) \cite{2000A&A...360L..39N}, (29) \cite{2010ApJS..186...63R}, (30) \cite{2000AJ....120..447K}, (31) \cite{2001AJ....121.3235K}, (32) \cite{2008ApJ...689.1295K}, (33) \cite{2009A&A...503..639T}, (34) \cite{1997ApJ...476..311K}, (35) \cite{1999ApJ...519..802K}, (36) \cite{2001ApJ...548..908L}, (37) \cite{2001ApJ...555..368S}, (38) \cite{2002AJ....124.1170D}, (39) \cite{2007MNRAS.378L..24B}, (40) \cite{2007A&A...473..257L}, (41) \cite{2008ApJ...675.1375L}, (42) \cite{2008A&A...491..311S}, (43) \cite{2008ApJ...673L.185B}, (44)  \cite{2000ApJ...541..390L}, (45)  \cite{2011MNRAS.416.1430N}, (46) \cite{2013ApJ...762...88M}, (47) \cite{2013ApJ...762..118W}.
\tablefoot{ 
\tablefoottext{a}{Optical spectral types.}
\tablefoottext{b}{From synthetic spectra.}
\tablefoottext{c}{Total dynamical mass of the system.}
\tablefoottext{d}{\cite{2003AJ....126.1526B} resolved the companion Gl 417 B as a tigh binary. \cite{2008ApJ...689.1295K} classified this object L4.5pec in the optical and suggest it is composed of a pair of L4 + L6/L6.5 or L4.5 + L6.5/L7 brown dwarfs using a relation between M$_{I}$ and spectral types.}
}
\end{table*}


\section{Observations and data reduction}

\subsection{Observations}
Targets were observed from October 19, 2006 to May 19, 2009 with the Spectrograph for INtegral Field Observations in the Near Infrared \citep[SINFONI, see][]{2003SPIE.4841.1548E, 2004Msngr.117...17B} mounted at the VLT/UT4. SINFONI  provides adaptive-optics assisted integral field spectroscopy in the near-infrared. It  is composed of a SPectrograph  for Infrared Faint Field Imaging (SPIFFI)  fed  by  a modified version of the Multi-Applications Curvature Adaptive Optics system \citep[MACAO, see][]{2003SPIE.4839..329B}.

We used the gratings J and H+K to cover the 1.1--1.4 and 1.45--2.45 $\mu$m ranges at a spectral resolution of $\sim$2000 and $\sim$1500 respectively (see Table~8). We observed AB Pic b, DH Tau B, GSC08047  B, and TWA 5B in AO-mode together with the SPIFFI 25 $\times$ 12.5 pre-optics to resolve the companions. Primaries provided a reference for the wavefront sensing. They were kept outside the narrow field of view (FoV) of the instrument (0.8''$\times$0.8''). Each sequence is composed of 5 to 8 acquisitions on the source with small dithering and one acquisition on the sky. A similar strategy was followed for the observations of HR7329 B and USCO CTIO 108B but using the  pre-optics providing spatial pixels of 100 $\times$ 50 mas. The instrument was used with the 250 $\times$ 125 pre-optics in seeing-limited mode (NOAO) for the isolated targets (2M0141, KPNO Tau 4, 2M0345, Cha1109, OTS 44) and for Gl 417 B. Sequences were composed of 4 to 8 exposures each. Between exposures, small offsets were applied to dither the source over the 8$\times$8'' field of view.  Finally, we also observed B and G type stars following a AB pattern right after our science targets to correct our spectra from telluric features.

 \begin{figure}[t]
   \centering
   \includegraphics[width=\linewidth]{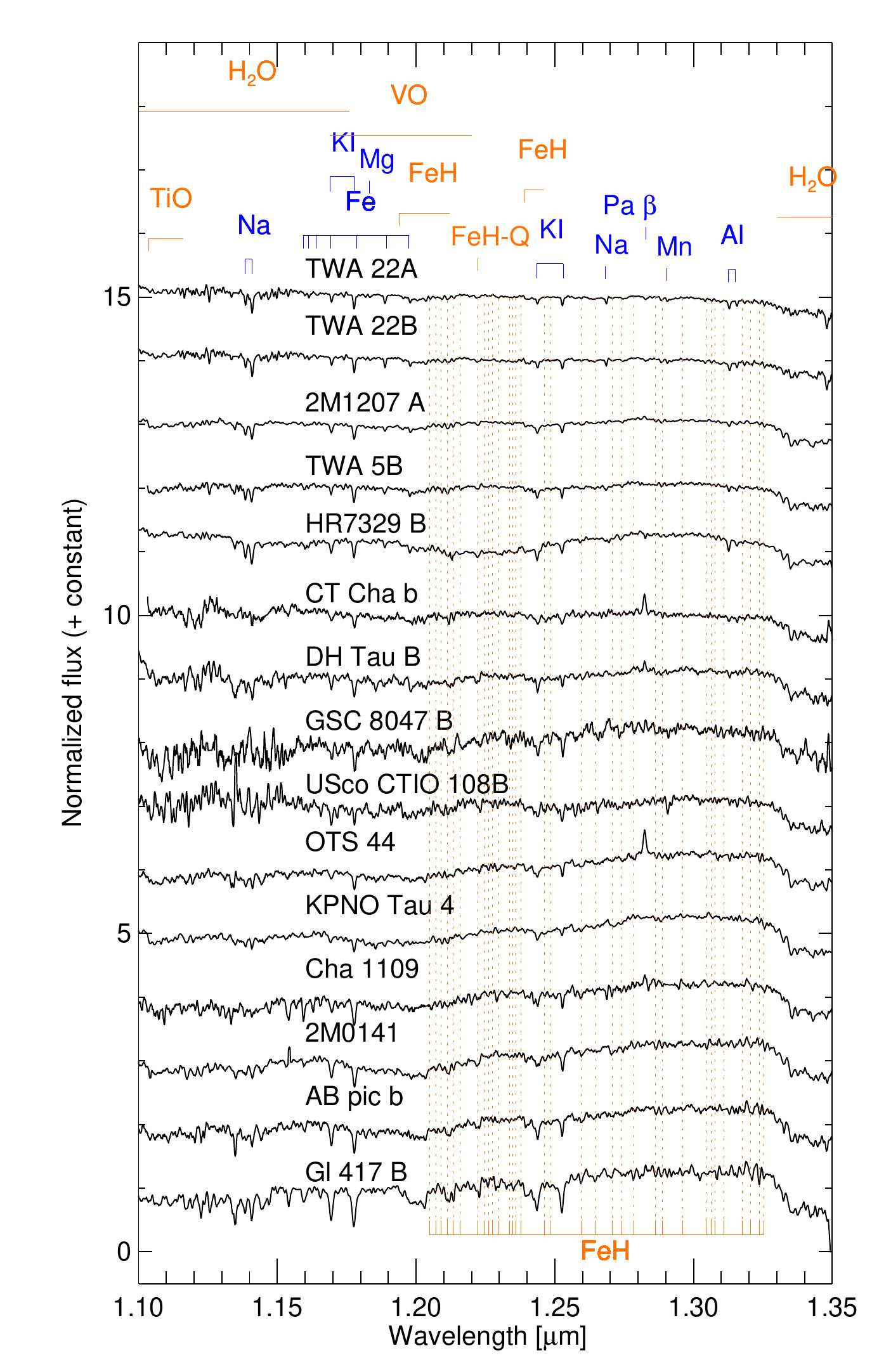}
      \caption{J band spectra (1.1--1.35 $\mu$m) of the young M6--L4.5 objects observed  with SINFONI. Molecular absorptions are indicated in green and atomic features are reported in blue.}
         \label{FigJband}
   \end{figure}

   \begin{figure}[t]
   \centering
   \includegraphics[width=\linewidth]{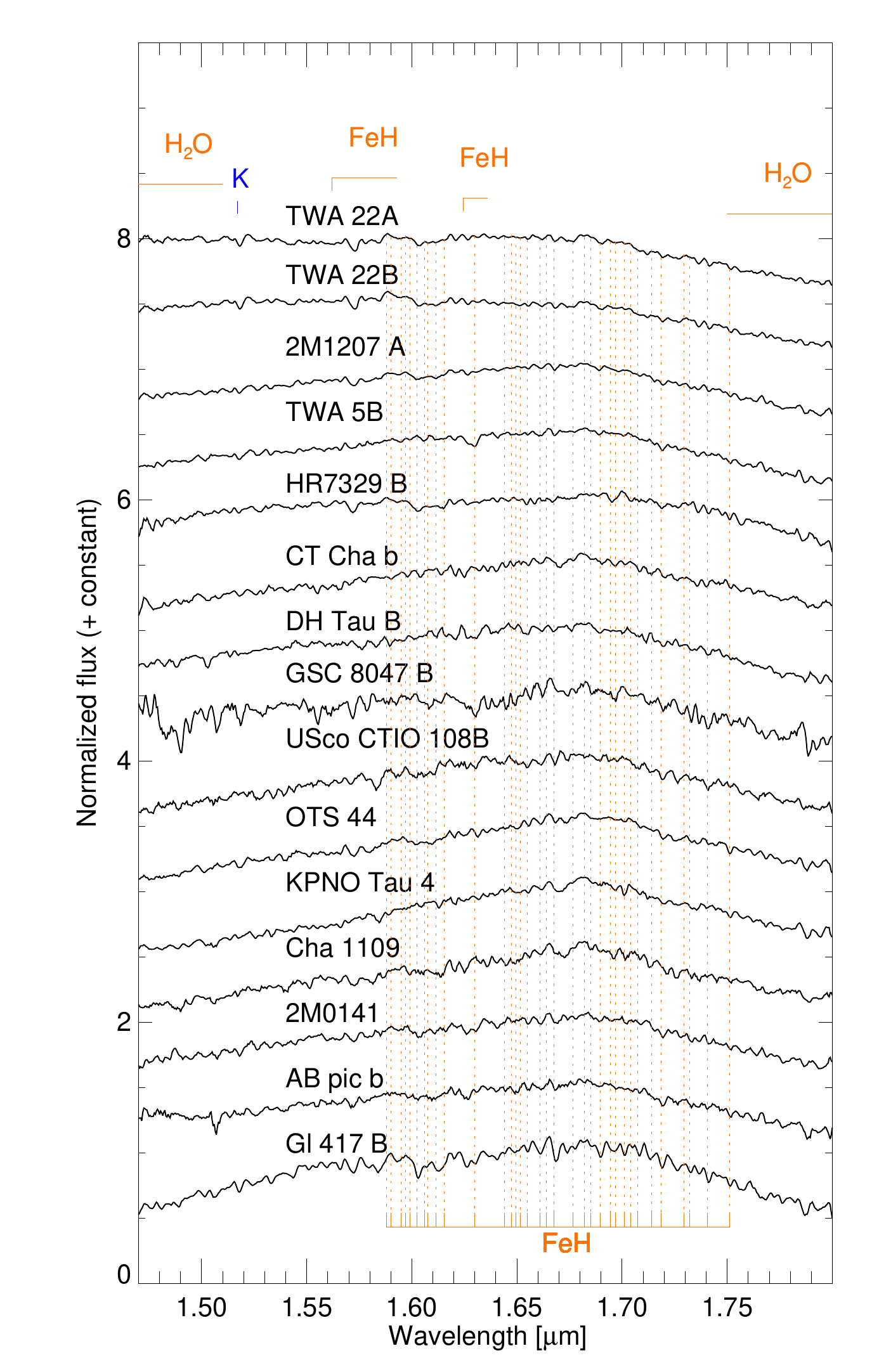}
      \caption{Same as Fig. \ref{FigJband} but for the H band (1.47--1.80 $\mu$m).}
         \label{FigHband}
   \end{figure}

   \begin{figure}[t]
   \centering
   \includegraphics[width=\linewidth]{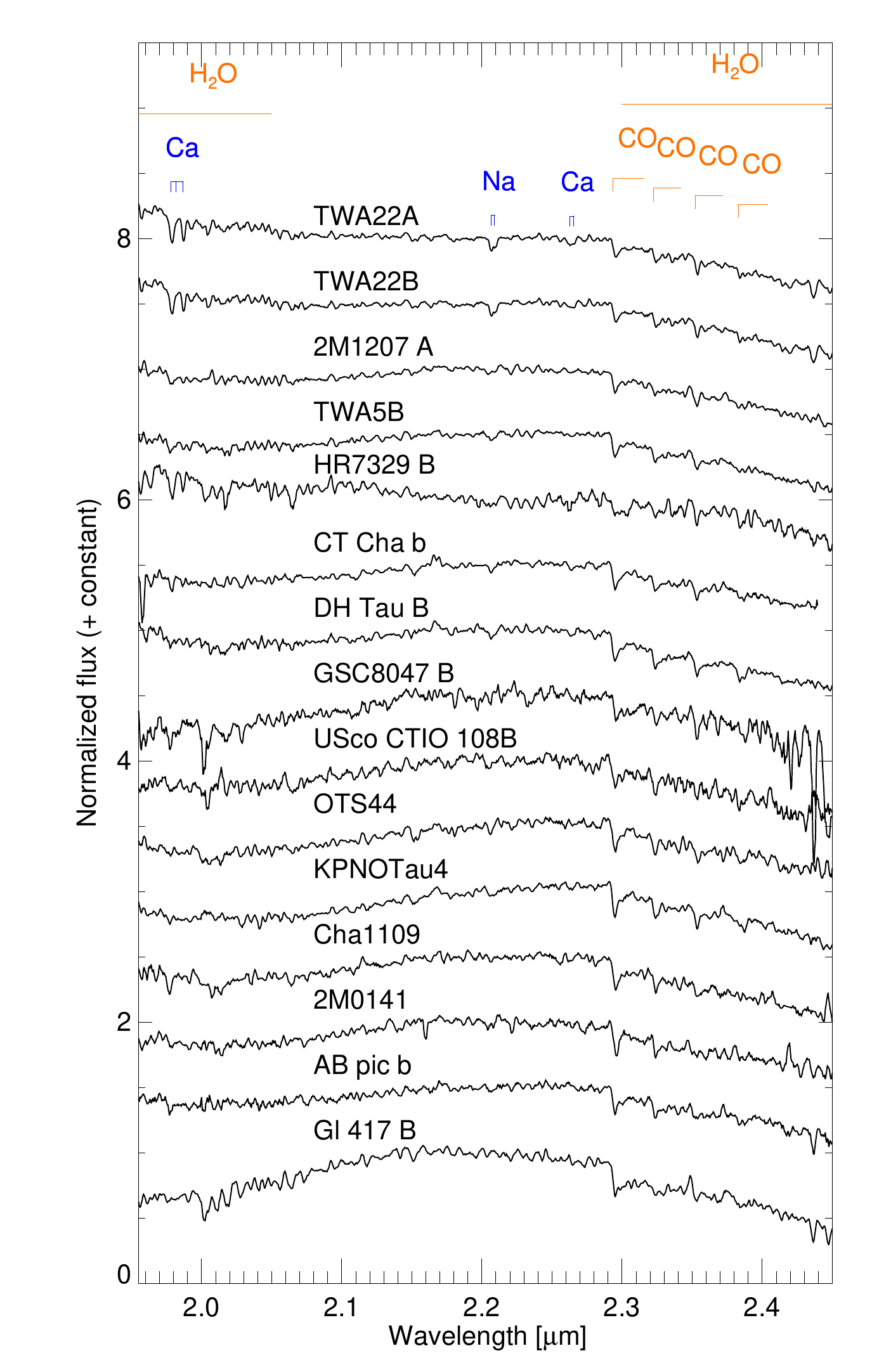}
      \caption{Same as Fig. \ref{FigJband} but for the K band (1.955--2.45 $\mu$m). }
         \label{FigKband}
   \end{figure}

\subsection{Data reduction}
\label{subsec:datared}
We reduced homogeneously our dataset with the ESO data reduction pipeline version 1.9.8 \citep{2006NewAR..50..398A}. We used in addition custom scripts to suppress several electronic effects (odd even effect affecting slitlet n$^{\circ}$25, negative rows) affecting our raw images.  The pipeline successfully carried out bad pixels detection, flat fielding, distortion coefficient computation, and wavelength calibration using calibration frames acquired the day following the observations.  In NOAO-mode, a synthetic sky frame was built from the median of the input object frames. Individual datacubes were reconstructed from sky-subtracted object frames and merged into a final mozaicked datacube. Datacubes of telluric standard stars (STD) were obtained in a similar way.
We noticed that raw science frames with integration time greater than 100s were affected by a fluctuation of the  mean  flux count level (up to 15\% of the maximum flux level of the source) of the SPIFFI detector. This fluctuation  also appears in the dark frames taken during day-time. It can not be removed by the mean of sky or dark-subtraction and is amplified when the pipeline correct the on-off (or sky-object) frames by the detector gain. This effect \textit{can severely bias the spectral slope of the spectra, and consequently the atmospheric parameter estimates of the objects}. We then carefully removed this fluctuation in the raw frames computing the median flux level in six non-iluminated columns on the left of the chip before applying any pipeline recipes. 
The quality of the results was monitored by eye and using the ESO quality control parameters\footnote{The ESO quality control parameters are criteria used on data to know the instrument status, monitor  it over time, and to evaluate the data quality. They can be found at \texttt{http://www.eso.org/\-obs\-er\-ving/dfo/\-qua\-li\-ty/SIN\-FONI/qc\-/qc1.html}. We used them to check the consistency of the number of non-linear pixels found, the distortion solution, the median inter-slitlet distances on the detector, and the dispersion solution for each sources and setup.}. Residual sky-variations were estimated and subtracted across the FoV in the datacubes of GSC08047 B, AB Pic b, and 2M0141. Finally, we corrected all the datacubes obtained in AO-mode from the wavelength-dependent shift of the sources in the FoV induced by the atmospheric refraction. 

     \begin{figure}[t]
   \centering
   \includegraphics[width=\linewidth]{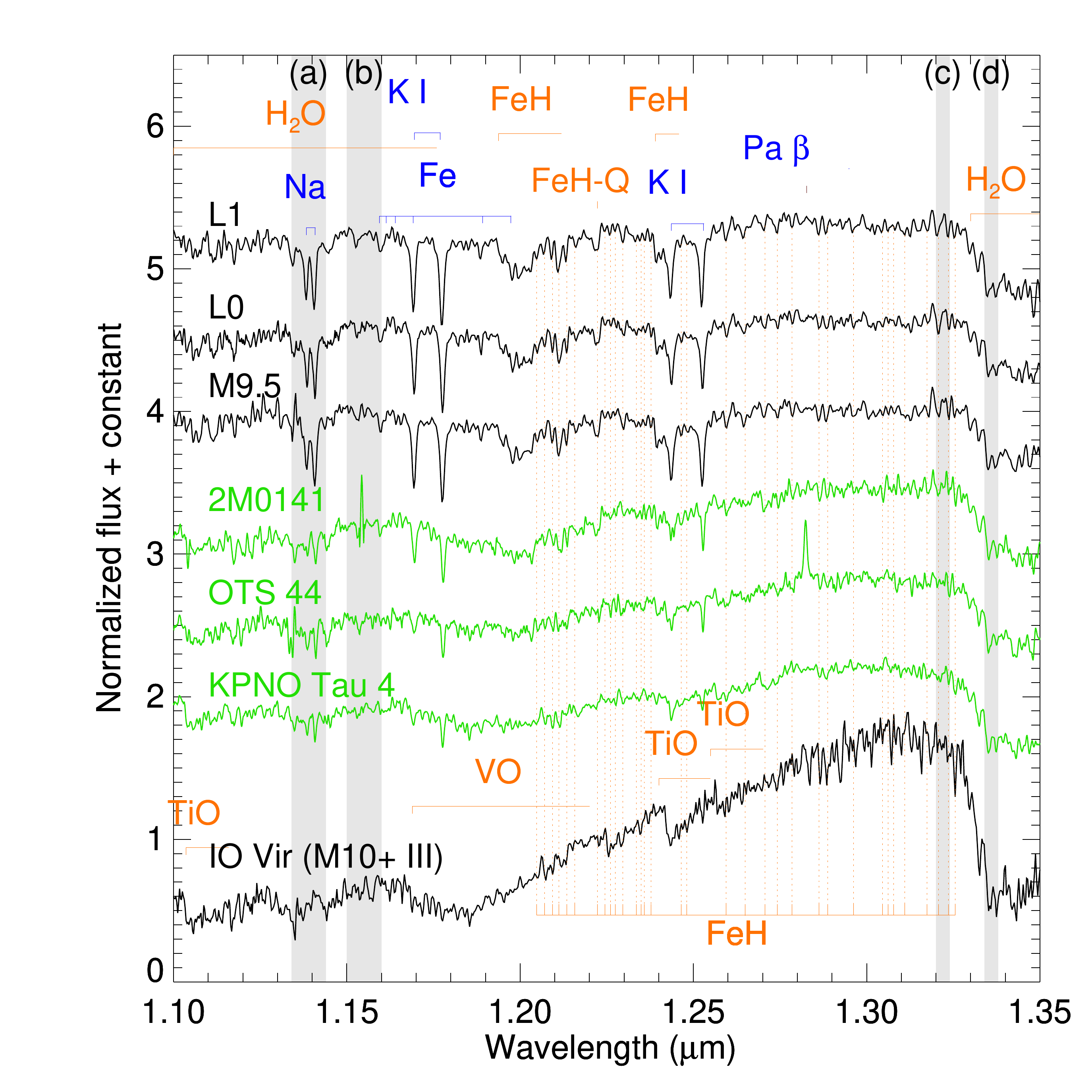}
      \caption{Comparison of the J band SINFONI spectra (red) of 2M0141 (L0), OTS44 (M9.5), and KPNO Tau 4 (M9.5) to R$\sim$2000 templates (black) of  M9.5--L1 field dwarfs (BRI B0021-0214, C05; 2M0345, ML03; 2MASS J14392836+1929149, C05) classified in the optical, and to the spectrum of a very late type giant (IO Virginis, R09). The flux ratio  inside the gray zones \textit{(a)} and \textit{(b)} defines the Na index defined by A07 and used in part 3.1.2. The flux ratio of gray zones \textit{(c)} and \textit{(d)} defined by G03 measures the depth of the water band longward of 1.3 $\mu$m. It can be used to classify young M dwarfs but appears slightly age-dependant at the M--L transition.}
         \label{AgeseqJ}
   \end{figure}

  \begin{figure}[t]
   \centering
   \includegraphics[width=\linewidth]{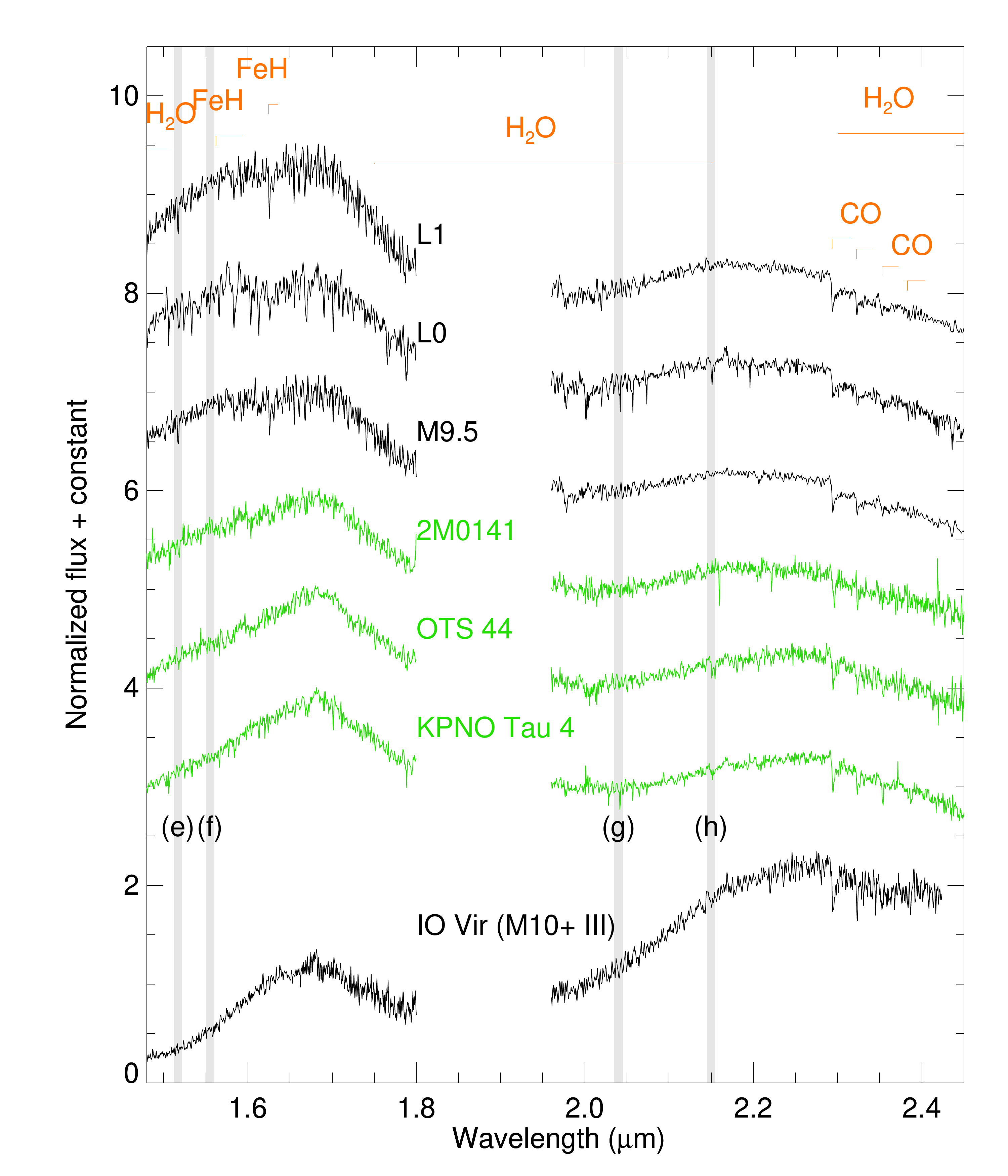}
      \caption{Same as Figure \ref{AgeseqJ} for the H and K bands. Flux ratio of the (e) and (f) zones, and of the (g) and (h) zones define the spectral indices H$\mathrm{_{2}}$O--1.5$\mu$m and H$\mathrm{_{2}}$O 2 used in Fig. \ref{Figindices}. They vary only slightly with the age from 1 Myr and can be used to confirm the spectral type of our targets.}
         \label{AgeseqHK}
   \end{figure}

\subsection{Spectral extraction}
The J band datacubes of TWA 5B, CT Cha b, and HR7329 B are contaminated by speckles crossing rapidly the FoV overlaid on a smooth flux gradient arising from their host stars. A similar pattern is present into the H+K band datacubes of these sources and  in those of DH Tau B, and GSC08047B.  We then developed and apply a custom spectral extraction algorithm to evaluate the impact of these contributions to the final spectra, close to that used by \cite{2011ApJ...733...65B} to decontaminate the datacubes of HR8799b. We build and subtracted a model of the companion flux distribution at each wavelength using the telluric standard star datacube as an input point spread function.  The residual datacube was then smoothed using median filtering in the spatial plane to create a model of the smooth gradient (or halo). The model was subtracted from the original datacube. This step was repeated until reaching a suitable fit. We finally applied a modified version of the spectral deconvolution algorithm \citep{2002ApJ...578..543S, 2007MNRAS.378.1229T} on the datacubes of DH Tau B, GSC08047B, CT Cha b, and TWA 5B to suppress all the remaining low-frequency chromatic features. The algorithm produces final datacubes cleaned from most of the contamination and shows that the contribution of the speckles to DH Tau B, GSC08047B, and TWA 5B's final spectra is neglectable. We applied the procedure to the J and H+K band datacubes of HR7329 B but noticed a residual flux level likely to be the difraction spikes of the primary star. Some residual contamination could remain in the spectrum of this object.

We also re-extracted the J and H+K band spectra from the datacubes of the tight binary TWA 22AB obtained by \cite{2009A&A...506..799B}  on February 02, 2007 using an upgraded version of the CLEAN-based algorithm presented in the aforementioned paper, and recently applied on the datacubes of the Z CMa system (Bonnefoy et al. 2013, submitted). The new algorithm now conserves the spatial information during the extraction and reduces differential flux losses that limited the spectral analysis so far. The algorithm converges in less than two iterations. The extraction error is estimated to be less than 5\%. 

Finally, we integrated the source flux in each final datacube over apertures selected to optimize the signal to noise ratio without introducing differential flux losses. Some of our spectra were suffering from strong sky-line residuals that were flagged using the spectral atlas of \cite{2000A&A...354.1134R} and removed. Spectra of B-type  standards star were corrected from their intrinsic features and divided by black bodies at the appropriate temperature \citep{1991Ap&SS.183...91T}. G-type standard star spectra were divided by reference spectra (smoothed at R=1500 in the H+K band) of G-type stars from the IRTF spectral library\footnote{\texttt{http://irtfweb.ifa.hawaii.edu/\-$\sim$spex/\-IR\-TF\_Spec\-tral\_Li\-bra\-ry/}}  \citep[][hereafter R09]{2009ApJS..185..289R}. The science spectra were then divided by standard star spectra, normalized to the continuum, and averaged to get master spectra. Finally,  we connected the J and H+K band spectra of the targets with accurate NIR magnitudes (i.e. 2M0141, 2M0345, 2M1207 A, Cha1109, KPNO Tau 4, OTS44, DH Tau B, USco CTIO 108B, and Gl417B) using appropriate filter pass bands and a model spectrum of Vega \citep{1985A&A...151..399M, 1985IAUS..111..225H}. The 1.1-2.5 $\mu m$ spectra of 2M0141 and 2M1207 A closely match the low resolution (R$\sim$75--120) JHK band spectra of the SpecXPrism library\footnote{\url{http://pono.ucsd.edu/~adam/browndwarfs/spexprism/}}, thus validating the scaling procedure. Finally, the good agreement between the pseudo continuums of the  J+H and K band low resolution (R$\sim$300--440) spectra\footnote{\url{http://harbor.scitec.kobe-u.ac.jp/~yitoh/index.html}} of DH Tau B gathered by  \cite{2005ApJ...620..984I} and of the SINFONI spectrum enable to check that the spectral extraction have altered the spectral slope. 


  \begin{figure}
   \centering
   \includegraphics[width=\columnwidth]{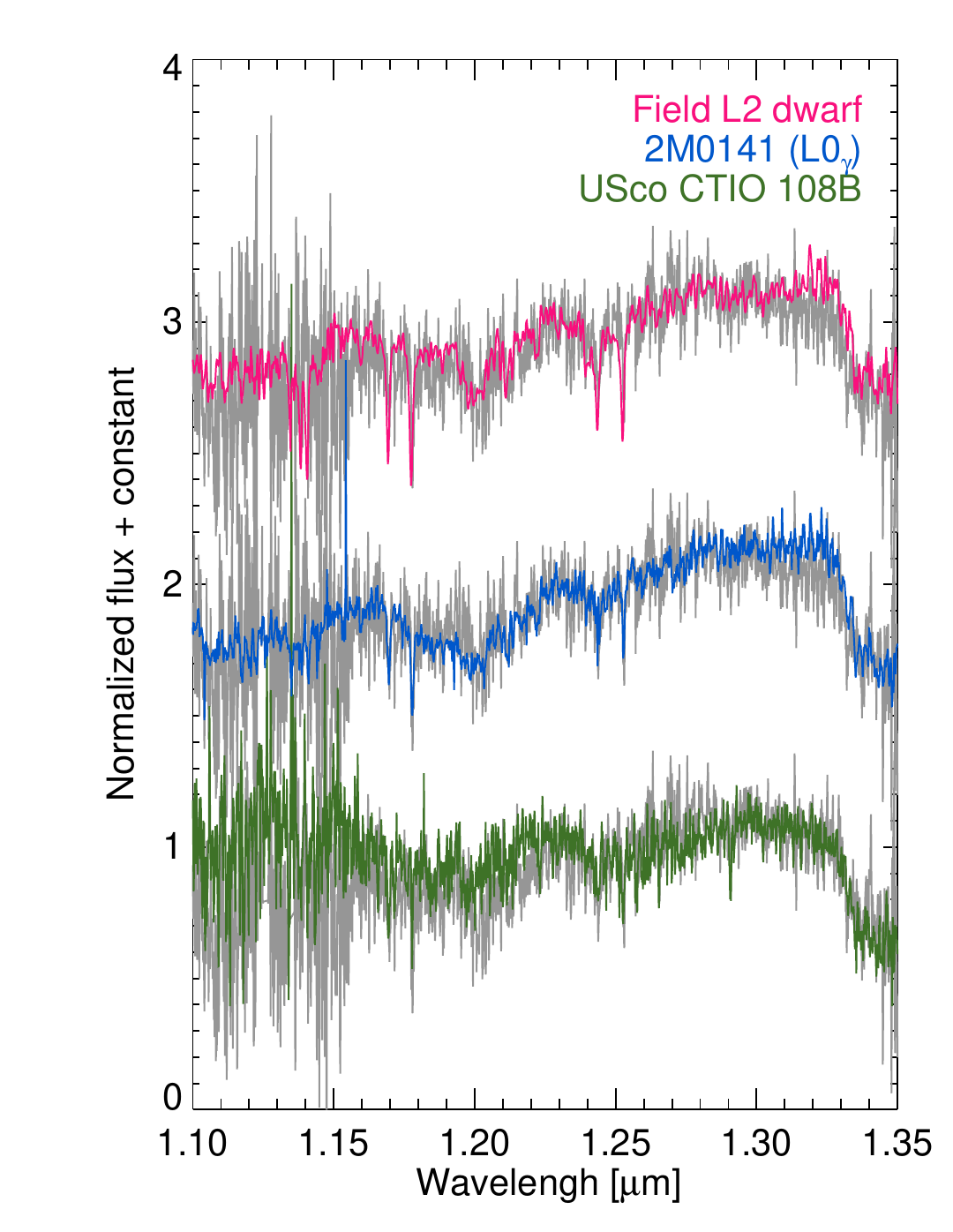}
      \caption{J-band spectrum of the companion to GSC 08047-00232 (grey) compared to the spectrum of the young L0 dwarf 2MASS J01415823-4633574 \citep{2006ApJ...639.1120K}, of the old field L2 dwarf 2MASS J0015448+351603 \citep{2000AJ....120..447K}, and of the young M9.5 companion USco CTIO 108B \citep{2008ApJ...673L.185B}.}
         \label{FigGSC8047BJ}
   \end{figure}

   \begin{figure}[t]
   \centering
   \includegraphics[width=\linewidth]{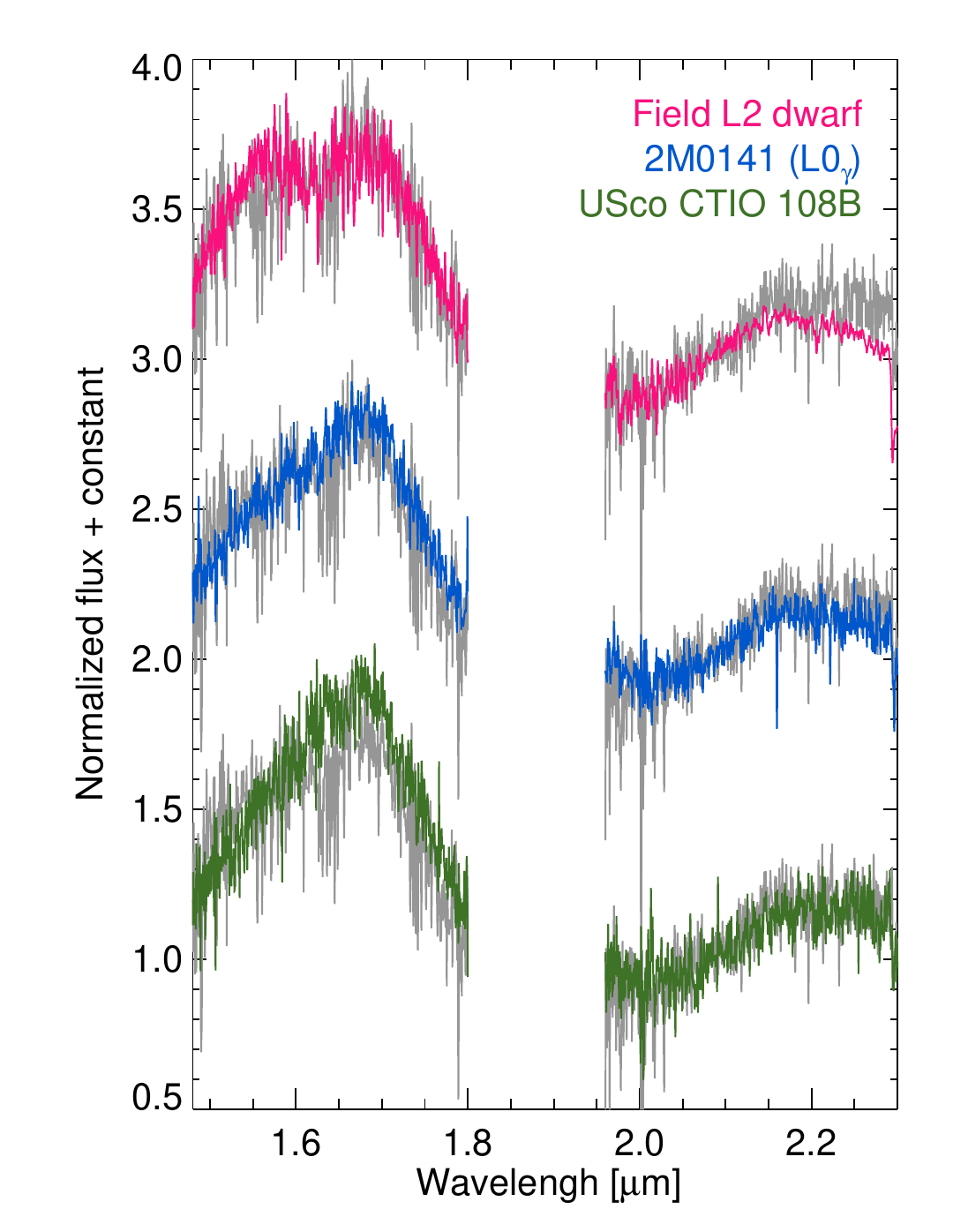}
      \caption{Same as Fig. \ref{FigGSC8047BJ} but for the H+K band.}
         \label{FigGSC8047BHK}
   \end{figure}


\section{Results}
\label{section:results}
\subsection{Empirical analysis}
\label{empanalysis}
\subsubsection{Line identification}
\label{lineID}
The SINFONI spectra of young dwarfs are presented in Figs. \ref{FigJband}, \ref{FigHband}, and \ref{FigKband}. Apart for GSC08047B, all the spectra have sufficient SNR (30 to 150 from 1.20 to 1.32 $\mu$m; 30 to 80  from 1.50 to 1.60 $\mu$m) to allow studying weak atomic and molecular features and their subtle variations with the object age and spectral type. 

The J band spectra are dominated by two broad molecular absorptions of H$_{2}$O  shortward of 1.17 $\mu$m and longward of 1.33 $\mu$m \citep{1967ApJS...14..171A}, and FeH at 1.194 $\mu$m and 1.239 $\mu$m \citep{1987ApJS...65..721P}. We also retrieve several narrow absorptions of FeH in the spectra of TWA 22 AB, TWA 5B, KPNO Tau 4, OTS 44, 2M0141, USco CTIO 108B, CT Cha b, HR7329 B, AB Pic b and Gl 417 B that were previously identified in the spectra of field M--L dwarfs \citep{2003ApJ...582.1066C}. 

The doublets of Na I and K I at 1.138 $\mu$m, 1.169 $\mu$m, 1.177 $\mu$m, 1.243 $\mu$m, and 1.253 $\mu$m represent the overwhelming atomic features in the J band. The spectra of TWA22 A \& B contain as well Mg I (1.183 $\mu$m), Fe I (1.159, 1.161, 1.164, 1.169, 1.178, 1.189, 1.197 $\mu$m), Na I (1.268 $\mu$m), Mn I (1.290 $\mu$m), and Al I (1.3127 and 1.3154 $\mu$m doublet) lines. Al I absorptions are still present in the spectra of 2M1207 A, HR7329 B, CT Cha b,  and TWA 5B. They are not found in field dwarfs spectra of L dwarfs (see Fig. 1 of ML03) and tend to increase with age in the spectra of M6 dwarfs \citep[e.g. Fig. 3 of][]{2010ApJS..186...63R}. These lines already suggest that these objects are M dwarfs, in agreement with their optical spectral types (see Tab. \ref{table:1}).

J-band spectra of the M9.5 dwarfs OTS 44, KPNO Tau 4, and of the L0 dwarf 2M0141 look like those obtained by \cite{2004ApJ...617..565L}, \cite{2004ApJ...600.1020M}, and K06. They are compared in Fig. \ref{AgeseqJ} to spectra of M9.5--L1 field dwarfs and to the spectrum of a late M giant \citep[IO Virginis,][]{2004ApJ...600.1020M}. The three young dwarfs show features intermediate between those of the giant and of the dwarfs. KPNO Tau 4 (Taurus, $\sim 1$ Myr) displays a marked absorption of TiO  at 1.103 $\mu$m (0--0 band of $\Phi$ ( $b$ $^{1}\Pi$--$d$ $^{1}\Sigma$)), found in the spectra of the giant (R09), and diminished in the spectrum of the $\sim$ 3 Myr OTS 44 and of 2M0141 \citep[likely to be $\sim$30 Myr old, see the discussion in Fig. 2 of][]{2009ApJ...703..399L}. The FeH bands at 1.194 $\mu$m, 1.222 $\mu$m, and 1.239 $\mu$m, strong in the spectra of field dwarfs, decrease slightly from 1 Myr to 3 Myr, and are absent in the spectrum of IO Virginis. K06 invoked the broad depression of VO, from 1.17 to 1.20 $\mu$m in the giant spectrum, to explain the blue slope that surrounds the K I doublet seen in our young objects spectra. The K I lines at 1.169, 1.177, and 1.253 $\mu$m are significantly stronger  in the spectrum of 2M0141 (equivalent widths variation greater than 8, 14, and 5$\sigma$ respectively) than in those of KPNO Tau 4 (and also of OTS 44). The  lines at 1.169 $\mu$m and 1.253 $\mu$m are the most sensitive to the age as they appears weaker in the spectrum of the 1 Myr dwarf than at 3 Myr (see also table \ref{table:3b}). Na I (1.138 $\mu$m) and K I (1.169/1.177 $\mu$m and 1.243/1.253 $\mu$m) doublets are weaker in the spectra of these objects than in the spectra of field dwarfs analogues (equivalent widths difference greater than 5$\sigma$). Finally, the very red slope of the young object spectra  have been interpreted by K06 as a lowering of the  collision-induced absorption (CIA) of H$_{2}$ which usually cover the JHK band \citep[see the left panels of Fig. 5 of][]{1997A&A...324..185B}.  The J band spectra of Cha1109,  USco CTIo 108 B, and AB Pic B have similar features, consistent with being young $\geq$ M9.5 dwarfs. Finally, the DH Tau B spectrum has a bluer continuum, weaker H$_{2}$O and alkali lines than KPNO Tau 4 and OTS44. These differences could be interpreted by an increase of the effective temperature. This put an upper limit of M9.5 for the spectral types of this companion. 

Our H+K band spectra exhibit several features characteristic of young objects. The prominent one is the triangular shape of the H band \citep[e.g.][K06]{2001MNRAS.326..695L}. The FeH absorptions at 1.583--1.591, 1.625 $\mu$m  \citep{2001ApJ...559..424W}, and over the 1.588--1.751 range \citep{2003ApJ...582.1066C} that are clearly seen in the spectra of old field M-L dwarfs are absent here. The water absorptions from 1.48 to 1.51 $\mu$m that deeply carves the band appear stronger in the spectra of the young objects classified in the optical when compared to field dwarfs counterparts. The K I line at 1.517 $\mu$m, the Ca I absorptions around 1.98 $\mu$m, and the Na I doublet at 2.206 and 2.209 $\mu$m are the strongest atomic lines in the H+K band spectra of M5--L1 field dwarfs (C05, ML03). Their non-detection in our spectra (apart for TWA 22) suggests that our objects are either later than $\sim$M5 or/and young. 

We analyzed the spectral features of OTS 44, KPNO Tau 4, and 2M0141 in their H and K bands following the same method as for the J band (Fig. \ref{AgeseqHK}). We note the progressive reddening of the band, the strengthening of  the absorption from 1.3 to 1.6 $\mu$m, and the reduction of the negative slope  from 2.2 to 2.3 $\mu$m,  that can all be attributed to cooler line forming regions and reduced CIA of H$\mathrm{_{2}}$ \citep[][K06]{1997A&A...324..185B, 2001ApJ...556..357A}. The depth of the water bands from 1.70 to 1.80 $\mu$m, and from 1.95 to 2.15 $\mu$m is identical in the spectra of young objects (see Fig. \ref{AgeseqHK}). These peculiar features are also present in the spectra of Cha 1109, USco CTIO 108B and AB Pic b, and confirm their late type nature as well as their young age. 

To conclude, the spectra of OTS44 and  CT Cha b have a noticleable Paschen $\beta$ emission line. A similar -- but weaker -- line might also be present in the spectra of 2M1207A,  and DH Tau B. This line has already been observed in the spectra of GQ Lup b\footnote{\cite{2007ApJ...656..505M} and \cite{2009ApJ...704.1098L} do not find this line in spectra of this companion taken with two other integral-field spectrographs at different epochs.} \citep{2007A&A...463..309S} and GSC 06214-00210 b \citep{2011ApJ...743..148B}. We study the characteristics and possible origin of this line for OTS 44  in a forthcoming paper (Joergens, Bonnefoy et al.  in prep).
     
\subsubsection{Comparison to spectral templates}
\label{classoptical}
Spectral types inferred from the comparison of our spectra to those of field dwarfs would be systematically too late because of the strengthening of water absorptions and the effect of reduced CIA at young ages \citep{1999ApJ...525..440L, 2004ApJ...617..565L}. To get accurate spectral types, we then attempted to classify our targets following the approach of \cite{2004ApJ...617..565L} and based on the comparison to young dwarfs spectra classified at optical wavelengths. 

The spectra of DH Tau B  and CT Cha b were compared to low-resolution (R$\sim$100) templates of young (1--3 Myr) optically classified M5-M9 members of the Taurus  and IC348 \citep{2007AJ....134..411M}  deredened by their extinction values \citep{1999ApJ...525..466L, 2002ApJ...580..317B, 2004ApJ...617.1216L, 2007A&A...465..855G}. We used in addition medium resolution (R$\sim$2000) spectra of M5-M7 Taurus and  $\rho$ Ophiucus objects \citep{2009ApJ...697..824A, 2010ApJS..186...63R} and low-resolution  (R$\sim$300--350) J band spectra of 1--3 Myr old objects obtained by \cite{2003ApJ...593.1074G}. Finally, we also used spectra of OTS 44, Cha1109, and KPNO Tau 4 as templates of M9.5 dwarfs. We assume A$\mathrm{_{V}}$=0 for KPNO Tau 4 and OTS 44 \citep{2002ApJ...580..317B, 2007ApJS..173..104L}, and A$\mathrm{_{V}}$=1.1 for Cha1109 \citep{2005ApJ...635L..93L} for the following analysis of the spectra.

We used templates of older objects for the targets AB Pic b, USco CTIO 108B, HR 7329 B, and GSC8047B. We then selected a sample of optically classified spectra from the $\sim$5-11 Myr \citep{2012ApJ...746..154P} old  Upper Scorpius \citep{2003ApJ...593.1074G, 2004ApJ...610.1045S, 2008MNRAS.383.1385L, 2009ApJ...697..824A}, and the $\sim$8 Myr old TW Hydrae association \citep{2009ApJ...697..824A, 2010ApJS..186...63R}.  We used in addition the SINFONI spectra of TWA 5B and 2M1207 A as templates of M8.5-M9 and M8.25 dwarfs respectively.  

   \begin{figure}[ht]
   \centering
   \includegraphics[width=\linewidth]{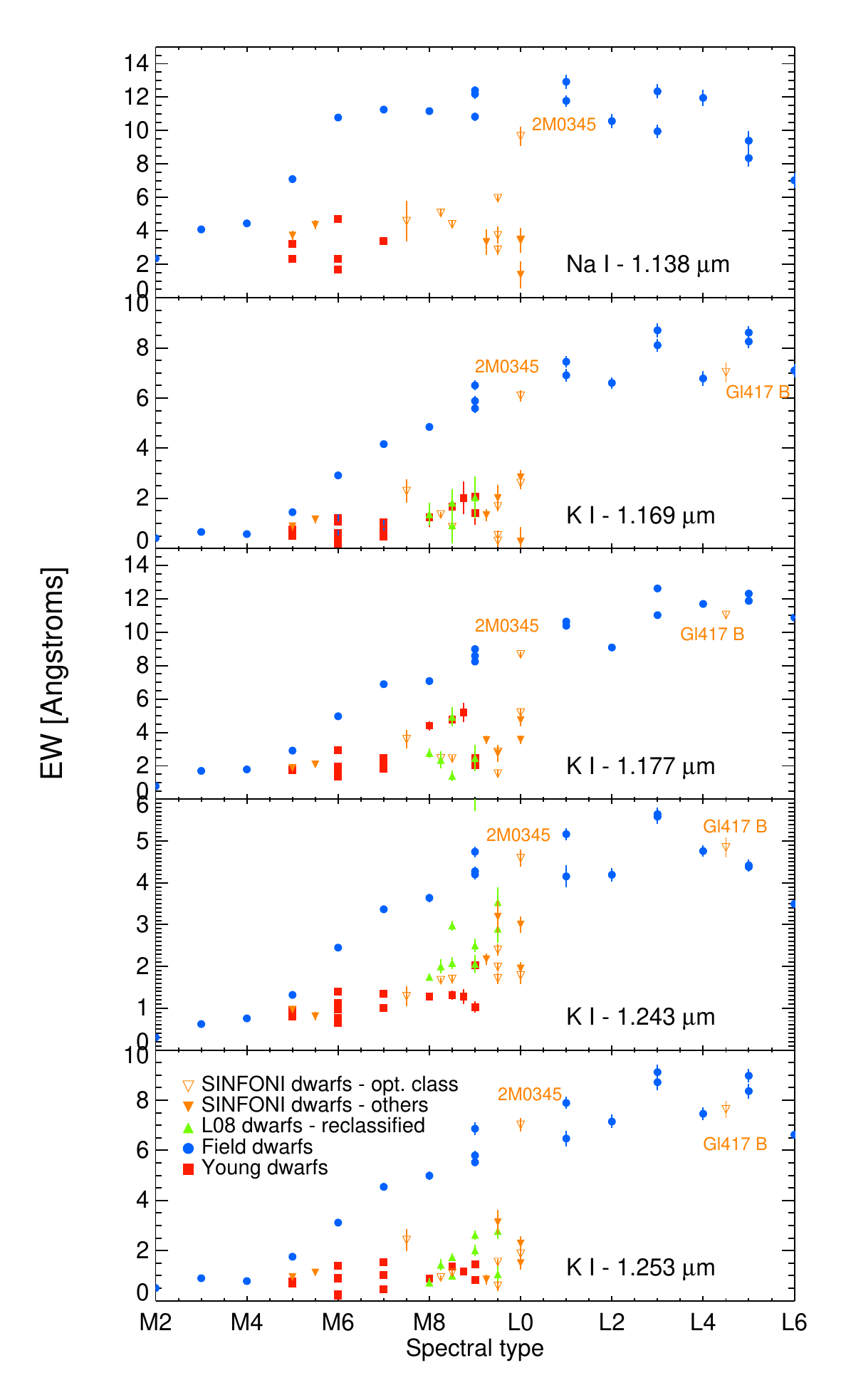}
      \caption{Equivalent widths (EW) of gravity sensitive lines of Na I (1.138 $\mu$m) and K I (1.169 $\mu$m, 1.177 $\mu$m, 1.243 $\mu$m, 1.253 $\mu$m) computed our the SINFONI spectra and on reference spectra (see Fig. \ref{Figindices} for details). Spectra of objects not reported in this plot suffered from too low SNR or artefacts that prevented computing the EW.}
         \label{FigEW}
   \end{figure}

Finally  we used for all our targets the SINFONI spectrum of the young L0 dwarf 2M0141 \citep{2006ApJ...639.1120K} and a SpecX \citep{2003PASP..115..362R} spectrum of the moderately young \citep[20-300 Myr, see][]{2010arXiv1004.3965Z} L3 dwarf G196-3 B \citep{2004ApJ...600.1020M, 2009AJ....137.3345C}.  Although these sources could have reduced water absorption bands compared to younger counterparts, they can be used to place upper limits on the spectral types of our objects. 

At low resolving powers, the JHK band of DH Tau B is best fitted by  M8.5--M9.25 dwarfs with A$_{V}$ $\lesssim$ 2. The comparison with J band spectra at R$\sim$300 clearly has features intermediate between those found in M8.5 and M9.5 spectra. Finally, we noticed that the K I doublets and FeH absorptions are deeper than in the medium resolution spectrum of the M9 Upper Sco member DENIS161103-242642. These features are known to increase at higher ages. Their strength anticorrelates with the spectral types  in M-L field dwarfs spectra. This suggests, together with the red slope, that the companion is later than M9.  We then adopt a spectral type M$9.25\pm0.25$ for DH Tau B. Using the typical colors of young objects given in Tab. 13 of \cite{2010ApJS..186..111L} and the conversion formulae between $\mathrm{E(J-Ks)}$ and A$\mathrm{_{V}}$ of \cite{1998ApJ...500..525S}, we find A$\mathrm{_{V}}$=1.16$\pm$1.05 mag. This value is in good agreement within the error bars with the extinction found for the primary star \citep[A$\mathrm{_{V}}$=0--1.5 mag, see][]{1989AJ.....97.1451S, 2001ApJ...556..265W, 2005ApJ...620..984I} and used by \cite{2006ApJ...649..894L} to derive new estimates for the companion mass. We then de-reddened our spectrum by this amount for the analysis presented in the next sections.

The NIR spectral slope of CT Cha b is roughly compatible with that of M7.25--M8.5 Taurus dwarfs and A$\mathrm{_{V}\leq3.25}$ mag. The J band deredened by  A$\mathrm{_{V}\leq}$ 1 is well fitted by R$\sim$300 spectra of M8 dwarfs members of  IC 348 (IC 348-363, IC 348-355; see G03). However, the H+K band seems compatible with A$\mathrm{_{V}=}$5.2$\pm$0.8 found by C08. We then preliminary adopt M8$^{+0.5}_{-0.75}$ dwarfs and confirm that the NIR spectral slopes are affected by circum(sub)stellar material. 

As noted by B10, the J and H+K band spectra of AB Pic b are clones of 2M0141. The strength of the 1.3 $\mu$m water absorption is comparable to that of field L1 dwarfs and remains lower than that of G196-3 B. This then put an upper limit of L1 on the spectral type of this companion. The spectrum also appears later than an M9 dwarf of Upper Sco. AB Pic b is then confirmed to be a L0$\pm$1 dwarf. Since the spectra are not showing heavy reddening, we assume in addition $\mathrm{A_{V}}$=0.27$\pm$0.02 mag of the primary star \citep{2009ApJ...694.1085V}. 

The J band spectra of GSC08047 B and USco CTIO 108B have close shapes (Fig. \ref{FigGSC8047BJ}). They fall intermediate between the spectrum of 2M0141 and that of a M9 Upper Sco dwarf. We also find a correct match of the GSC08047 B spectrum with that of a L2 field dwarf \citep[2MASS J00154476+3516026, see][]{2000AJ....120..447K} from the NIRSPEC brown dwarf library\footnote{\texttt{http://www.astro.ucla.edu/\-$\sim$mclean/\-BDSSarchive/}} (see Fig. \ref{FigGSC8047BJ}). The H+K band spectra perfectly fit that of  2M0141. The H-band spectrum of GSC08047B is also well reproduced by that of the L2 dwarf  (see Fig. \ref{FigGSC8047BHK}). Nevertheless, the K band continuum $\geq$2.15 $\mu$m of 2M0141 give a much better fit than the L2 field dwarf do for this companion. Since the evolution of this feature is closely related to the surface gravity \citep{2001ApJ...556..357A}, we conclude that GSC08047 B is more likely a young M9.5$\pm$0.5 dwarf. Our conclusion is consistent with the membership of GSC08047 A to the 30 Myr old Colomba association \citep{2008hsf2.book..757T}. We assign a similar spectral class to USco CTIO 108B from this comparison. This classification is consistent with the one derived at optical wavelengths \citep{2008ApJ...673L.185B}.  

\begin{figure}[t]
   \centering
   \includegraphics[width=\linewidth]{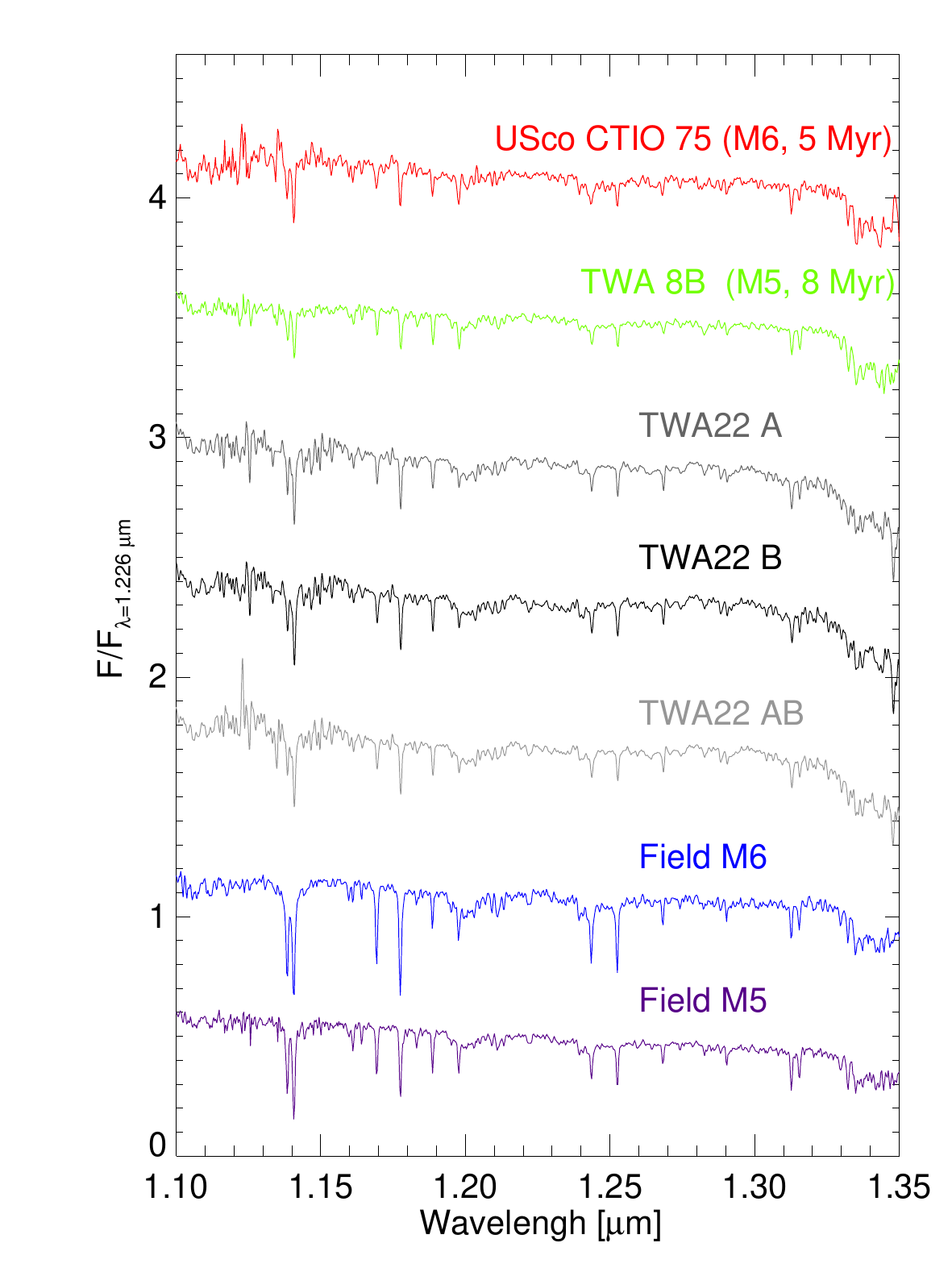}
      \caption{J band spectra of TWA 22A and B, and TWA 22AB compared to young SpecX spectra of young M5-M6 \citep{2010ApJ...715..561A} and old field M5--M6 dwarfs \citep{2005ApJ...623.1115C}.}
         \label{TWA22J}
   \end{figure}

\begin{figure}[t]
   \centering
   \includegraphics[width=\linewidth]{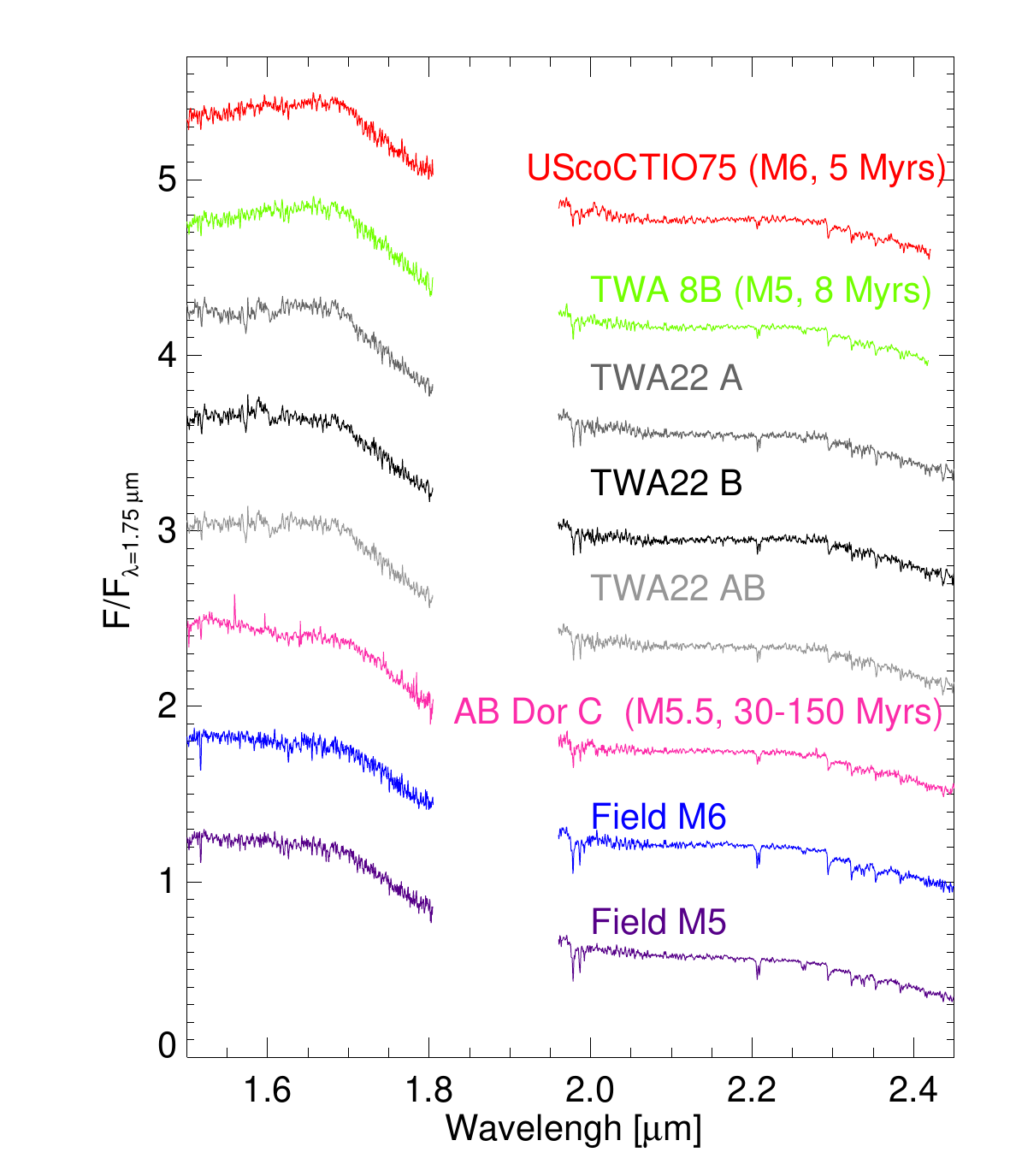}
      \caption{Same as Fig.  \ref{TWA22J} but for the H+K band. We also add here the spectrum of the M5.5 dwarf AB Dor C \cite{2007ApJ...665..736C} which is currently estimated to be 50--150 Myr old \citep{2005Natur.433..286C, 2005AN....326.1033N, 2005ApJ...628L..69L, 2008A&A...482..939B}.}
         \label{TWA22HK}
   \end{figure}

The new J band spectra of TWA 22A \& B have water absorptions and spectral slopes similar to those of M5--M7 templates of young and old field dwarfs at R$\sim$2000 (see Fig \ref{TWA22J}). Strengths of FeH features are intermediate between those of field dwarfs and young (5--8 Myr) M5-M6 dwarfs spectra. The alkali lines are diminished compared to those of M6-M7 field dwarf. They are only slightly weakened compared to a field M5.  The water band longward of 1.3 $\mu$m and the 1.2 $\mu$m FeH absorptions are slightly deeper in the spectrum of TWA 22B. The H+K band spectrum better matches those of young objects (Fig. \ref{TWA22HK}). H band spectra are more peaky than in the spectrum of the M5.5$\pm$1 dwarf \citep{2007ApJ...665..736C} AB Dor C \citep[30--150 Myr;][]{2005Natur.433..286C, 2005AN....326.1033N, 2005ApJ...628L..69L, 2008A&A...482..939B}. The age-sensitive K I absorptions at 1.517 $\mu$m, the Ca I lines around 1.98 $\mu$m, and the Na I doublet at 2.206--2.209 $\mu$m  are intermediate between those of young and old M6--M7 dwarfs but are similar to those of a field M5. The K band of TWA 22 B is a bit redder than the one of TWA 22 A. This leads us to revise the classification to M5$\pm$1 for TWA 22A and  M5.5$\pm$1 for TWA 22B. We conclude that all features can be explained if the binary components are older than 5 Myr but have not reached their final contraction radius yet. This is in agreementwith results of \cite{2013ApJ...762...88M} which confirm that TWA 22 is a high probability member (99\% probability) of the $\beta$ Pictoris moving group. This is also consistent with the conclusions of \cite{2013ApJ...762..118W} which further rule out the membership to TW Hydrae originaly proposed by \cite{2003ApJ...599..342S} based on new parallaxes measurements for members of this association. 

The spectrum of Gl 417 B does not show the triangular H band profile observed in the spectra of the low gravity L dwarfs G196--3 B and 2MASS 0501--0010 \citep[L4$\mathrm{_{\gamma}}$,][]{2009AJ....137.3345C, 2010ApJ...715..561A}. On the contrary, it is well reproduced by the spectra of  2MASS J15065441+1321060 \citep[L3,][]{2000AJ....120.1085G} and GD 165 B \citep[L4, ][]{1999ApJ...519..802K}.  This is in good agreement with the high surface gravities found from spectral modelling \citep{2009A&A...503..639T}. 

To conclude, we also compare the spectra of 2M1207 A, TWA 5B, OTS 44, KPNO Tau 4, and 2M0141 to templates using the same criteria. We find spectral types in agreement (at the 1$\sigma$ level) with those derived in the optical (see column 1 of Tab. \ref{table:3}). We assign a spectral type L0$\pm$1 for Cha1109 given the good overall fit to 2M0141 and the K I lines at 1.169 and 1.253 $\mu$m that appear deeper than in the spectrum of OTS 44.

\subsubsection{Indices and equivalent widths}
\label{section:indices}
\begin{figure}[ht]
   \centering
   \includegraphics[width=\columnwidth]{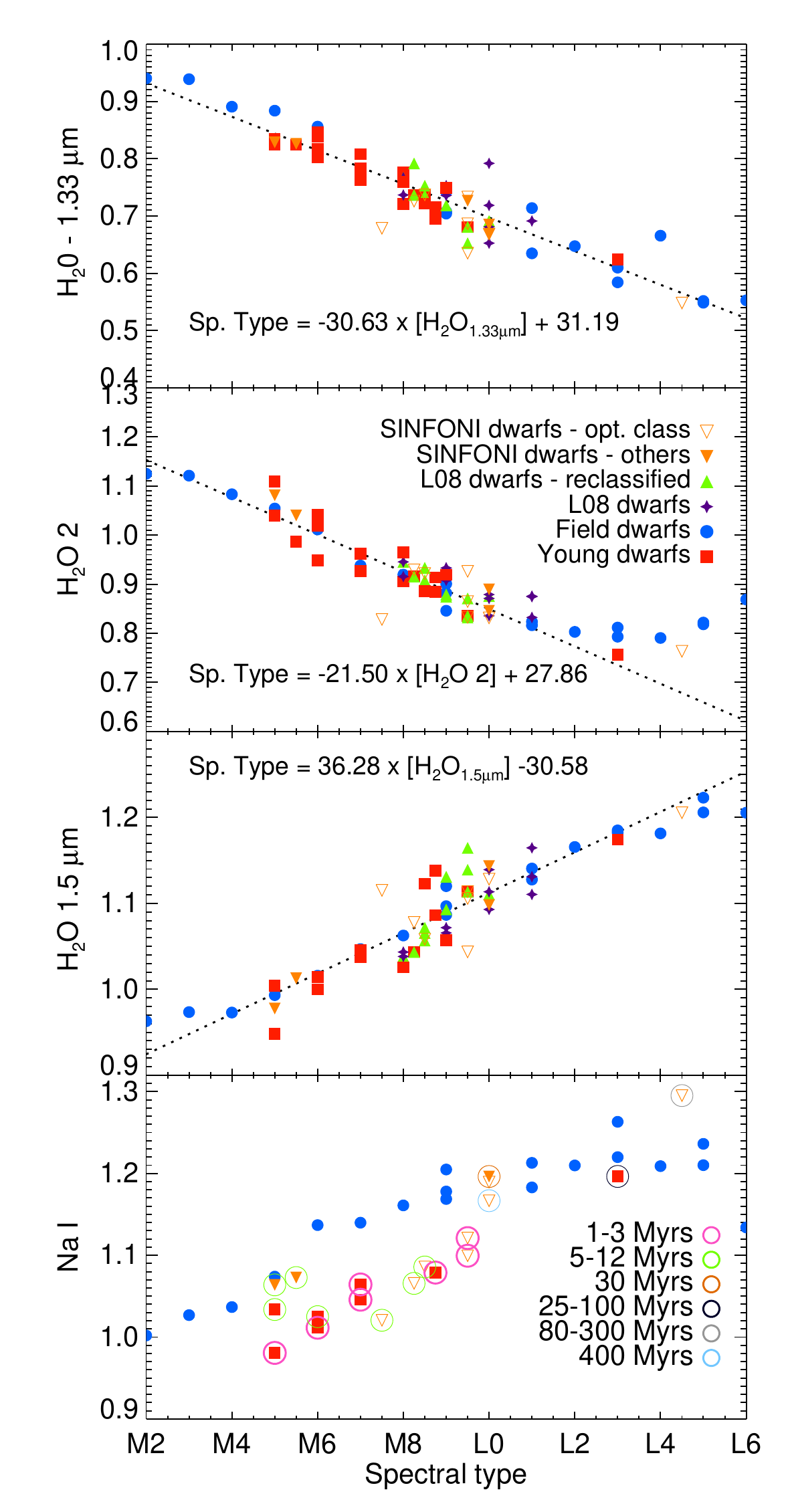} 
      \caption{Spectral indices used for the classification and age estimation computed on SINFONI spectra of dwarfs classified in the optical (orange triangles) and in the NIR (filled orange triangles), on medium resolution spectra of young dwarfs classified in the optical (red filled squares; the references associated to the spectra are  given in section \ref{classoptical}), and on field dwarfs spectra from the IRTF library \citep[blue filled dots; ][]{2005ApJ...623.1115C, 2009ApJS..185..289R}. We also overlaid the indices values for the  young late dwarfs found and classified by \cite{2008MNRAS.383.1385L} (violet stars) and our proposed alternative classification using the method described in section \ref{classoptical} (green filled inverted triangles; see Appendix \ref{Lodieu2008an}).}
         \label{Figindices}
   \end{figure}

We tested the age dependency of several spectral indices defined in the literature on the set of young optically classified dwarfs  used in the previous section and on field dwarfs spectra of C05 degraded at R=300.  We finally selected three indices that show a moderate scatter even for very young (1--3 Myr) objects  (Fig. \ref{Figindices}). 

The first two indices, $H_{2}O_{1.33\mu m}$ and $H_{2}O\:2$, defined by  \cite{2003ApJ...593.1074G}  and \cite{2004ApJ...610.1045S}, measure the strengthening of water absorptions at 1.32 and 2.04 $\mu$m. The zones used to compute the indices are indicated in Fig. \ref{AgeseqJ} and \ref{AgeseqHK}. Despite these lines deepen with  age, the indices follow linear trends at the M--L transition and can be used to constrain the spectral type of our objects within two subclasses (see Fig. \ref{Figindices}): 

\begin{equation}
Sp.\:type = -30.63\:(\pm 1.51) \times [H_{2}O_{1.33\mu m}] +31.19\:(\pm 1.13) 
\end{equation}

\begin{equation}
Sp.\:type = -21.50\:(\pm 2.15) \times [H_{2}O\:2] +27.86\:(\pm 2.01)
\end{equation}

We also computed the H$\mathrm{_{2}}$O index ($H_{2}O_{1.5\mu m}$) defined by \cite[][hereafter A07]{2007ApJ...657..511A} slightly modified (defined as $\mathrm{\langle F_{\lambda = 1.55-1.56}\rangle / \langle F_{\lambda = 1.512-1.522}\rangle}$) to avoid noisy regions around 1.48 $\mu$m and found the following relation:

\begin{equation}
Sp.\:type = 36.28\:(\pm 2.70) \times [H_{2}O_{1.5\mu m}] -30.58\:(\pm 2.91) 
\end{equation}

where  Sp. types values of 0, 10, and 15 correspond to spectral types M0, L0, and L5 respectively.

 Indices values and associated spectral types derived from SINFONI spectra and computed the same way are reported in Table \ref{table:3b}. The mean spectral types derived from the indices values and the adopted ones are reported in Table \ref{table:3}. The indices are also derived for the late type dwarfs of L08 and reported in Fig. \ref{Figindices} considering the classification of L08 (violet stars), and the spectral classes we find using the method and the associated optically classified templates of section \ref{classoptical} (green filled triangles).  

We also studied the behavior of the Na I spectral index of A07 ($\mathrm{\langle F_{\lambda = 1.150-1.160}\rangle / \langle F_{\lambda = 1.134-1.144}\rangle}$)  in the lower-right panel of Fig. \ref{Figindices}.  Interestingly, AB Pic b and 2M0141 fall on the field dwarf sequence (blue dots) along with 2M0345 \citep[$\sim$400 Myr, ][]{2007MNRAS.378L..24B} while KPNO Tau 4 and OTS 44 have lower indices values. It confirms that this index, jointly used with water band indices, is a valuable tool to identify new 1--3 Myr clusters members.  

Finally, we used our spectra smoothed at R$\sim$1700 to derive the equivalent widths (EW) with associated errors of Na I and K I lines in the J band (see Appendix \ref{EWcomp} for more details).  As expected, the young objects have lower EW than in field dwarfs (Fig. \ref{FigEW}). The EW computed for  2M0345 and Gl 417B  and  for field dwarfs analogues are similar. This means that the K I and Na I  lines of L0 and L4.5 dwarfs can be used to set upper limits of $\sim$ 600 and $\sim300$ Myr on the age of the sources.

\begin{table*}[t]
\begin{minipage}[ht]{\linewidth}
\caption{Spectral indices and equivalent widths. Spectral types associated to the indices are reported in brackets. Values of 5 and 10 correspond to spectral types M5 and L0 respectively.}
\label{table:3b}
\centering
\begin{tabular}{l*{9}{c}}     
\hline\hline
Object & H$\mathrm{_{2}}$O & H$\mathrm{_{2}}$O & H$\mathrm{_{2}}$O 2 & Na & EW (Na I) & EW(K I)  & EW(K I) & EW(K I) & EW(K I) \\
		& 1.3 $\mu$m &  1.5 $\mu$m &        &    &  1.138 $\mu$m &  1.169 $\mu$m  & 1.177 $\mu$m & 1.243 $\mu$m & 1.253 $\mu$m \\	
	&	&	&	&	&	($\mathrm{\AA}$) &	($\mathrm{\AA}$) &	($\mathrm{\AA}$) &	($\mathrm{\AA}$) &	($\mathrm{\AA}$) \\
\hline
2M1207 A & 0.73 (8.83) & 1.08 (8.60) & 0.93 (7.87) & 1.07 & 5.08$\pm$0.20 & 1.36$\pm$0.11&2.44$\pm$0.11&1.67$\pm$0.06&0.92$\pm$0.09\\
OTS 44 & 0.69 (10.05) & 1.11 (9.69) & 0.86 (9.37) & 1.12 & 3.74$\pm$0.51 & 0.53$\pm$0.19&2.83$\pm$0.19&1.99$\pm$0.11&0.59$\pm$0.19\\
KPNO Tau 4 & 0.64 (11.58) & 1.11 (9.69) & 0.83 (10.01) & 1.10 & 2.88$\pm$0.29 & 0.30$\pm$0.23&1.54$\pm$0.24&1.71$\pm$0.13&-0.49$\pm$0.19\\
2M0141 & 0.67 (10.67) & 1.13 (10.41) & 0.83 (10.01) & 1.19 & 3.47$\pm$0.57 & 2.62$\pm$0.26&5.19$\pm$0.25&1.78$\pm$0.20&1.88$\pm$0.23\\
TWA 5B & 0.74 (8.52) & 1.06 (7.88) & 0.92 (8.08) & 1.09 & 4.40$\pm$0.28 & 0.85$\pm$0.16&2.44$\pm$0.20&1.70$\pm$0.10&1.12$\pm$0.11\\
Cha 1109 & 0.68 (10.36) & 1.14 (10.78) & 0.85 (9.58) & -- & 1.38$\pm$0.80 & 0.28$\pm$0.58&3.57$\pm$0.26&1.95$\pm$0.16&2.28$\pm$0.29\\
Gl 417 B & 0.55 (14.34) & 1.21 (13.31) & 0.76 (--) & 1.30 & -- & 7.03$\pm$0.40&11.02$\pm$0.23&4.85$\pm$0.23&7.64$\pm$0.33\\
ABPic b & 0.67 (10.66) & 1.10 (9.33) & 0.89 (8.72) & 1.20 & 3.46$\pm$0.75 & 2.84$\pm$0.30&4.74$\pm$0.35&3.00$\pm$0.19&1.50$\pm$0.25\\
USco CTIO 108B   &   0.68 (10.36)  &  1.11 (9.69) &  0.83 (10.01)  & --  & 5.96$\pm$0.18 &  1.68$\pm$0.17 & 2.82$\pm$0.14 & 2.39$\pm$0.14  & 1.54$\pm$0.13\\
HR7329B   &   0.73 (8.83)  &  1.04 (7.15) &  0.93 (7.86)  & --  & 4.58$\pm$1.21 & 2.29$\pm$0.47 &  3.59$\pm0.54$ &  1.29$\pm$0.24 &  2.42$\pm$0.14\\
CT Cha b & 0.74 (8.52) & 1.05 (7.51) & 0.92 (8.08) & -- & -- & 1.36$\pm$0.22 & 1.38$\pm$0.25&0.95$\pm$0.17&0.62$\pm$0.25\\
DH Tau B & 0.71 (9.44) & 1.10 (9.32) & 0.91 (8.29) & -- & 3.38$\pm$0.76 & 1.33$\pm$0.24&3.54$\pm$0.25&2.17$\pm$0.13&0.85$\pm$0.20\\
GSC8047 B & 0.73 (8.83) & -- & -- & -- & -- & 2.01$\pm$0.53&2.75$\pm$0.52&3.19$\pm$0.39&3.14$\pm$0.48\\
TWA 22A & 0.83 (5.76) & 0.98 (4.97) & 1.08 (4.64) & 1.06 & 3.73$\pm$0.30 & 0.88$\pm$0.15&1.88$\pm$0.17&0.97$\pm$0.04&0.95$\pm$0.07\\
TWA 22B & 0.83 (5.76) & 1.01 (6.06) & 1.04 (5.50) & 1.07 & 4.37$\pm$0.27 & 1.17$\pm$0.15&2.08$\pm$0.16&0.81$\pm$0.10&1.13$\pm$0.09\\
\hline
2M0345 & 0.69 (10.05) & 1.08 (8.60) & 0.85 (9.58) & 1.17 & 9.67$\pm$0.57 & 6.09$\pm$0.22&8.67$\pm$0.17&4.59$\pm$0.20&7.02$\pm$0.27\\
\hline
\end{tabular}
\end{minipage}
\end{table*}

\begin{table}[t]
\begin{minipage}[t]{\columnwidth}
\caption{Near-infrared spectral types}
\label{table:3}
\centering
\begin{tabular}{ l l l l}     
\hline\hline
Object  & Continuum & Indices & Adopted \\
\hline
2M1207 A  & M8.5$\pm0.5$ &  M8.5$\pm$2 & M8.5$\pm$0.5 \\
OTS 44  & M9.5$\pm0.5$ &  L0$\pm$2 &  M9.5$\pm$0.5 \\
KPNO Tau 4  & M9.5$\pm0.5$ &  L0.5$\pm$2 &  M9.5$\pm$0.5 \\
2M0141  & L0$\pm1$ &  L0.5$\pm$2 &  L0$\pm$1 \\
TWA 5B & M8.5$\pm0.5$ &  M8.25$\pm$2 & M8.5$\pm$0.5 \\
USco CTIO 108B  &  M9.5$\pm0.5$ & L0$\pm$2 & M9.5$\pm0.5$ \\
HR7329 B  &  M8.5$\pm0.5$ & M8$\pm2$ & M8.5$\pm0.5$ \\
Cha 1109  & L0$ \pm1$ &  L0$\pm$2 & L0$\pm$1 \\
Gl 417 B  & L3.5$ \pm0.5$ &  L4$\pm$2 & L4$\pm$2 \\
AB Pic b  & L0$\pm1$ &  L0$\pm$2 &  L0$\pm$1 \\
CT Cha b  & M8$^{+0.5}_{-0.75}$ &  M8$\pm$2 & M8$^{+0.5}_{-0.75}$ \\
DH Tau B  & M9.25$\pm0.25$ &  M9$\pm$2 & M9.25$\pm$0.25 \\
GSC8047 B  & M9.5$\pm0.5$ &  M9$\pm$2 & M9.5$\pm$0.5 \\
TWA 22A  & M5$\pm1$ &  M5$\pm$2 & M5$\pm$1 \\
TWA 22B  & M5.5$\pm1$ &  M6$\pm$2 & M5.5$\pm$1 \\
2M0345  & L0$\pm$1 &  M9.5$\pm$2 &  L0$\pm$1 \\
\hline
\end{tabular}
\end{minipage}
\end{table}


\subsection{Comparison to synthetic spectra}
	\label{fitatmo}
	\subsubsection{The libraries}
We compared our spectra to synthetic spectra corresponding to the 2010 and 2012 releases of the BT-SETTL models\footnote{\url{http://phoenix.ens-lyon.fr/Grids/BT-Settl/}} \citep[][]{2011ASPC..448...91A, 2012EAS....57....3A} for 1000 K $\mathrm{\leq  \mathrm{T_{eff}} \leq }$ 4000 K, 3.5 dex $\leq$ log g $\leq$ 5.5 dex with 100 K and 0.5 dex increments respectively. Results were checked using the DRIFT-PHOENIX models \citep{2003A&A...399..297W, 2004A&A...414..335W, 2006A&A...455..325H, 2008ApJ...675L.105H}. The libraries and related grids of synthetic spectra are described in details in \cite{2013arXiv1302.1160B}. The DRIFT-PHOENIX models used the referenced solar abundances of \cite{1993A&A...271..587G}, the  BT-SETTL10 use the ones of \cite{2009ARA&A..47..481A}, while the BT-SETTL12 use the abundances of \cite{2011SoPh..268..255C}. We made the assumption that the objects have solar abundances. This is likely the case regarding recent metallicity measurements in Chameleon I \citep{2006A&A...446..971J, 2008A&A...480..889S}, Taurus \citep{2008A&A...480..889S}, Ursa Major group \citep{2009A&A...508..677A, 2012MNRAS.427.2905B}, and young associations \citep{2009A&A...501..965V, 2012MNRAS.427.2905B, 2013arXiv1301.5036B}.
We also analysed the TWA 22 A and B spectra using the BT-SETTL grids and the GAIA-COND library \citep{2005ESASP.576..565B}.

	\subsubsection{Fitting results}
\label{partatmoparam}
\begin{figure*}[t]
   \centering
   \includegraphics[width=18cm]{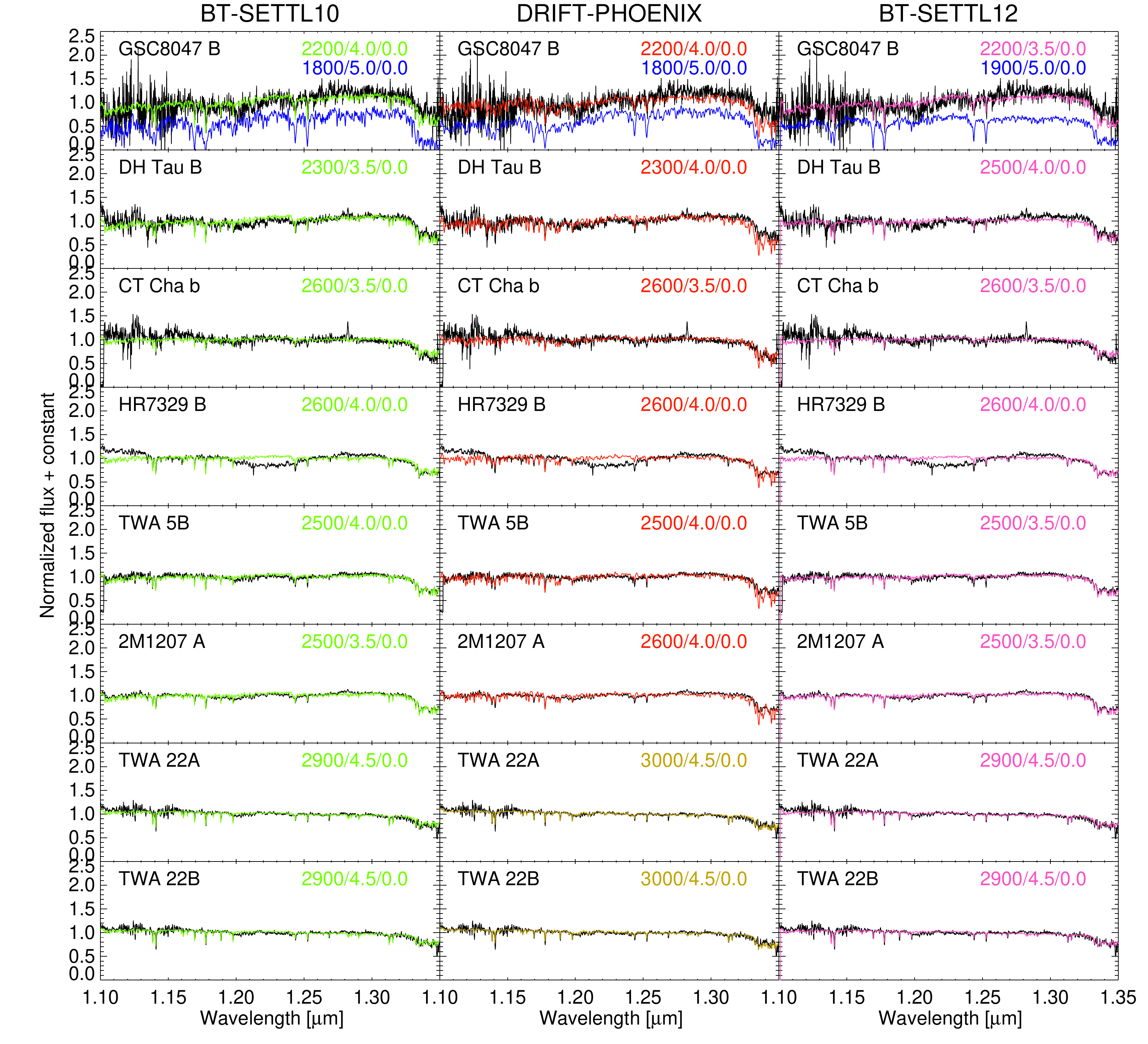}
      \caption{Best matched synthetic spectra from the BT-SETTL10 (green line),  the DRIFT-PHOENIX (red line), and the BT-SETTL12 (pink line) libraries to  J-band spectra of the SINFONI spectral library (black line).  The spectrum of DH Tau B was de-reddened by $\mathrm{A_{V}=1.16}$ mag. The corresponding atmospheric parameters are reported in each panel. 2900/4.5/0.0 corresponds to T$\mathrm{_{eff}}$=2900 K, log g =4.5 dex, and M/H=0.0 dex.}
         \label{FigsynthJhot}
   \end{figure*}

\begin{figure*}[t]
   \centering
   \includegraphics[width=18cm]{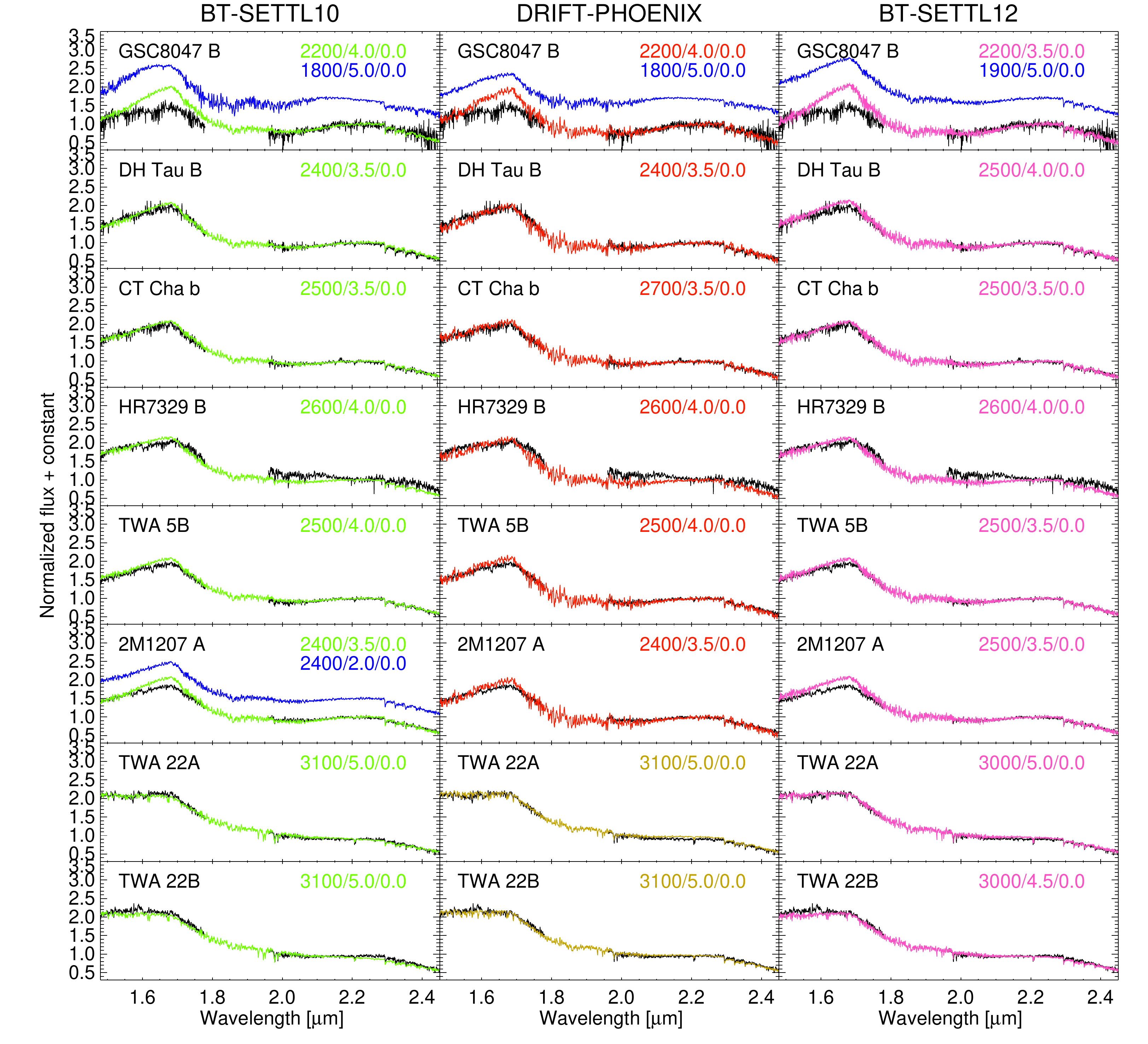}
      \caption{Same as Fig~\ref{FigsynthJhot} but for the H+K band. Alternative fitting solutions are overlaid in blue and are shifted for clarity (+0.5 in normalised flux unit).}
         \label{FigsynthHKhot}
   \end{figure*}

\begin{figure*}[t]
   \centering
   \includegraphics[width=18cm]{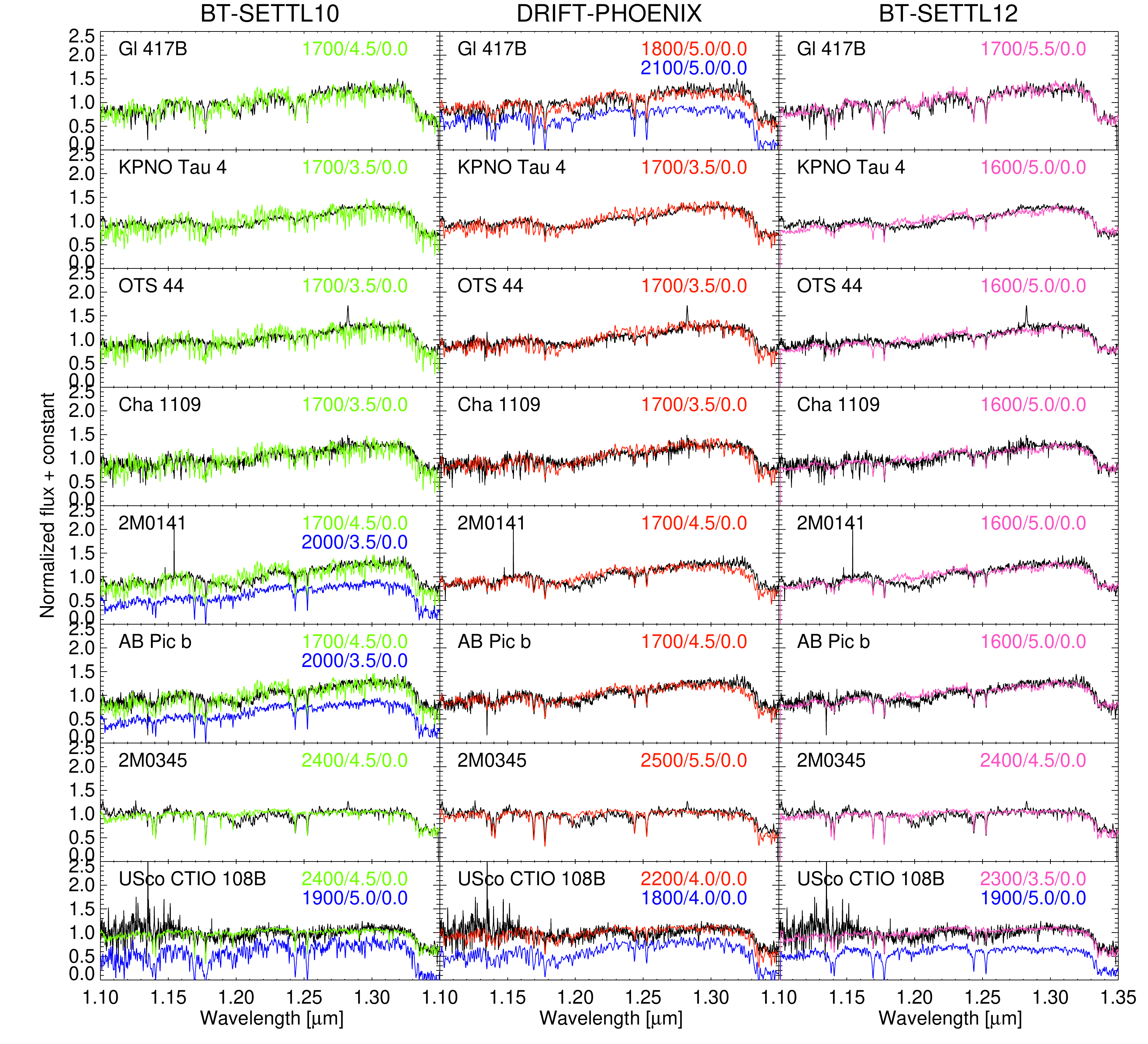}
      \caption{Same as Fig~\ref{FigsynthJhot} but for cooler objects. The spectrum of spectrum of Cha1109  was de-reddened by $\mathrm{A_{V}=1.1}$ mag. Alternative fitting solutions are overlaid in blue (shifted for clarity by -0.3 to -0.4 in normalised flux unit).}
         \label{FigsynthJcool}
   \end{figure*}

\begin{figure*}[t]
	\centering
	\includegraphics[width=18cm]{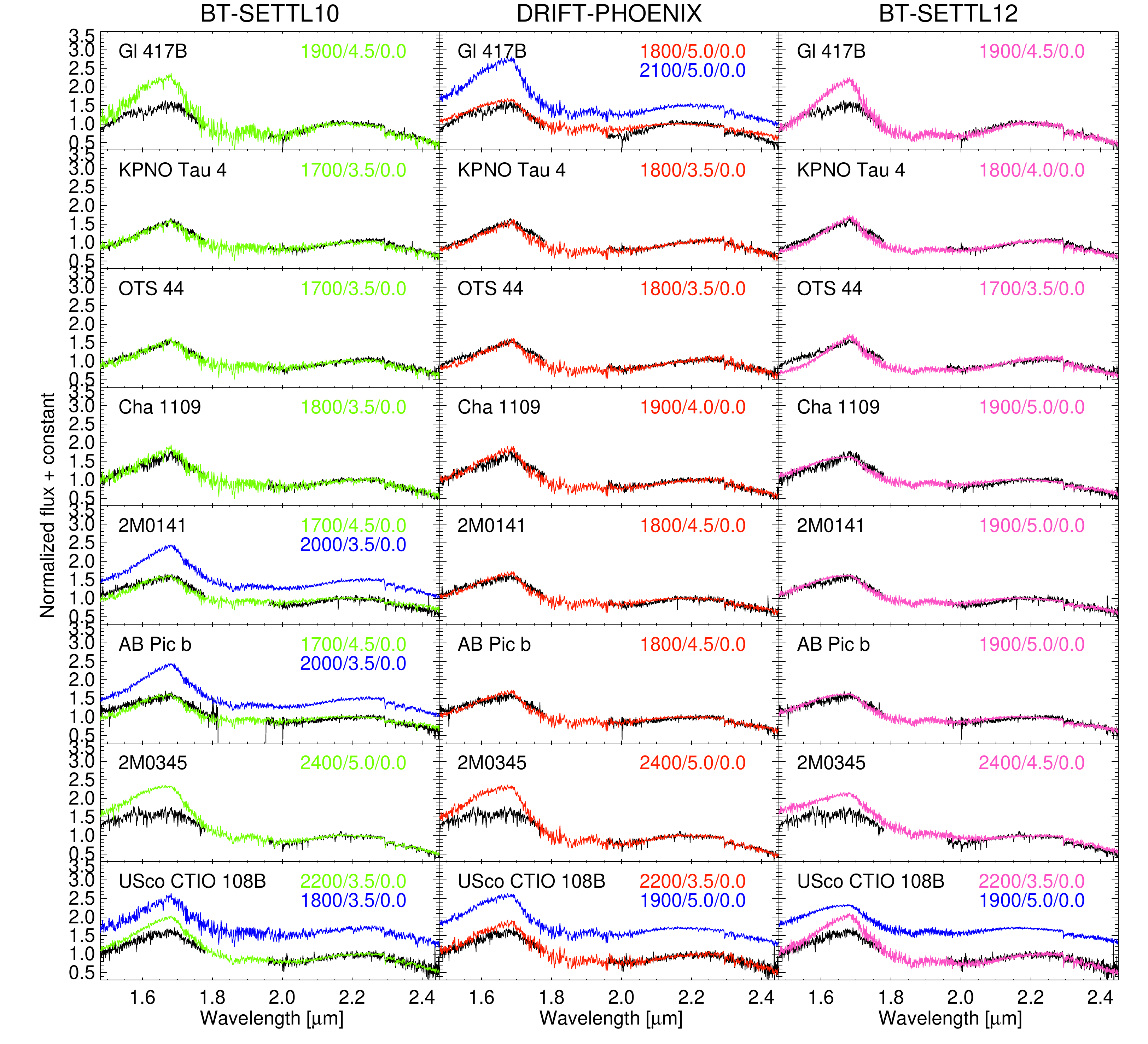}	
		\caption{Same as Fig~\ref{FigsynthHKhot} but for cooler objects. Alternative fitting solutions are overlaid in blue and are shifted for clarity (+0.5 to +0.7 in normalised flux unit).}
			\label{FigsynthHKcool}
	\end{figure*}

We applied for each empirical de-redenned spectrum the weighted least square method described in \cite{2007ApJ...657.1064M} to find the best matching synthetic spectrum over the J, H+K, and JHK bands. Fits were checked visually. The best $\mathrm{T_{eff}}$ and log g values are reported in Table~\ref{table:4}, and in Table~\ref{table:4bis} for TWA 22. The corresponding fits are plotted in Fig~\ref{FigsynthJhot}, \ref{FigsynthHKhot}, \ref{FigsynthJcool}, and \ref{FigsynthHKcool}. Bellow 2000 K, the least squares solution does not reflect the true best BT-SETTL10 fitting spectrum. This is due to  lines found over the whole spectral interval of the fit that appear only in these models . These lines originate from forests of narrow molecular  absorptions that usually make the pseudo-continuum \citep{2007ApJ...658.1217M, 2010ApJS..186...63R} and that are not Niquist-sampled by the models. This issue deasappears in 2012 of the BT-SETTL grid that benefit from a finer wavelength sampling.

BT-SETTL10 and DRIFT-PHOENIX models show several major improvements over the previous generation of models (AMES-DUSTY00, BT-SETTL08) we used for the analysis of the AB Pic b spectra \citep{2010A&A...512A..52B}. The slope of the pseudo-continuum in the J band is now self-consistently reproduced with the depth of the water absorption at 1.32 $\mu$m, the K I doublets at 1.169/1.177 $\mu$m and 1.243/1.253 $\mu$m. The triangular H-band shape typical of young objects is also better matched together with the K-band water absorptions and the CO overtones longward of 2.3 $\mu$m. The temperatures inferred from the J and H+K band fit are not necessarily the same. But the differences do not exceed 200 K. 

Moreover, the 1.1-2.5 $\mu$m spectra of Cha 1109, OTS 44, USCO CTIO 108B, GSC8047 B, and KPNO Tau 4 are very well reproduced by a single model. We found several alternative best fitting solutions for the J and HK band spectra of  AB Pic b, 2M0141, Gl 417 B, and 2M1207 A, and USCO CTIO 108B. But the corresponding spectra have too much blue slopes when considering the 1.1-2.5 $\mu$m (JHK) interval as a whole (\textbf{for} 2M0141), their H-band shape is too sharp between 1.55 and 1.70 $\mu$m (for AB Pic b, 2M0141, Gl 417 B), or the depth of their gravity-sensitive features at J band (Na I, K I) do not match (for 2M1207A).

The T$\mathrm{_{eff}}$=1600--1900 K derived for AB Pic b is now in agreement with that derived from the Lyon's group evolutionary models and from the JHK colors fit of \cite{2007ApJ...657.1064M}. $\mathrm{T_{eff}}$ estimates for TWA 5B and 2M1207 A are in-between the values found by \cite{2010ApJS..186...63R} fitting  medium and high-resolution J-band spectra of these sources.

The detailed structure and depth of the H$\mathrm{_{2}}$O band from 1.32 to 1.35 $\mu$m are better reproduced by the BT-SETTL models than by the DRIFT-PHOENIX models. These differences are related to the revised oxygen abundances \citep[50\%;][]{2009ARA&A..47..481A, 2011SoPh..268..255C} and molecular broadening damping constants in the BT-SETTL models and to differences in the greenhouse effects predicted by the two different cloud model approaches. Finally, both models still lack FeH bands around 1.2 $\mu$m \citep{2001ApJ...548..908L, 2010ApJS..186...63R}. Missing FeH absorptions could also be -- at least partially -- responsible for the non reproducibility of the H-band shape of evolved objects such as 2M0345 and Gl 417 B. The atmospheric parameters derived for these later objects are then still uncertain.  We compare our T$\mathrm{_{eff}}$ estimates to those derived using spectral type/ T$\mathrm{_{eff}}$  relationships valid for  young M and L0 dwarfs in section \ref{modeluncert}.

GAIA and BT-SETTL10 models fit the spectra of TWA22 A and B for $\mathrm{T_{eff}}$ compatible with their spectral types \citep[$T_{eff}\sim3120$ and 3060 K for A and B respectively using the conversion scale of][]{2003ApJ...593.1093L}. The mean surface gravity appears slightly higher than expected for 12 Myr objects.  Both the pseudo-continuum and the individual features are well fitted by the models further confirming the spectral extraction. 

\begin{table*}[ht]
\begin{minipage}[ht]{\linewidth}
\caption{Atmospheric parameters derived from the fit of our empirical spectra by synthetic spectra of the BT-SETTL10, BT-SETTL12,  and DRIFT-PHOENIX libraries in the J, H+K and JHK bands. 2000/3.5 means $\mathrm{\mathrm{T_{eff}}}$=2000 K and log g =3.5 dex for example. }
\label{table:4}
\centering
\begin{tabular}{llllllllll}     
\hline\hline
Objet 			&	\multicolumn{3}{c}{BT-SETTL10}					&	\multicolumn{3}{c}{BT-SETTL12}				 	&	\multicolumn{3}{c}{DRIFT-PHOENIX}	\\
			   		& 		J 			&		 H+K		&			JHK  			& 			 J		 &		 H+K			&			JHK 		&	 J		 &		 H+K			&			JHK 		\\
\hline
2M1207 A		&	2500/3.5	&	2400/3.5	&		2500/3.5		&	2500/3.5   &  2500/3.5  &  2500/3.5  &   2600/4.0	 &	2400/3.5		&	2600/4.0		\\
					&					&	2400/2.0	&							&	   &   &   &				 &						&								 \\
OTS 44			&	1700/3.5	&	1700/3.5	&		1700/3.5		&	1600/3.5  &   1700/3.5  &  1800/4.0  &  1700/3.5	 &	1800/3.5		&	1700/3.5				 \\						
KPNO Tau 4 	&	1700/3.5	&	1700/3.5	&		1700/3.5		&	1600/5.0  &  1800/4.0   & 1800/4.0  &    1700/3.5	&	1800/3.5		&	1700/3.5	\\
2M0141		&	1700/4.5	&	1700/3.5	&		1700/4.5		&	1600/5.0	&  1900/5.0  &  1700/5.5  &    1700/4.5	&	1800/4.5		&	1700/4.5					 \\
					&	2000/3.5	&	2000/3.5	&							&					&					 &	1800/4.0		&           			&						&									 \\
TWA 5B			&	2500/4.0	&	2500/4.0	&			--				&	2500/3.5  &  2500/3.5  &  --  &  2500/4.0	&	2500/4.0		&		--						 \\
Cha 1109		&	1700/3.5	&	1800/3.5	&		1700/3.5		&	1600/5.0		&		1900/5.0		&		1700/5.0   &  1700/3.5	&	1900/4.0		&	1700/3.5					 \\
Gl 417 B		&	1700/4.5	&	1700/4.5	&		1700/4.5		&	1700/5.5  &   1900/4.5  &  1800/4.0  &  1800/5.0	&	1800/5.0		&	1800/5.0					 \\
					&					&					&							&	&   &   & 2100/5.0	&	2100/5.0		&									 \\
AB Pic b			&	1700/4.5	&	1700/4.5	&			--			&  1600/5.0  &  1900/5.0  &  -- 	&	1700/4.5	&	1800/4.5		&		--					 	 \\
					&	2000/3.5	&	2000/3.5	&			--				&					 &					   &					&				&						&		--						 \\
CT Cha b		&	2600/3.5	&	2500/3.5	&			--				&	2600/3.5   &  2500/3.5   &   --  &    2600/3.5	&	2700/3.5		&		--						 \\	
USco CTIO 108B  &  2400/4.5  & 2200/4.0   & 2200/4.0   &	 2300/3.5   & 2200/3.5  & 2300/3.5  & 2300/3.5   &  2200/3.5  &  2200/3.5      \\
										 &  1900/5.0  &  1800/3.5  &  1900/5.0  &  1900/5.0  &  1900/5.0  & 1900/5.0  &  1800/4.0   &  1800/4.0  &  1800/4.0   \\
HR7329 B  &  2800/4.0  &  2800/4.0:  &  --  & 2600/4.0: & 2600/4.0: & --  &  2800/4.0:   &  2800/4.0:  &  --   \\
DH Tau B		&	2300/3.5	&	2400/3.5	&		2500/3.5		&	2500/4.0   &  2600/4.0  &  2500/4.0  &  2300/4.0	&	2400/3.5		&	2400/3.5					 \\		
GSC8047 B	&	2200/4.0	&	2200/4.0	&			--	    &  2200/3.5   &  2200/3.5  &  --  			&	2200/4.0	&	2200/4.0		&		--					 \\	
					&	1800/5,0	&	1800/5.0		&    --    &   1900/5.0   &   1900/5.0   &  --       &    1800/5.0   &   1800/5.0  &  --  \\		
2M0345		&	2400/4.5	&	2400/5.0	&		2400/4.5		&	2400/4.5  &  2400/4.5   & 2400/4.5   &   2500/5.5	&	2400/5.0		&	1800/5.5			 \\
\hline		
\end{tabular}
\end{minipage}
\end{table*}

\begin{table*}[t]
\begin{minipage}[ht]{\linewidth}
\caption{Atmospheric parameters derived from the fit of  the TWA 22A \& B spectra by synthetic spectra of the  GAIA-COND and BT-SETTL libraries in the J, H+K and JHK bands. 3100/5.0 means $\mathrm{\mathrm{T_{eff}}}$=3100 K and log g =5.0 dex for example.}
\label{table:4bis}
\centering
\begin{tabular}{ccccccc}     
\hline\hline
Objet 			&	\multicolumn{2}{c}{GAIA-COND}					&	\multicolumn{2}{c}{BT-SETTL10}				&	\multicolumn{2}{c}{BT-SETTL12}	 \\
			   		& 		J 			&		 H+K		  			& 			 J		 &		 H+K						& 			 J		 &		 H+K		\\
\hline
TWA22 A	& 2900/4.0-5.0	&		3100/5.0 &	2900/4.0-5.0		&		3100/5.0-5.5		&		2900/4.5  &  3000/5.0  \\
TWA22 B	&	2900-3000/4.5-5.0	&		3000-3100/4.0-5.5 &	2900-3000/4.0-5.5		&		2900-3100/4.0-5.5	  	&	2900/4.5   &   3000/4.5  \\
\hline
\end{tabular}
\end{minipage}
\end{table*}

\begin{table}[ht]
\begin{minipage}[ht]{\linewidth}
\caption{Adopted atmospheric parameters derived from the fit of our empirical spectra by synthetic spectra and compared to effective temperatures obtained from spectral types (see text).}
\label{table:4bis}
\centering
\begin{tabular}{llll}     
\hline\hline
Objet 			&	$\mathrm{T_{eff}}$	(K) 		&	log g  &  $\mathrm{T_{eff}}$ from\\
			   		& 																			&							&  Spectral types (K)			\\
\hline
2M1207 A		&		$2500\pm100$												&	$3.5\pm0.5$			&		2640	\\		
					&																				&							&			\\
OTS 44			&			$1700\pm100$					&	$3.5\pm0.5$ 		&	2300	\\					
KPNO Tau 4 	&		$1700\pm100$									& $3.5\pm0.5$		&	2300 	\\
2M0141		&		$1800^{+100}_{-200}$								&	$\mathrm{4.5\pm0.5}$		&	 2200	\\	
					&							 											&							&			\\
TWA 5B			&		$2500\pm100$	 											&	$4.0\pm0.5$	&	2570		\\
Cha 1109		&			$1800\pm100$	 											&	$4.0\pm0.5$	&	 2300		\\
Gl 417 B		&		$1800\pm100$	 											&	$5.0\pm0.5$			&		1660		\\
					&								 											&							&			\\	
AB Pic b			&			$1800^{+100}_{-200}$					&	$4.5\pm0.5$	&		2200	\\	
					&									 											&						&				\\									
CT Cha b		&			$2600\pm100$	 											&	$3.5\pm0.5$	&	2700	\\	
DH Tau B		&	$2400\pm100$	 											&	$3.5\pm0.5$			&	2350			\\		
GSC8047 B	&		$2200\pm100$	 											&	$4.0\pm0.5$		&	2300	\\		
USco CTIO 108B &		$2300\pm100$				&	$4.0\pm0.5$		&	2300	\\		
HR7329 B	&		$2600\pm100$	 											&	\dots		&	2570	\\		
2M0345		&		$2400\pm100$												&	$5.0\pm0.5$		&		2230 \\		
TWA22 A		&		$3000\pm100$	&		$4.5\pm0.5$ 	&		3125\\		
TWA22 B		&		$3000\pm100$	&		$4.5\pm0.5$		&		3064 \\		
\hline	
\end{tabular}
\end{minipage}
\end{table}

\subsection{Masses, luminosities, and radii}
\label{masslum}
Together with the age, the atmospheric parameters derived in part \ref{partatmoparam} enable to estimate new masses and radii for each of our targets using evolutionary models predictions of \cite{1998A&A...337..403B} and \cite{2000ApJ...542..464C}. When the target age was unknown, we only relied on the atmospheric parameters.  We note that these values do not depend on the predicted absolute fluxes for given passbands that rely on outdated atmospheric models (NEXTGEN, AMES-DUSTY00). 

 We considered here a conservative age range of 1--10 Myr for Taurus members given the age spread measurements of \cite{1995ApJS..101..117K}, \cite{2001AJ....121.1030H} and \cite{2003ApJ...590..348L}. We also choose an age range of 1--6 Myr for Chameleon I targets \citep{2007ApJS..173..104L}. We adopt an age of 3 to 11 Myr for USco CTIO 108 B \citep{2012AJ....144...55M, 2012ApJ...746..154P}. Finally, the age of 2M0345 is infered from the age estimation of Ursa Major \citep[300 to 600 Myr]{2002MNRAS.334..193C, 2003AJ....125.1980K}. The results are reported in table~\ref{Table:table5}.

\begin{table*}[t]
\begin{minipage}[t]{\linewidth}
\caption{Masses and radii predicted by evolutionary models of \cite{1998A&A...337..403B} and \cite{2000ApJ...542..464C} for the given age, T$\mathrm{_{eff}}$ and log g  estimates of the SINFONI sources.}
\label{Table:table5}
\centering
\begin{tabular}{lccc|ccc}     
\hline\hline
Objet			&	Age				&	Predited Mass	from $\mathrm{T_{eff}}$		&	Predicted radius	from  $\mathrm{T_{eff}}$ &	$\mathrm{L/L_{\odot}}$&	Predicted		 &	Semi-empirical  	\\	
					&					& 	&	 	&  							   &	mass	from $\mathrm{L/L_{\odot}}$				& radius			\\
					&(Myr)         &   (M$\mathrm{_{Jup}}$)	&	(R$\mathrm{_{Jup}}$)  &   & (M$\mathrm{_{Jup}}$)	  &   (R$\mathrm{_{Jup}}$) \\
\hline					
2M1207 A		&	7.5-9.1			&	24$\pm$5				&	2.4$^{+0.4}_{-0.8}$  &	--	&	--	&	--	\\					
TWA 5B			&	7.5-9.1			&	24$\pm$5				&	2.4$^{+0.4}_{-0.8}$	 &	--	&	--	&		--\\					
CT Cha b		& 2$\pm$2		&	30$^{+5}_{-11}$		&	2.6$^{+1.2}_{-0.2}$	\\			
DH Tau B		&	0.5-10			&	15$^{+7}_{-4}$ 		&	2.6$^{+0.1}_{-0.2}$	 &	--	&	--	&	--	\\
HR7329  B     &   8-20          &    35$^{+20}_{-15}$   &   2.7$\pm$0.2   &   --	&	--	&	--	\\
TWA22	A		&	8-34			&	75$^{-20}_{+35}$		&	3.4$^{+0.9}_{-1.2}$	 &	--	&	--	&	--	\\
TWA22	B		&	8-34			&	75$^{-15}_{+35}$		&	3.4$^{+0.9}_{-1.2}$	 &--		&	--	&	--	\\
2M0141		&	10-40			&	13$_{-2}^{+1}$			&	1.5$\pm$0.1    &	--	&	--	&	--	\\
					&	--					&	11$^{+12}_{-6}$		&	1.5$^{+0.3}_{-0.1}$	 &	--	&		--&	--	\\		
GSC8047 B	&	10-40			&	22$^{+4}_{-7}$			&	1.6$^{+0.3}_{-0.1}$	 &	--	&		--&	--	\\		
USco CTIO 108 B	 &	3-11		&	16$^{+9}_{-4}$			&	2.0$\pm$0.1		 &	-3.19$\pm$0.05		&	10--15		&	1.6$\pm$0.1		\\
OTS 44			&	1-6				&	6$_{-1}^{+3}$			&	1.7$^{+0.3}_{-0.1}$ & 	-2.90$\pm$0.08		&		9--17				&	3.8$^{+0.6}_{-0.5}$	\\				
KPNO Tau 4	&	0.5-10			&	4$_{-1}^{+8}$			&	2.0$^{+0.1}_{-0.4}$	 & -2.49$\pm$0.01		&		15--27				&	6.1$^{+0.5}_{-0.4}$	\\					
Cha 1109		&	1-6				&	7$^{+4}_{-2}$			&	1.7$^{+0.4}_{-0.1}$	 & -3.23$\pm$0.03		&		5-13					&	2.4$^{+0.4}_{-0.3}$	 \\				
AB Pic b			& 10-40			&	13$_{-2}^{+1}$			&	1.5$\pm$0.1		&  -3.74$\pm$0.07		&		10-14				&	1.4$\pm$0.2			\\
2M0345		&	300-600		&	75$^{-15}_{+35}$		&	1.11$\pm$0.06		&	-3.58$\pm$0.03		&		55-73				&	0.9$\pm$1				\\				
Gl 417B			&	80-250			&	30$\pm$10				&	1.2$\pm$0.1	&  -3.84$\pm$0.03		&		23-37				&	1.3$\pm$0.1		\\
\hline					
\end{tabular}
\end{minipage}
\end{table*}

Following this method, OTS 44, KPNO Tau 4, and Cha 1109 have masses in the planetary-mass range; DH Tau B, 2M0141, USco CTIO 108 B, and AB Pic b have masses at the
deuterium-burning mass limit; 2M1207A, TWA 5B, GSC8047B,  HR7329 B, CT Cha b, and Gl 417 B are brown dwarfs; and TWA 22A and B have masses close to the substellar boundary. The age estimate derived for 2M0141 is still compatible with this source being a member of young nearby associations. This object lies at the deuterium-burning boundary in case it is a true 30 Myr old Tucana-Horologium member. The new masses of OTS 44 and KPNO Tau 4 are lower than previously estimated ($\sim$ 15 M$_{Jup}$) by \cite[][]{2004ApJ...617..565L} and \cite{2002ApJ...580..317B} respectively. The mass of KPNO Tau 4  is however in agreement with  10 M$\mathrm{_{Jup}}$ reported in \cite[][]{2006ApJ...649..306K}. Masses of 2M1207 A,  TWA 5B, DH Tau B, GSC8047 B, 2M0141, Cha1109, AB Pic b, and Gl 417 B agree with values found in past studies (reported in Table \ref{table:1}).

The predicted total mass of TWA 22 AB  considering the spectroscopic temperatures of the components, their surface gravities, and their estimated  luminosities now \textit{agrees within error bars} with the dynamical mass of the system \citep{2009A&A...506..799B} \textit{considering a conservative age range of 8-34 Myr for the} $\beta$ \textit{Pictoris moving group} \citep{2008A&A...480..735F}.

We derived the empirical luminosity of USco CTIO 108 B, AB Pic b, 2M0141, Cha1109, KPNO Tau 4, and OTS 44 using their absolute flux and K-band bolometric corrections recently estimated by \cite{2010ApJ...714L..84T} for 2M0141 and KPNO Tau 4, and assuming $\mathrm{M_{bol_{\odot}}=4.75\:mag}$.  We use  these luminosities and the spectroscopic $\mathrm{T_{eff}}$ (Table \ref{Table:table5}) to derive semi-empirical radii estimates using the Stephan-Boltzmann law ($\mathrm{L=4 \pi R^2 \sigma T_{eff}^4}$).  The newly estimate radius of AB Pic b agree with the evolutionary models predictions. On the contrary, the semi-empirical radii of 1--3 Myr old objects appears to be 1.4 to 3 times larger than those inferred from the models.  \\

\subsection{Conversion scales}
 The inconsistency on the radii can be explained as an underestimation of the spectroscopic temperatures compared to the predictions of evolutionary models for the observed luminosities. We  compared the spectroscopic $\mathrm{T_{eff}}$ derived from our sample of SINFONI spectra and from spectra of young M and L members of the Upper Scorpius star forming region obtained by L08 to the \cite{2009ApJ...702..154S} conversion scales and to additional scales built for young objects (Figure \ref{Figconvscales}). The atmospheric parameters associated to each object of L08 are  reported in Appendix \ref{Lodieu2008an}. \cite{2003ApJ...593.1093L} proposed a temperature scale intermediate between those of field dwarfs and of giant stars for young dwarfs with optical spectral types earlier than M9. The scale was built so that members of Taurus and IC348 fall parallel to \cite{1998A&A...337..403B} and \cite{2000ApJ...542..464C} isochrones in H-R diagrams at the cluster ages. \cite{2002ApJ...580..317B} and \cite{2008ApJ...675.1375L} later extrapolated the previous scale for M9.5 ($\mathrm{T_{eff}}$=2300 K) and L0 ($\mathrm{T_{eff}}$=2200 K) subclasses using the temperature differences predicted for mature M9 and L0 dwarfs.  

\begin{figure*}
\includegraphics[width=18cm]{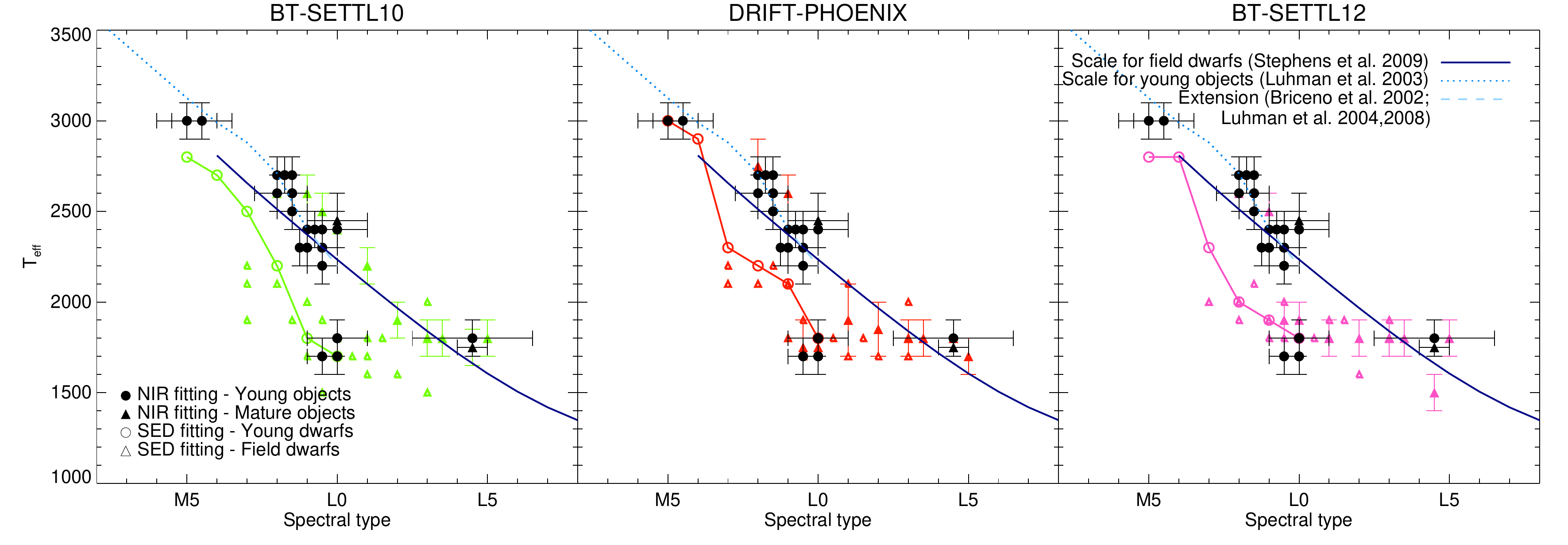}
\caption{Spectroscopic effective temperatures of field objects (squares) and young objects (dots) compared to those derived from $\mathrm{T_{eff}}$ vs spectral type conversion scales valid for old field dwarfs \citep[black line]{2009ApJ...702..154S} and young objects \citep[dashed and/or dotted lines;][]{2003ApJ...593.1093L,2002ApJ...580..317B,2004ApJ...617..565L, 2008ApJ...675.1375L}. The spectroscopic temperatures were found comparing the near-infrared spectra (IRTF, NIRSPEC, SINFONI) to synthetic spectra of the DRIFT-PHOENIX and BT-SETTL libraries (filled symbols), or fitting spectral energy distributions built from the typical  colors of young objects \citep[open circles;][]{2010ApJS..186..111L} or from the photometry of field dwarfs \citep[open triangles;][]{2006ApJ...651..502P}.}
\label{Figconvscales}
\end{figure*}

Predicted temperature/spectral types for young objects earlier than M9 agree with  the Luhman's scale. Effective temperatures of Gl 417 B and 2M0345 are compatible with the ones predicted using the conversion scale of field dwarfs. However, the  effective temperatures of  KPNO Tau 4, OTS 44, 2M0141, Cha1109 (classified in the optical), and AB Pic b fall 300-500 K lower than predictions. The objects of L08 confirm the shift. M9.5 dwarfs like GSC8047 B and USCO CTIO 108 B and their younger counterparts OTS 44,  KPNO Tau 4, and Cha1109 fall on two distinct temperature ranges.  This suggests either a strong change in  effective temperature of young M9.5-L0 dwarfs with the age, or that atmosheric models fail to provide realistic effective temperatures. We  investigate the later hypothesis in the following section.

\section{Discussion}
\subsection{Infrared excesses and extinctions}
\label{extinc}
\begin{figure}[t]
   \centering
   \includegraphics[width=\columnwidth]{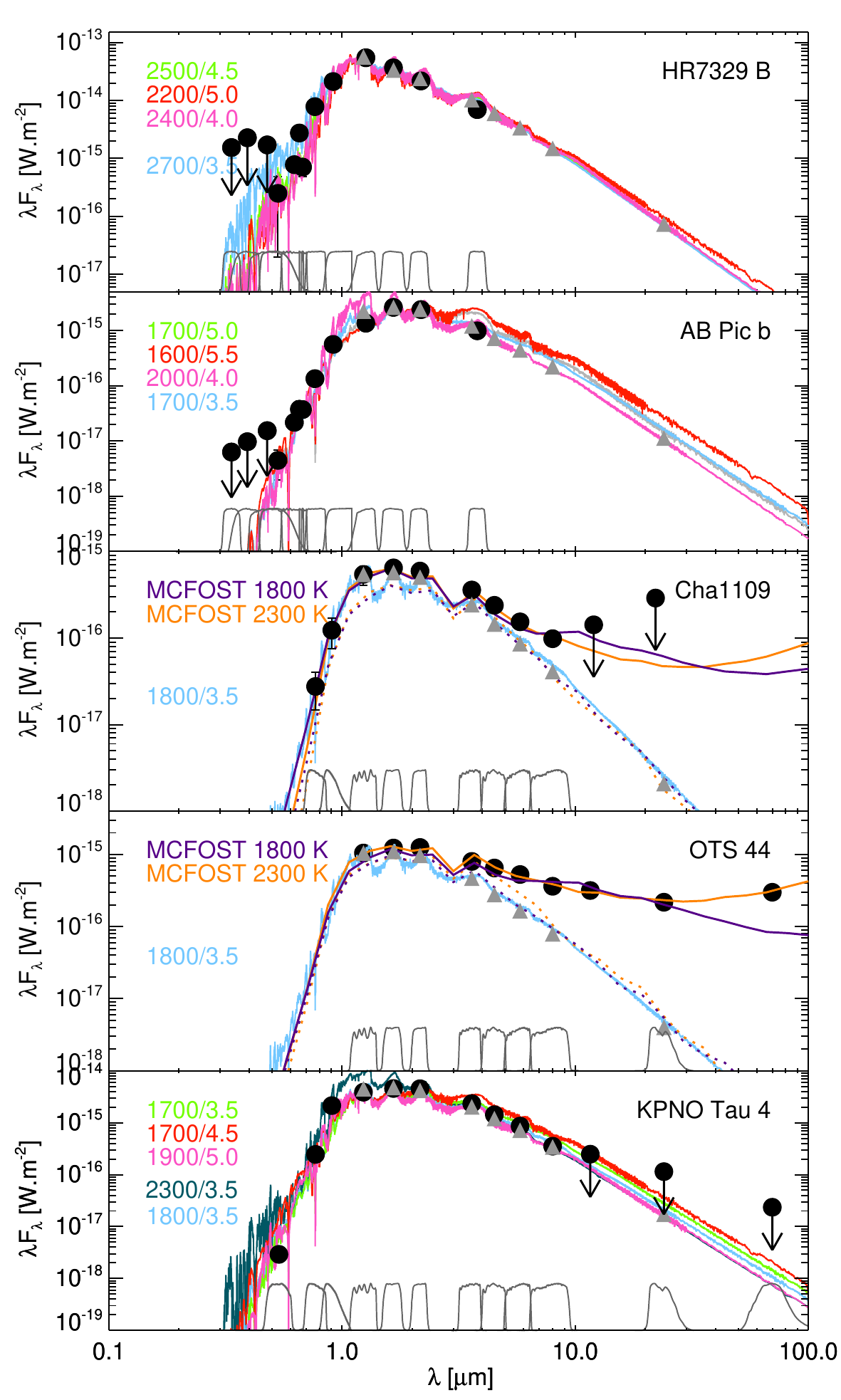}
      \caption{Comparison of the SED (filled  circles) of several young M9.5-L0 dwarfs with best-fitted DRIFT-PHOENIX (red),  BT-SETTL10 (green), and BT-SETTL12 (magenta) spectra. The SED of Cha1109 is de-reddened by $\mathrm{A_{V}=1.1\:mag}$. The typical photometry of M9.5 (for Cha1109, OTS 44, and KPNO Tau 4) and L0 dwarfs (for AB Pic b) is overlaid (triangles) along with the corresponding fitting solution found using the BT-SETTL10 grid (light blue).}
         \label{SEDs}
   \end{figure}

Excesses can impact semi-empical radii estimates reported in section \ref{masslum} as they tend to decrease the luminosity and can affect the JHK spectral slope that is also sensitive to $T_{eff}$.  Excesses and extinctions can be reasonably excluded for AB Pic b (and for 2M0141) given the proximity and the age of this companion. It is however a source of concern for the younger and more distant objects KPNO Tau 4, OTS 44, and Cha1109. \cite{2002ApJ...580..317B} found $A_{V}=0$ for KPNO Tau 4 comparing the optical spectrum of the source to those of unextincted field dwarfs analogs. \cite{2005ApJ...620L..51L} later concluded that the source does not exhibit excess emission from circumstellar material based on the mid-infrared spectral energy distribution (SED) fitting and given $A_{V}=0$. The analysis also relied on the good agreement between the mid-infrared photometry with that of a blackbody at the temperature of the source found extrapolating the $\mathrm{T_{eff}}$/spectral type conversion scales (section \ref{modeluncert}). The lack of significant disk-contribution  was also confirmed by \cite{2007A&A...465..855G} from a full-SED analysis but required a higher extinction ($A_{V}=2.45$).  The extinction was also re-evaluated by \cite{2006ApJ...649..306K} to $A_{V}=4$ based on the fit of the optical+NIR SED with redenned field dwarfs templates. \cite{2005ApJ...620L..51L} later used the SED of KPNO Tau 4 considering A$_{V}$=0 together with a disk model to conclude that OTS 44 was affected by significant mid-IR excess only longward of 3 $\mu$m. \cite{2005ApJ...635L..93L}  made a similar analysis for Cha1109 and reached similar conclusions. We compared the infrared SED of KPNO Tau 4  (see Figure \ref{SEDs}) to the typical SED of a young M9.5 source built taking the average of the typical empirical infrared colors of young M9 and L0 dwarfs reported in Table 13 of \cite{2010ApJS..186..111L} and normalizing the J-band flux to the one of the object.  The object SED were built from the photometry found in the literature (mostly from HST, 2MASS, and \textit{Spitzer}\footnote{\textit{Spitzer} magnitudes were converted to flux densities using the Gemini converter: \url{http://www.gemini.edu/?q=node/11119}.})
This comparison excludes redennings corresponding to $A_{V}>1.8$ mag  \cite[using the redenning law of ][with R=3.1]{2003ARA&A..41..241D, 2003ApJ...598.1017D, 2003ApJ...598.1026D} and confirms that the source does not have significant excess. The procedure was repeated for OTS 44 and Cha 1109. We found that the JHK band photometry of OTS 44 could be affected by a reddening corresponding to $A_{V} \sim 2.0$ mag. Cha1109 near-infrared photometry is not affected by significant additional excess once an additional extinction of $A_{V}=1$ mag is considered.  Theses conclusions are consistent with the SED classes reported in \cite{2008ApJ...675.1375L} and  \cite{2010ApJS..186..111L} that were found computing infrared slope excesses and that rely on the same input standard photometry.  To conclude, we also compared the JHKL' band photometry of AB Pic b to the standard colors of young L0 dwarfs. The photometry was converted to fluxes using a flux calibrated spectrum of Vega \citep{1985A&A...151..399M, 1985IAUS..111..225H} and the NaCo filter passbands. We find that the SED is not reproduced by the  typical flux unless it  is normalized using the H band flux. AB Pic b looks $\sim$0.5-1.0 mag fainter in the J band.  Once calibrated in flux, the JHK band spectrum of the companion displays the same features as 2M0141 while the J band appears shifted to lower fluxes. The origin of the under-luminosity remains unclear. It could be due to a systematic error on the photometry (see section \ref{subsec:comp}), to variability (section \ref{rembias}), and/or to peculiar atmospheric properties. 

KPNO Tau 4 and OTS 44 are still fitted by synthetic spectra at $\mathrm{T_{eff}}$=1800-1900 K and log g=3.5 dex once the JHK spectra have been corrected from an extinction of $A_{V}=1.8$ mag. The best fitting solution for Cha1109 changes to $T_{eff}=2200$ K and log g=3.5 dex if we deredden the spectrum by $A_{V}$=1.1+1.8 mag. This temperature is in better agreement with the one expected for a young L0 dwarf (see section \ref{modeluncert}). We estimate log $L/L_{\odot}$=-2.40 for KPNO Tau 4, -3.15 for Cha1109, and -2.82 for OTS 44 considering the magnitudes corrections ratio of \cite{1998ApJ...500..525S} and these extreme $A_{V}$. These new spectroscopic $\mathrm{T_{eff}}$ and luminosities  brings semi-radii estimates of Cha1109 to $R=1.8^{+0.3}_{-0.2} R_{Jup}$)  in agreement with the predictions from evolutionary models ($R=1.5^{+0.6}_{-0.3} R_{Jup}$). This is not the case for KPNO Tau 4 and OTS 44 ($R=5.7^{+0.7}_{-0.6}  R_{Jup}$ and $R=3.5^{+0.8}_{-0.9}  R_{Jup}$ respectively).

\begin{table*}[t]
\caption{Parameters of the disk models used to represent the spectral energy distribution model  of  Cha1109 and OTS 44, and considering two different photospheres for the objects. $a_{max}$  is the maximum grain size, $h_{0}$ is the scale height at a reference radius of 100 AU. $R_{in}$ is the inner-edge radius of the disk. $\beta$ parametrizes the flaring (see text).}
\label{tab:fitSED}
\centering
\begin{tabular}{l|cccccc|cccccc}     
\hline\hline
\multirow{2}{*}{}		&		$a_{max}$		&		$M_{dust}$		&		$h_{0}$			&		$R_{in}$		&		$\beta$			&		R		&		$a_{max}$		&		$M_{dust}$		&		$h_{0}$			&		$R_{in}$		&		$\beta$			&		R	\\	
				&		($\mu$m)		&		($M_{\odot}$)		&		(AU) 		&		(AU)		&			&		($R_{Jup}$)	&		($\mu$m)		&		($M_{\odot}$)		&		(AU) 		&		(AU)		&			&		($R_{Jup}$) \\ 
				\hline
Object		& \multicolumn{6}{|c|}{$T_{eff}=$2300K/log g=3.5}		&  \multicolumn{6}{c}{$T_{eff}=$1800K/log g=3.5}	 \\	
\hline
Cha1109	&	1.0		&	 	$10^{-4}$	&		11		&	 0.02		&		1.2			&		2.0	&	1.0		&	 	$10^{-4}$   &  15 & 0.02		&		1.125		&		2.1	\\
OTS44		&	1.0		&	 	$10^{-4}$	&		15		&	 0.04		&		1.2		    &		3.6	&   1.0		&	 	$10^{-4}$   &	15 &	0.02		&	     1.050	    &   	3.2	\\		  
\hline	
\end{tabular}
\end{table*}

   We used a posteriori the radiative transfer code MCFOST \citep{2006A&A...459..797P} based on the Monte-Carlo method to model and fit SED of Cha1109 and OTS44 and further investigate the impact of disk emission on the near-infrared spectra.  MCFOST considers disks as axisymmetric flared density structures with scale heights $h(r) =h_{0}(r/r_{0})^{\beta}$. $r$ is the radial coordinate in the equatorial plane, $h_{0}$ is the scale height of the disk at $r_{0}=100$ AU, and $\beta$ parametrize the flaring. The disk models have a cylindrical geometry characterized by a inner-edge radius $r_{in}$. The dust grain distribution follows a power law  with a maximum grain size $a_{max}$. We used two BT-SETTL10 spectra, at $\mathrm{T_{eff}}$=2300 K and  log g=3.5 and at $\mathrm{T_{eff}}$=1800 K and log g=3.5, as guess for the photospheric emission for each of the object.  The scaling factor of the synthetic spectra were let as free parameters in the fit. The parameters of the best fitting models are reported in table \ref{tab:fitSED} and SED models are overlaid in Figure \ref{SEDs}. Following this approach, the radius of Cha1109 agrees with evolutionary models predictions while that of OTS 44 still remains too large for the two choices of photospheres. We confirm that excesses are not biasing our spectral analysis.

 \subsection{On the classification of late-M and early-L dwarfs}
\label{classcaveats}
The spectroscopic temperatures of our later type sources also encouraged us to question the validity of the \textit{absolute} classification of M9.5-L0 dwarfs. The very late M-type Mira used to classify OTS 44, Cha 1109, KPNO Tau 4  at optical wavelengths \citep[][and ref therein]{2002ApJ...580..317B, 2007ApJS..173..104L,  2008ApJ...675.1375L} are very often variables. This is well illustrated in the near-infrared for the dust-enshrouded \citep{1976PASP...88..535C, 2000A&A...353..322A}  M10+III star IO Virginis which optical spectrum was used to classify KPNO Tau 4 \citep{2002ApJ...580..317B}. We compared  the J and H+K band spectra of this star taken from \cite{2002A&A...393..167L}, \cite{2009ApJS..185..289R}, and obtained reducing SINFONI data collected on January 25, 2008. The slope and the main spectral features (VO, TiO, $\mathrm{H_{2}O}$) are variable. 

The method proposed by \cite{2009AJ....137.3345C},  to classify young L dwarfs, and based on the comparison to field dwarfs templates, is promising. We note, however, that  our $\mathrm{T_{eff}}$ estimation for 2M0141 (classified as $L0_{\gamma}$) is the same as found for G 196-3B ($L3_{\beta}$) by \cite[][found adjusting a black body on the SED.]{2010arXiv1004.3965Z}.

\subsection{Behaviour of  atmospheric models at the M-L transition}
	\label{modeluncert}	
	
The spectroscopic temperatures strongly rely on the good representation of the NIR spectral slopes of the atmospheric models. While the near-infrared spectra of late-M and early-L dwarfs are extremely sensitive to the greenhouse effect due to silicate cloud formation, the  small spread on effective temperatures estimates derived from the yet very different cloud model approaches of the DRIFT-PHOENIX and BT-SETTL (2010 and 2012) models seems to show that the understanding of cloud physics converges, and that this wavelength range is little sensitive to differences in cloud modelling as soon as the main ingredients (mixing, condensation, sedimentation, and supersaturation) are accounted for. Nevertheless, DRIFT-PHOENIX models have recently been tested by  \cite{2011A&A...529A..44W}  on a set of low resolution (R=75-200)  spectra of a various set of young and evolved  M and L dwarfs over 0.5-2.5 $\mu m$.  They notably fitted a low resolution (R$\sim$120) spectrum of 2M0141 and found effective temperatures and surface gravities in good agreement with our values. Their fit also pointed out an underprediction of the temperatures  at the M-L transition with respect to the scale of \cite{2009ApJ...702..154S} valid for field dwarfs and built from empirical luminosities and radii predicted by evolutionary models.

\begin{figure*}[t]
   \centering
   \includegraphics[width=18cm]{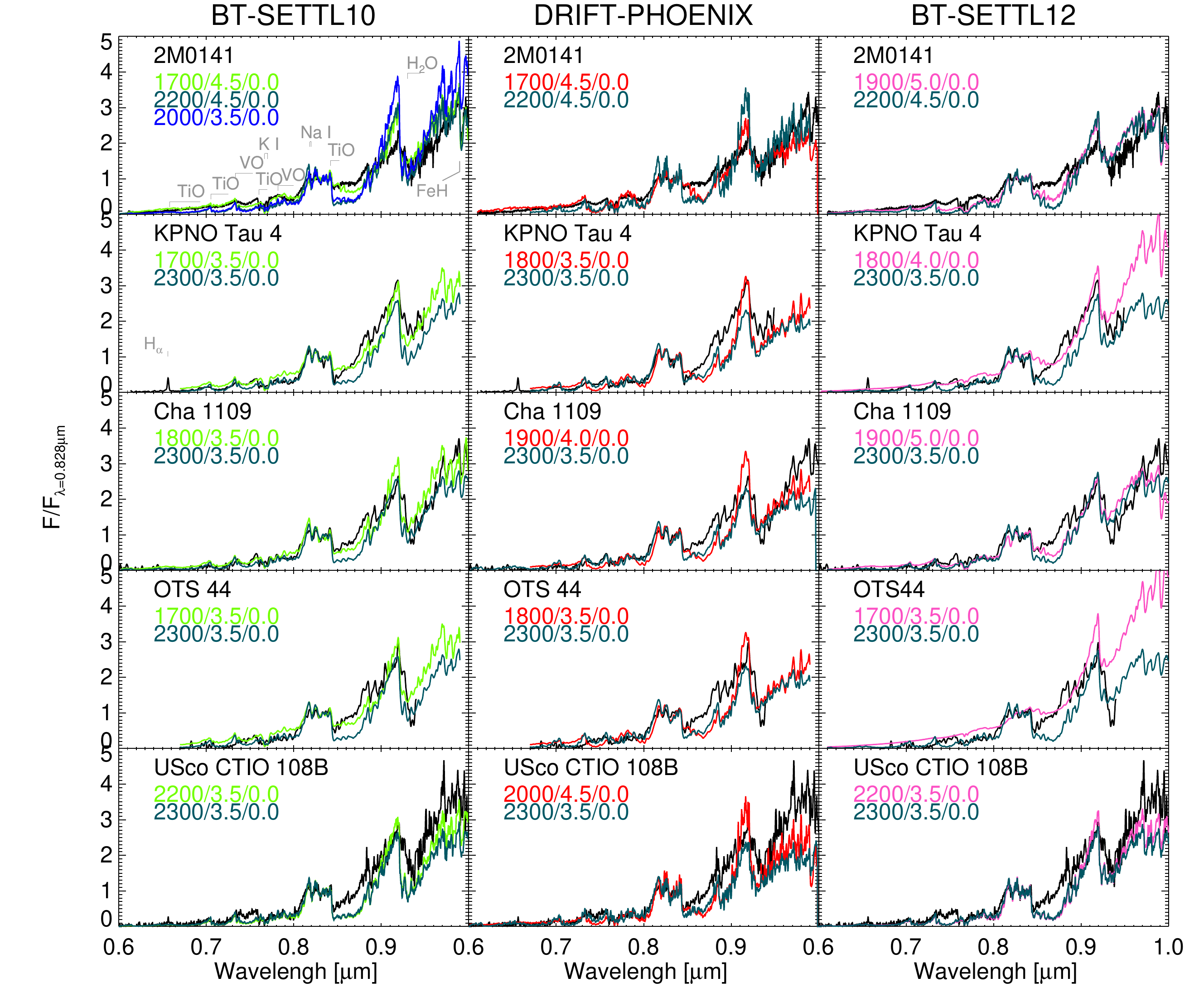}
      \caption{Normalized optical spectra of 2M0141, KPNO Tau 4, Cha1109, OTS 44, and USCO CTIO 108B compared to the best fitting BT-SETTL10 (green; left), DRIFT-PHOENIX (red; middle), and BT-SETTL12 (magenta; right) spectra found in the near-infrared. The spectrum of Cha1109 is de-redenned by $A_{V}$=1.1 mag. Synthetic spectra at the temperature predicted  by conversion scales are overlaid in blue-green. We also show the alternative fitting solution found for 2M0141 using  BT-SETTL10 spectra (blue).}
         \label{spec_optical}
   \end{figure*}

\begin{figure*}[t]
   \centering
   \includegraphics[width=18.5cm]{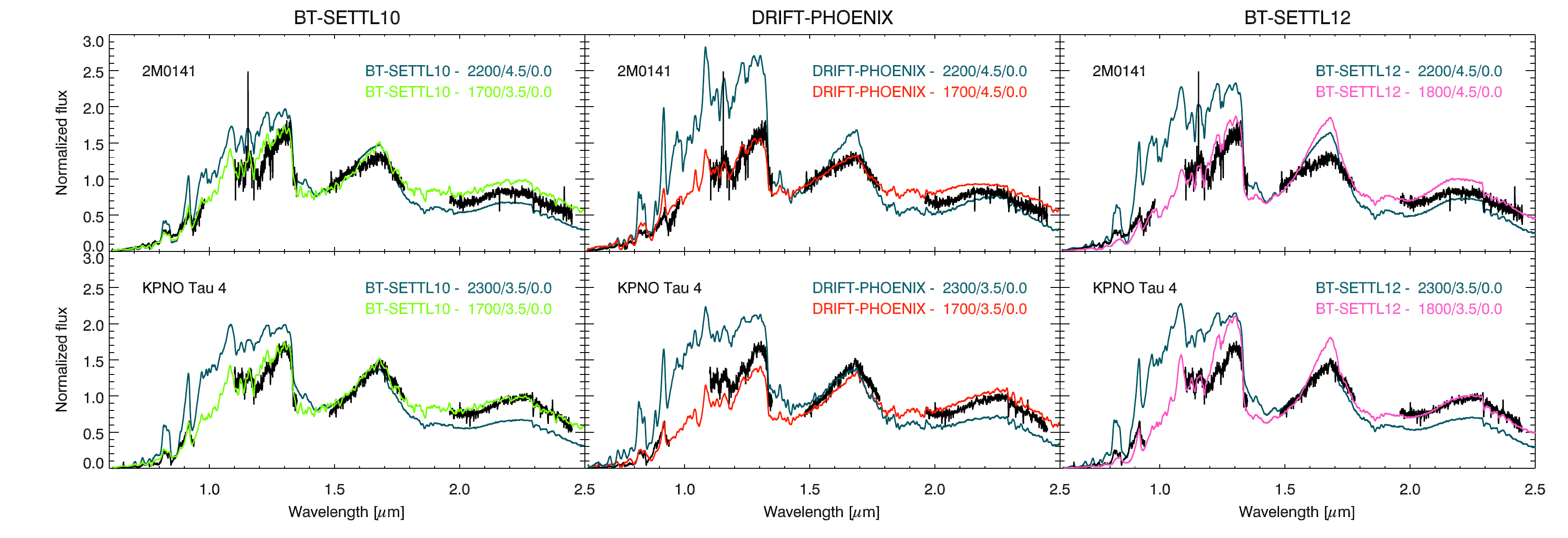}
      \caption{0.6-2.5 $\mu$m spectra of 2M0141, and KPNO Tau 4 normalized over the considered wavelength range and compared to best fitted BT-SETTL10 (green line; shifted by +0.5  normalized flux unit) and DRIFT-PHOENIX (red; shifted by -0.5 normalized flux unit) spectra. A spectrum at $\mathrm{T_{eff}}$ corresponding to predictions from conversion scales ang log g=3.5 dex (grey-blue) is overlaid for comparison.}
         \label{spec_opticalNIR}
   \end{figure*}

We decided to make additional tests to evaluate the robustness of the atmospheric models using complementary material found in the litterature, the results presented above, and the three synthetic spectral libraries:\\
\begin{enumerate}
\item We plotted the BT-SETTL10 and BT-SETTL12 spectra over the 1.1-2.5 $\mu$m range around $\mathrm{T_{eff}=1600-2000}$ K and confirmed that the spectral slope  vary  rapidly with the temperature and the surface gravity  between 1600 K and 2000 K. 	In this wavelength range, the computation of the mean fraction and the size distribution of grains inside the cloud require a large number of iterations. For some of the spectra, the models could have failed to converge (and then to predic the correct dust fraction and size distribution) and could then explain this rapid change. The conclusions derived by \cite{2011A&A...529A..44W} also certainly apply to the BT-SETTL models, e.g. the model underestimates the dust
opacity for $\mathrm{T_{eff}}$ above $\sim$1900 K, resulting in too blue spectral slopes, and underestimated effective temperatures. 
\item We built for the first time the 0.55-3.8 $\mu$m spectral energy distribution of AB Pic b and HR7329 B from avaliable near-infrared photometry \citep[][ this work]{2011MNRAS.416.1430N} and archival Hubble data (see Appendix \ref{OpticPhot}). We compared these SED and that of KPNO Tau 4  to BT-SETTL and DRIFT-PHOENIX models. The fit enable to test the global reproduciliby of atmospheric models.  The procedure involved  finding the scaling factor ($R^{2}/d^{2}$. R is the object radius and d the distance to the Earth) that made the synthetized photometry (expressed as a surfacic flux) best matching the observed values (found together with the best $\mathrm{T_{eff}}$ and log g). Fitting solutions are shown in Figure \ref{SEDs}. Models successfully reproduce the \textit{Spitzer} fluxes, the near-infrared photometry (\textit{2MASS}), and the optical fluxes of KPNO Tau 4 for $\mathrm{T_{eff}}$=1700-1800K and log g =3.5-4.5 dex.  Synthetic SEDs at $\mathrm{T_{eff}}$=2200-2300 K can not reproduce simultaneously the  \textit{Spitzer} fluxes and the photometry shortward of 1.5 $\mu$m. The best fitted dilution factor at these $\mathrm{T_{eff}}$ corresponds to non physical radii (R$<$ 1 $R_{Jup}$). The photometry of AB Pic b and HR7329 B are best matched for $\mathrm{T_{eff}=}$1700-2000 K and log g=3.5-5.5, $\mathrm{T_{eff}=}$2200-2500 K and log g=4.0-5.0, respectively. We find corresponding radii of $\mathrm{7.11 \pm 0.03\:R_{Jup}}$ for KPNO Tau 4, $\mathrm{1.65 \pm 0.05\:R_{Jup}}$ for AB pic b, and $\mathrm{3.05 \pm 0.03\:R_{Jup}}$ for HR7329 B. Temperatures are in good agreement with those reported  in Table \ref{table:4}. These radii of KPNO Tau 4 and AB Pic b are however systematicaly higher than the mean values reported in Table \ref{Table:table5}. This can be interpreted as an overestimation of the scaling parameter (due possibly to a slight under-estimation of the effective temperature) or of the empirical luminosity. 
\item We repeated the procedure on the \textit{2MASS+Spitzer} colors of typical young M5-L0 dwarfs reported in \cite{2010ApJS..186..111L}. The fits are visually of good quality but are less constrained for dwarfs earlier than M9. The corresponding temperatures are ploted in Figure \ref{Figconvscales}. We also find a drop of  the effective temperature at the M-L transition, but for earlier spectral types with respect to the drop found from near-infrared spectra. 
\item We fitted the normalised optical spectra (see Figure \ref{spec_optical}) of 2M0141, Cha1109, OTS44, and KPNO Tau 4, USco CTIO 108B reported in \cite{2009AJ....137.3345C},  \cite{2008ApJ...675.1375L}, and \cite{2008ApJ...673L.185B}. We found effective temperatures of 1700-1900 K for  2M0141, 2300 K for Cha1109, 2400-2500 K for OTS44, 2200-2400 K for KPNO Tau 4, and 2000-2200 K for USco CTIO 108 B. The new temperatures found for the 1-3 Myr old object better agree with those inferred from the conversion scales. The visual inspection releals that synthetic spectra at $T_{eff}$=2200 and 2400 K and 3.5 $\leq$ log g $\leq$ 5.5 provide a good representation of the optical spectra  of the objets shortward of 0.8 $\mu$m, but fail to represent correctly the spectra from 0.8 to 1 $\mu$m (Figure \ref{spec_optical}). We also fitted the 0.6-2.5 $\mu$m spectra of KPNO Tau 4 and 2M0141 with the three set of models (Figure \ref{spec_opticalNIR}). Spectra were built scaling the JHK and the optical spectra to the low resolution SpecXprism spectra of these sources \citep{2006ApJ...639.1120K, 2007AJ....134..411M}.  The spectrum of 2M0141  is fitted at $T_{eff}$=1700-1800 K/log g=4.5 by the three models. We find alternative solutions at $T_{eff}=$1900 K/log g=3.5 and $T_{eff}=$2000 K/log g=4.5 with BT-SETTL10. They are similar to those found by \cite{2006ApJ...639.1120K} with older DUSTY-type atmospheric models. But these solutions are not confirmed with DRIFT-PHOENIX.  KPNO Tau 4 is well fitted by DRIFT-PHONIX and BT-SETTL10 models at $T_{eff}$=1700-1800 K/log g=3.5. BT-SETTL12 models give a poor fit to the spectrum of this source.
\item We compared synthetic spectra to the spectral energy distribution (optical, when available + \textit{2MASS} JHK + \textit{Spitzer} fluxes) of M7-L7 field dwarfs studied in \cite{2006ApJ...651..502P}. Best fitting solutions were checked visually and were discarded, if needed. The temperatures of the best fitted models are reported in Figure \ref{Figconvscales}. BT-SETTL spectra provide the best fits. DRIFT-PHOENIX fit well the optical and  the JHK fluxes but tend to predict too much high fluxes longward of 3 $\mu$m.  Both models gives $\mathrm{T_{eff}}$ estimates compatible with the \cite{2009ApJ...702..154S} scale for dwarfs later than L2.  The temperature of M7-L0 dwarfs fall $\sim$300 K lower than the aforementioned scale. However, our fits are less constrained at these temperatures. 
\item We finally fitted the J, H, K, and JHK band spectra of several late-M and early-L dwarfs of the IRTF and NIRSPEC libraries of \cite{2003ApJ...596..561M} and \cite{2005ApJ...623.1115C}. The BT-SETTL spectra reproduce the JHK pseudo-continuum slopes together with the depth of the main (atomic and molecular) absorptions. The H band flux is however badly reproduced and lead to inconsistent temperatures if fitted alone (as for 2M0345). The DRIFT-PHOENIX and BT-SETTL12 models predict too low temperatures for M9.5-L1 dwarfs. The effective temperatures corresponding to the best fitted spectra do not exceed 200 K for the different models for a given object. Spectroscopic temperatures agree with $\mathrm{T_{eff}/spectral\:type}$  conversion scales of field dwarfsif we decide to fit the K band spectrum only. This issue in the modeling of the M-L transition of field-dwarfs with the BT-SETTL models will be discussed in forthcoming paper (Allard et al. 2013, in preparation.)
\end{enumerate}

These tests demonstrate the ability of the three  atmospheric models to predict effective temperatures from the near-infrared spectra of young objects consistent with those found fitting the corresponding spectral energy distributions. Nevertheless, local non-reproducibilities in the optical shows that the atmosphere models and the related dust cloud models are not fully valid. Therefore, problems in the modeling of the dust might still affect the temperature predictions derived from fitting near-infared spectra alone.  For this reason, the analysis demonstrate that the companions that will be discovered by planet imager instruments (VLT/SPHERE, Gemini/GPI, Subaru/SCexAO, LBT/LMIRCam) should be characterised using complementary spectrophotometric observations completing the wavelength coverage accessible from these instruments.

\subsection{Binarity and variability}
\label{rembias}
An overluminosity caused by unresolved binarity or possible variability could  explain the inconsistency of evolutionary models predictions for the measured $\mathrm{T_{eff}}$ and luminosities of Cha1109, OTS44, and KPNO Tau 4. The multiplicity level of the Taurus and Chameleon targets has not been constrained for the whole separation/mass range, or is even unknown. \cite{2006ApJ...649..306K} put constraints at large separations ( $\gtrsim$20 AU) for KPNO Tau 4. As far as we know, the radial velocities of this source has not been monitored over a large temporal baseline to set additional limits at inner separations yet. No multiplicity studies has been conducted for the remaining Chameleon I targets. Then, future radial velocity and AO monitoring are needed.

 \cite {2002AJ....124.1001C} do not find significant photometric variability for OTS44.  The comparison of the low resolution spectra of KPNO Tau 4 \citep[][this work]{2006ApJ...646.1215L, 2007AJ....134..411M} do not reveal variability of the spectral lines (once smoothed at a homogeneous spectral resolution of $\sim$75). Howevever, the J band of the \cite{2006ApJ...646.1215L} spectrum is shifted to higher fluxes compared to the two the one of \cite{2007AJ....134..411M}. These two spectra were obtained in the J and H band \textit{simulaneously}. Therefore, photometric and/or spectroscopic monitoring is needed to understand if this shift comes from true  variability or rather from the instrument. Variability is also an issue for Cha1109.

\subsection{Initial conditions}
The  ``hot-start" evolutionary models of low mass stars and brown dwarfs used in section \ref{masslum} are known to be affected by several sources of uncertainties at very young ages  \citep{2002A&A...382..563B}. Their predictions have only been marginally tested given the difficulties to provide independent measurements of the physical properties of the objets (radii, masses, atmospheric parameters,...) at a high accuracy level \citep[see the review of \citet{2007prpl.conf..411M} and][]{2007AJ....133.2008K, 2008A&A...481..141L, 2009A&A...506..799B}. The fact that the deviation of semi-empirical radii with respect to models predictions is decreasing with age support the idea of a sensitivity of evolutionary tracks predictions to initial conditions.

2M1207 A, Cha1109, OTS 44, are known to be surrounded by disks.  KPNO Tau 4 has been reported as a possible accretor by \cite{2005ApJ...626..498M} although some accretion signatures could be of photospheric origin. Finally, \cite{2009ApJ...696.1589H} estimated an accretion rates of -11.9 $log M_{\odot}\:yr^{-1}$ for 2M1207 A and suggested this brown-dwarf could experience accretion variability.  

``hot-start" evolutionary models of \cite{2000ApJ...542..464C} suppose the object evolution starts from a hot, large, and non adiabatic sphere. These models do not account for accretion of disk materials and related effects on the initial energy budget of the object. \cite{2009ApJ...702L..27B} investigated the effect of accretion on the evolution of brown dwarfs and low-mass stars and demonstrated that radii and luminosities depends on the accretion. The new ``Warm-start" evolutionary models \citep{2012ApJ...745..174S, 2013arXiv1302.1517M} are also starting to explore the influence of initial conditions (initial entropy) for planetary-mass  objects. We used the ``warm-start" models of \cite{2012ApJ...745..174S} on OTS44, AB Pic b, KPNO Tau 4, and Cha1109 in order to determine which range of initial entropies and masses can reproduce the spectroscopic temperatures and the luminosities of these objects. Radii and $\mathrm{T_{eff}}$ for the object ages, masses of 1-15 $\mathrm{M_{Jup}}$, and initial entropies ranging from 8 to 13 Boltzmann constant  per baryon ($\mathrm{k_{b}}$/baryon), were kindly provided by the authors of the models and combined to derive luminosity predictions. The luminosity and temperature of KPNO Tau 4 fall outside of the bonds of the models. The spectroscopic temperature and luminosity of OTS 44  and Cha1109 can not be reconciled for the given range of  initial entropies and masses covered by the models. Therefore, this re-inforce the hypothesis that a bias exists in the temperature or luminosity determination of these objects. The models gives a lower limit on the mass of  OTS44 and Cha1109 of 8 an 6 $\mathrm{M_{Jup}}$ using the observed luminosity as input. Conversely, the temperature and luminosity of AB Pic b and USCO CTIO 108B are simultaneously reproduced by the models. The models give a lower limit of 10 and 8.5  $\mathrm{M_{Jup}}$ on the masses of these two companions respectively.

\subsection{Comparison with previous studies}
\label{subsec:comp}

The work presented here extend the analysis made by \cite{2003ApJ...593.1074G} to spectral types M9.5-L0 and enable to study in more details gravity-indicator in the near-infrarted spectra of young M9.5 and L0 dwarfs initiated by \cite{2004ApJ...600.1020M}, \cite{2007ApJ...657..511A}, and \cite{2006ApJ...639.1120K}. Our sample do not evidence any difference between the spectra of young  companions and isolated objects (taken from the library, or from the literrature). In particular, the near-infrared spectra of AB Pic b  ($\sim$ 30 Myr) and of the isolated 2M0141 are nearly identical in each individual bands, and also possibly simultaneously over the 1.1-2.5 $\mu$m range (see below). 2M0141 represents, along with 2MASS J035523.37+113343.7 \citep{2013AJ....145....2F},  the second young field L dwarfs identified in the field \citep{2009AJ....137.3345C} known to share  similarities with a young planetary mass companion.

\cite{2012A&A...540A..85P} also recently analysed  the SINFONI spectra of DH Tau B, 2M1207A, 2M0141, AB Pic b, GSC0847 B, and TWA 5 B\footnote{Note that we  published online some of our results at \url{http://tel.archives-ouvertes.fr/docs/00/55/54/96/PDF/THESE_enligne_MickaelBonnefoy.pdf} and \url{http://tel.archives-ouvertes.fr/docs/00/55/54/96/ANNEX/slides_soutenance_optim.pdf} a long time before the submision of their paper.}. In comparison, we obtained the spectra of five more objects here and re-reduced all  SINFONI data presented in \cite{2010A&A...512A..52B}, \cite{2012A&A...540A..85P}, and \cite{2008A&A...491..311S} using the latest data  analysis tools in order to make a complementary analysis.  

\cite{2012A&A...540A..85P} find a spread in effective temperatures from the fit of the J, HK, and JHK  spectra of the objects with synthetic spectra of a given amotpheric models, and also considering different models. We do not retrieve the spread in effective temperatures derived from the different bands using the same version of the BT-SETTL10 models. To our knowledge, the authors did not accounted for the fluctuations of the mean flux count level in science frames with long data integration time (see section \ref{subsec:datared}). The effective temperature estimates are tied to the conservation of the overal shape of the pseudo-continuum. This bias in the data analysis might then explain why the authors reach such conclusions.

Furthermore, we believe the JHK band spectra of AB Pic b, GSC0847 B,  and TWA 5 B can not be built from the individual J and H+K band spectra of these sources because of the lack of accurate near-infrared photometry.  Our new analysis of the NaCo J, H, and K-band data of AB Pic b reported in \cite{2005A&A...438L..29C}, and used in \cite{2012A&A...540A..85P} lead to a different near-infrared photometry (see Appendix \ref{ABPicbPhot}).   We provide in addition a new photometry of AB Pic b from wide-field  J and $\mathrm{K_{s}}$ archive data of the system in Appendix \ref{ABPicbPhot}. We find a 0.5 to 1 magnitude spread in these bands that can not be explained by the different -- but close -- associated photometric systems. This spread do not lead to a revision of the mass estimate for the object and do not affect the fit of the SED significantly (section \ref{modeluncert}. The corresponding near-infrared colors of AB Pic b ( H-$\mathrm{K_{s}}$=0.64 mag from the H+K SINFONI spectrum) are better compatible with those of other early-L type companions \citep[see Figure 2 of ][]{2013arXiv1302.1160B}. This confirm the results found independently by Biller et al. 2013 (submitted) from NICI data. The JHK spectrum of the source built from the new SofI photometry perfectly match the one of 2M0141. The spread in the photometry alone introduces an uncertainty of 300 K on the fit of the JHK band with the SETTL models. Similarly, the  combinaison of avaliable photometry  on GSC8047 B (see Appendix \ref{ABPicbPhot}) introduces an uncertainty of 200 K on the fit of the JHK band. 


\section{Conclusions}
We have obtained high-quality 1.1-2.5 $\mu$m medium resolution spectra of 16 young low mass objects.  We identified several features indicative of surface gravity intermediate between those of mature field dwarfs and late-type giants. Spectra were compared to empirical templates of young objects classified at optical wavelengths to re-assigned infrared spectral types in an homogeneous way. We selected and used 3 spectral indices slightly sensitive to the age to confirm the classifications. We then estimate DH Tau B is a M9.25$\pm$0.25, GSC8047B is a M9.5$\pm$0.5, and AB Pic b is a L0$\pm$1. 

We used three state-of-the art libraries of synthetic spectra BT-SETTL10, BT-SETTL12, and DRIFT-PHOENIX to derive atmospheric parameters estimates for each of the sources.  These libraries were selected as they account for the formation and the gravitational settling of dust grains in the photospheres of the objects. We found that the new atmospheric models  better represent the infrared spectra of  young M-L type dwarfs with respect to the previous generation of BT-SETTL models and to AMES-DUSTY models. Models notably reproduce well the shape of the pseudo-continuum of  the J, H and K band for close $\mathrm{T_{eff}}$ and log g, including the strength of water-band absorptions. 

We also  conducted a similar analysis on the re-extracted 1.1-2.5 $\mu$m spectra of the individual components of the tight binary TWA 22 AB.  We find that TWA 22 A and TWA 22 B are M5$\pm$1 and M5.5 $\pm$1 dwarfs respectively.  Spectra also have features intermediates between those of 5-8 Myr old M5-M6 dwarfs and those of mature field dwarfs analogues. We used the BT-SETTL10, BT-SETTL12,  and GAIA-COND models grids to constraint $\mathrm{T_{eff}=3000\pm100}$ K and log g=$4.5\pm0.5$ dex for both components.  Assuming the most recent age estimate for the system, the new $\mathrm{T_{eff}}$ estimates brings  the total system mass mass predicted by evolutionary models in agreement with the dynamical mass reported in \cite{2009A&A...506..799B}. \\

Nevertheless,  $\mathrm{T_{eff}}$ derived from atmospheric models indicate a quick decrease of the effective temperature at the M-L transition at young ages not observed for field dwarfs analogues.  We used empirically determined bolometric corrections of young M9.5 and L0 dwarfs to derive new luminosity estimates for objects with known distances. These luminosities were combined to the spectroscopic $\mathrm{T_{eff}}$ estimates to compute semi-empirical radii.  M9.5 members of 1-3 Myr clusters and star forming region have radii two to three times bigger than those predicted using evolutionary models. This inconsistency  is equivalent to an overprediction of the temperatures of 300 to 500 K  by the models with respect to the ones determined spectroscopically. We then ultimately examined the possible biases related to spectral modeling and concluded that:

\begin{itemize}
\item infrared excess and extinction can not solve the issue for all the late-type objects.
\item bias might still remain in these atmospheric models. The synthetic spectral libraries reproduce the near-infrared spectra and the spectral energy distribution of the young M9.5-L0 sources  simultaneously, for close temperatures. 0.6-2.5 $\mu$m spectral continua can also reproduced by a single BT-SETTL and DRIFT-PHOENIX synthetic spectrum at similar T$_{eff}$. However, the main absorptions in the optical are not reproduced coherently. Atmospheric models also fail to reproduce near-infrared spectra of mature dwarfs with increased surface gravities.
\end{itemize}

Our library  demonstrated the needs for empirical templates of young and late-type dwarfs for the development, on the long term, of an homogeneous classification scheme for young objects that account for surface gravity, cloud properties, temperature,and metallicity. It  could already be used to characterise low mass young companions that will be detected with planet imagers (Subaru/HiCIAO, VLT/SPHERE, and Gemini/GPI). In the later case, the analysis show that additional measurements outside the NIR range should be conducted routinely in order to identify excesses and to confirm the analysis derived from atmospheric models.  

\begin{appendix}
\section{Details on the equivalent widths computation}
\label{EWcomp}
The EW and associated error bars are computed following the method of \cite{1992ApJS...83..147S}. The pseudo-continuum is fitted using a third order Legendre polynomial locally around the line ( from $\lambda_{1}$ to $\lambda_{2}$, and from $\lambda_{3}$ to $\lambda_{4}$) and interpolated at the feature wavelengths (between $\lambda_{2}$ to $\lambda_{3}$). The reference wavelengths used to compute the EW displayed in Fig. \ref{FigEW} are reported in Tab. \ref{table:appA}.

\begin{table}
\begin{minipage}{\columnwidth}
\caption{Reference wavelengths used for the computation of equivalent widths of Na I and K I lines.}
\label{table:appA}
\centering
\begin{tabular}{lllll}     
\hline\hline
Line  & $\mathrm{\lambda_{1}}$ & $\mathrm{\lambda_{2}}$ & $\mathrm{\lambda_{3}}$ & $\mathrm{\lambda_{4}}$ \\
		&		($\mu m$) & ($\mu m$) & ($\mu m$) & ($\mu m$) \\
\hline
Na I 	-- 1.138 $\mu$m &  1.1270 & 1.1360 & 1.1420 & 1.1580 \\
K I -- 1.169 $\mu$m & 1.1560 & 1.1670 & 1.1710 & 1.1760 \\
K I -- 1.177 $\mu$m  & 1.1710 & 1.1750 & 1.1805 & 1.1820 \\
K I -- 1.243 $\mu$m  & 1.2390 & 1.2415 & 1.2455 & 1.2490 \\
K I -- 1.253 $\mu$m  & 1.2455 & 1.2500  & 1.2550  & 1.2670 \\
\hline
\end{tabular}
\end{minipage}
\end{table}

\section{New analysis of Lodieu et al. 2008 spectra}
\label{Lodieu2008an}
We re-assigned near-infrared spectra to the sample of young late-type dwarfs members of Upper Sco presented in \cite{2008MNRAS.383.1385L} using the method detailed in section \ref{classoptical} and based on optical standards. We report our results in Table \ref{tab:revLodieu}. Each spectrum from this library was also compared to synthetic spectra from the BT-SETTL libraries and to the DRIFT-PHOENIX grid. The best fitted temperatures and surface gravities are reported in Table \ref{tab:revLodieu}. The atmospheric parameters are in agreement with the values reported in Section \ref{partatmoparam} for dwarfs at similar spectral types.

 \label{App:Lod}
 \begin{table*}[t]
\begin{minipage}[ht]{\linewidth}
\caption{Revised spectral types (based on optical standards) and atmospheric parameters for the sample of near-infrared spectra of Upper Scorpius dwarfs used in section \ref{section:results}.}
\label{tab:revLodieu}
\begin{center}
\renewcommand{\footnoterule}{}  
\begin{tabular}{lllll}
\hline \hline 
Object					&		Spectral type		&	New spectral type 	&	$\mathrm{T_{eff}}$	&  log g  \\
							&									&									&		(K)						&				\\
\hline
UScoJ155419-213543	&	M8	&	M8			&	$2700_{-200}^{+100}$	& $4.0\pm0.5$ \\
UScoJ154722-213914	&	L0		&  M8.25	&	$2700\pm100$	&	$4.0\pm0.5$ \\
UScoJ160648-223040	&	M8	&	M8.25	& $2700\pm100$	& $4.0\pm0.5$  \\
UScoJ160847-223547	&	M9	&	M8.5			&	$2700\pm100$	&	$3.5\pm0.5$  \\
UScoJ160830-233511	&	M9	&	M8.5	&		$2600\pm100$	&	$3.5\pm0.5$  \\
UScoJ161047-223949	&	M9	&	M8.5			&		$2600\pm100$	&	$4.0\pm0.5$  \\
UScoJ160723-221102	&	L1		& M8.5		&$2500^{+200}_{-100}$&  $4.0\pm0.5$ \\
UScoJ160603-221930	&	L2		& M8.75		&	$2300\pm100$ &  $4.0\pm0.5$  \\
UScoJ160606-233513	&	L0		& M9&	$2400\pm100$   &		$3.5\pm0.5$   \\
UScoJ160714-232101	&	L0		& M9&	$2300\pm100$   &  	$3.5\pm0.5$  \\
UScoJ160818-223225	&	L0		& M9&	$2300\pm100$	&	$4.0\pm0.5$ \\
UScoJ160828-231510	&	L1		& M9&	$2300\pm100$	  &	$3.5\pm0.5$ \\
UScoJ160737-224247	&	L0		& M9.5&  $2300\pm100$   &  $3.5\pm0.5$  \\
UScoJ160727-223904	&	L1		& M9.5	&	$2300\pm100$	&	$4.0\pm0.5$  \\
UScoJ160737-224247	&	L0		& M9.5		&	$2300\pm100$  & 		$4.0\pm0.5$  \\
UScoJ161302-212428	&	L0		& M9.5	& $2400\pm100$  &		$4.0\pm0.5$  \\
UScoJ161228-215936	&	L1		& L0	&	$1800\pm100$	&$3.5\pm0.5$ \\
UScoJ161441-235105	&	L1		& L0 &	$1800\pm100$	&$3.5\pm0.5$  \\
UScoJ163919-253409	&	L1		& L0	&$1700\pm100$	&$3.5\pm0.5$ \\
UScoJ160918-222923	&	L1		& L0	& $1700\pm100	$	&$3.5\pm0.5$ \\
\hline
\end{tabular}
\end{center}
\end{minipage}
\end{table*}

\section{New near-infrared photometry for AB Pic b and  GSC08047-00232 B.}
\label{ABPicbPhot}
AB Pic was observed on March 19th, 2006 with the SofI near-infrared  spectrograph and imaging camera at NTT \citep{1998Msngr..91....9M}. Four exposures of 20 and 40$\times$1.2 seconds were recorded using the J and $\mathrm{K_{s}}$ band filters of the instrument, respectively. We reduced the data using the ESO $Eclipse$ software \citep{1997Msngr..87...19D}. The companion is resolved in both datasets.  The stars saturate the detector and does not enable to compute directly a contrast for the companion. We removed most of its flux subtracting a radial profile. A few dozen of unrelated sources are detected in the field of view (up to 6.9') of the final mozaic images. We then chose to perform relative astrometry using ten of the brightest sources with accurate 2MASS photometry. The results are reported in Table \ref{tab:photABPicb}. The error combine quadraticaly the mean error bar associated to the 2MASS magnitudes of the reference stars, and the scatter in magnitude found for the different references considered.

The discrepancy of this photometry with respect to the one reported in \cite{2005A&A...438L..29C} led us to re-analyse the NaCo data presented in this paper. We reported a new NaCo J, H, and $\mathrm{K_{s}}$ photometry of the companion with more realistic error bars in Table \ref{tab:photABPicb}.

We also estimated the magnitude of the companion in the L' band of NaCo and reported it in table \ref{tab:photABPicb}. We used for that purpose the contrast value reported in \cite{2013arXiv1302.5367R} \citep[but considering the measured transmission factor of the NaCo neutral density filter; see ][]{2013arXiv1302.1160B}. We assumed that the WISE W1 photometry and  L' band magnitude of the star are close. This seems to be the case since the W1-$K_{s}$ color of AB Pic corresponds well to L'-$\mathrm{K_{s}}$ colors of stars at the same spectral types \citep{1996A&AS..119..547V}.

GSC08047-00232 B was also observed with SofI in the J and $K_{s}$ bands on October 11, 2005.  We built calibrated images from the four raw frames recorded in each band. The photometry of the companion was extracted the same way as for AB Pic b. The near-infrared photometry of the companion is summarized in Table \ref{tab:photGSC08047B}.

 \begin{table}[t]
\begin{minipage}[ht]{\linewidth}
\caption{Revised and new near-infrared photometry of AB Pic b.}
\label{tab:photABPicb}
\begin{center}
\renewcommand{\footnoterule}{}  
\begin{tabular}{lllllll}
\hline \hline 
UT Date &  Band					&	Instrument & 	$\lambda_{c}$	&  $\mathrm{\Delta \lambda}$&  	Magnitude  \\
						&		  & 		& 				($\mu$m)  & ($\mu$m)  & 	   		     			          \\
\hline
17/03/2003  &  J 	& NaCo									&		1.265  &  0.250		&	$15.77\pm0.36$  \\
17/03/2003  & H	 & NaCo										&		1.660  &  0.330		&	$14.64\pm0.15$  \\
25/09/2004  & H	& NaCo										&		1.660  &  0.330		&	$14.89\pm0.18$  \\
08/01/2006  & H 	& NaCo									&		1.660  &  0.330		&	$14.65\pm0.17$   \\
17/03/2003  & $\mathrm{K_{s}}$	& NaCo  		&		2.180  & 0.350		&	$14.09\pm0.07$   \\
26/11/2009 & L'						&		NaCo    &	  3.80		&		0.62   &   $13.20\pm0.10$     \\
19/03/2006  & J  & SofI 		&		1.247  &  0.290 &  $15.20\pm0.31$  \\   
19/03/2006  &$\mathrm{K_{s}}$ & SofI 		&		2.162  &  0.275 &  $13.65\pm0.20$   \\   
\hline
\end{tabular}
\end{center}
\end{minipage}
\end{table}

\begin{table}[t]
\begin{minipage}[ht]{\linewidth}
\caption{Near-infrared photometry of GSC08047-00232 B.}
\label{tab:photGSC08047B}
\begin{center}
\renewcommand{\footnoterule}{}  
\begin{tabular}{llllll}
\hline \hline 
Band					&	Instrument & 	$\lambda_{c}$	&  $\mathrm{\Delta \lambda}$ &  	Magnitude & Ref.	  \\
						&		  & 		($\mu$m)		&	($\mu$m) 	 & 	   		 &     			          \\
\hline
J	& Adonis									&		1.253  & 0.296		&	$16.25\pm0.25$ &  1  \\
H	& Adonis										&		1.643  &  0.353		&	$15.20\pm0.18$  &  1\\
$\mathrm{K}$			&  Adonis 		&		2.177  & 0.378		&	$14.9\pm0.2$  &  1 \\
$\mathrm{K}$			&  Adonis 		&		2.177  & 0.378		&	$15.0\pm0.3$  &  2 \\
J	& NaCo									&		1.265  &  0.250		&	$15.90\pm0.13$ &  3  \\
H	& NaCo										&		1.660  &  0.330		&	$15.45\pm0.20$  &  3\\
$\mathrm{K_{s}}$			& NaCo  		&		2.180  & 0.350		&	$14.75\pm0.13$  &  4 \\
J  & SofI 		&		1.247  &  0.290 &  $16.56\pm0.21$ &  4  \\   
$\mathrm{K_{s}}$ & SofI 		&		2.162  &  0.275 &  $15.17\pm0.34$  &  4 \\   
\hline
\end{tabular}
\end{center}
\end{minipage}
\textbf{References:} (1) - \cite{2003A&A...404..157C} ; (2) - \cite{2004A&A...420..647N} ; (3) - \cite{2005A&A...438L..29C}; (4) - This work
\end{table}

\section{Optical fluxes of young companions}
\label{OpticPhot}
AB Pic b and HR7329 B were observed with the Hubble/Wide Field Camera 3 instrument  on and January 24 and February 16, 2012, respectively. We used flux-calibrated images provided by the ESA Hubble Science Archive system\footnote{\url{http://archives.esac.esa.int/hst/}} of the sources corresponding to observations performed with 10 filters with central wavelengths from 3354.8 $\AA$ to 9158.8 $\AA$. We removed the stellar halo using a radial profile of the source or a low-pass filter and integrated the flux of the companion in 5-pixels width circular appertures. The companions were too faint to be detected shortward of 4.77 $\mu$m. We then rather estimated upper limits on the flux at these wavelengths. Results are reported in Table \ref{tab:photoptic}.

 \begin{table}[t]
\begin{minipage}[ht]{\linewidth}
\caption{Hubble/WFC3 flux densities for young companions and isolated brow-dwarfs of our sample with avaliable photometric data longward \textbf{of} 3 $\mu$m.}
\label{tab:photoptic}
\begin{center}
\renewcommand{\footnoterule}{}  
\begin{tabular}{lllllll}
\hline \hline 
Filter					&	$\lambda_{c}$	&   $\mathrm{\Delta \lambda}$&  	$\mathrm{\lambda \times F_{\lambda}}$  & $\mathrm{\lambda \times F_{\lambda}}$     \\
			&    &   &   AB Pic b  &  HR7329 B	  \\
						&		 	($\AA$)		&		($\AA$)  &($10^{-17} \mathrm{W.m^{-2}}$)	& ($10^{-15} \mathrm{W.m^{-2}}$) \\ 
\hline
F850LP 	& 9158.8		&		467.53  	  & $55.98\pm1.61$   & $21.17\pm0.19$  \\
F775W		&  7648.5   &  419.11		&		$13.41\pm0.56$  & $7.81\pm0.12$     \\
F673N		&		6765.9  &  41.92  &  		$3.71\pm0.91$  & $0.71\pm0.22$     \\
F656N 		&		6561.4  & 41.58  &  $3.75\pm1.65$  & $2.74\pm0.18$     \\
F625W		&     6241.3   &  451.03   &  $2.17\pm0.30$  & $0.78\pm0.20$     \\
F555W     &   5308.1   &  517.14   &  $0.44\pm0.25$  & $0.25\pm0.23$    \\
F475W		&  4773.7   &   421.33   &  $\geq 1.53$   & $\geq 1.71 $    \\
F390W		&   3924.4   &  291.22   &  $\geq 0.97$   & $\geq 2.27 $    \\
F336W     &   3354.8   &  158.44   &   $\geq 0.63$  & $ \geq 1.54 $       \\
\hline
\end{tabular}
\end{center}
\end{minipage}
\end{table}

\end{appendix}

\begin{acknowledgements}
We thank the ESO Paranal staff for performing the service-mode observations and for their support on the data reduction. We acknowledge partial financial support from the \textit{Agence National de la Recherche} and the \textit{Programmes Nationaux de Planétologie et de Physique Stellaire} (PNP \& PNPS), in France. C. Pinte acknowledges funding from the European Commission's 7$^\mathrm{th}$ Framework Program (contract PERG06-GA-2009-256513) and from Agence Nationale pour la Recherche (ANR) of France under contract ANR-2010-JCJC-0504-01. G. Chauvin acknowledges funding from  Agence Nationale pour la Recherche (ANR) of France through project grant ANR10-BLANC0504-01. We thank in particular Christiane Helling, Sorren Witte, and Peter Hauschildt for developping and providing the DRIFT-PHOENIX grid. We are gratefull to David Spiegel, Adam Burrows, and Gabriel-Dominique Marleau for giving us complementary predictions from the  ``warm-start" models and/or discussing the validity of the models. We also thank Kevin Luhman, Kelle Cruz, Mark Marley, Beth Biller, and Agnes Lèbre for fruitfull discussions about the characterization of young objects, the photometry of AB Pic b,  and on the spectroscopic properties of Mira variables.  We are grateful to Emily Rice, Nadya Gorlova, Katelyn Allers, Laird Close, Victor Bejar, and August Muench for providing their spectra. We also acknowledge  Ian S. McLean, Michael C. Cushing, John T. Rayner, Yoichi Itoh, Nicolas Lodieu, and Kelle Cruz for providing a free online access to their spectral libraries. This research has also benefited from the SpeX Prism Spectral Libraries, maintained by Adam Burgasser at \url{http://pono.ucsd.edu/~adam/browndwarfs/spexprism}. 
\end{acknowledgements}

\bibliographystyle{aa}
\bibliography{SINFlib_Bonnefoy}

\begin{thebibliography}{199}
\expandafter\ifx\csname natexlab\endcsname\relax\def\natexlab#1{#1}\fi

\bibitem[{{Abuter} {et~al.}(2006){Abuter}, {Schreiber}, {Eisenhauer}, {Ott},
  {Horrobin}, \& {Gillesen}}]{2006NewAR..50..398A}
{Abuter}, R., {Schreiber}, J., {Eisenhauer}, F., {et~al.} 2006, New Astronomy
  Review, 50, 398

\bibitem[{{Ali} {et~al.}(1995){Ali}, {Carr}, {Depoy}, {Frogel}, \&
  {Sellgren}}]{1995AJ....110.2415A}
{Ali}, B., {Carr}, J.~S., {Depoy}, D.~L., {Frogel}, J.~A., \& {Sellgren}, K.
  1995, \aj, 110, 2415

\bibitem[{{Allard} {et~al.}(2001){Allard}, {Hauschildt}, {Alexander},
  {Tamanai}, \& {Schweitzer}}]{2001ApJ...556..357A}
{Allard}, F., {Hauschildt}, P.~H., {Alexander}, D.~R., {Tamanai}, A., \&
  {Schweitzer}, A. 2001, \apj, 556, 357

\bibitem[{{Allard} {et~al.}(2011){Allard}, {Homeier}, \&
  {Freytag}}]{2011ASPC..448...91A}
{Allard}, F., {Homeier}, D., \& {Freytag}, B. 2011, in Astronomical Society of
  the Pacific Conference Series, Vol. 448, 16th Cambridge Workshop on Cool
  Stars, Stellar Systems, and the Sun, ed. C.~{Johns-Krull}, M.~K. {Browning},
  \& A.~A. {West}, 91

\bibitem[{{Allard} {et~al.}(2013){Allard}, {Homeier}, {Freytag},
  {Schaffenberger}, \& {Rajpurohit}}]{2013arXiv1302.6559A}
{Allard}, F., {Homeier}, D., {Freytag}, B., {Schaffenberger}, W., \&
  {Rajpurohit}, A.~S. 2013, ArXiv e-prints

\bibitem[{{Allard} {et~al.}(2012){Allard}, {Homeier}, {Freytag}, \&
  {Sharp}}]{2012EAS....57....3A}
{Allard}, F., {Homeier}, D., {Freytag}, B., \& {Sharp}, C.~M. 2012, in EAS
  Publications Series, Vol.~57, EAS Publications Series, ed. C.~{Reyl{\'e}},
  C.~{Charbonnel}, \& M.~{Schultheis}, 3--43

\bibitem[{{Allers} {et~al.}(2007){Allers}, {Jaffe}, {Luhman}, {Liu}, {Wilson},
  {Skrutskie}, {Nelson}, {Peterson}, {Smith}, \&
  {Cushing}}]{2007ApJ...657..511A}
{Allers}, K.~N., {Jaffe}, D.~T., {Luhman}, K.~L., {et~al.} 2007, \apj, 657, 511

\bibitem[{{Allers} {et~al.}(2010){Allers}, {Liu}, {Dupuy}, \&
  {Cushing}}]{2010ApJ...715..561A}
{Allers}, K.~N., {Liu}, M.~C., {Dupuy}, T.~J., \& {Cushing}, M.~C. 2010, \apj,
  715, 561

\bibitem[{{Allers} {et~al.}(2009){Allers}, {Liu}, {Shkolnik}, {Cushing},
  {Dupuy}, {Mathews}, {Reid}, {Cruz}, \& {Vacca}}]{2009ApJ...697..824A}
{Allers}, K.~N., {Liu}, M.~C., {Shkolnik}, E., {et~al.} 2009, \apj, 697, 824

\bibitem[{{Alvarez} {et~al.}(2000){Alvarez}, {Lan{\c c}on}, {Plez}, \&
  {Wood}}]{2000A&A...353..322A}
{Alvarez}, R., {Lan{\c c}on}, A., {Plez}, B., \& {Wood}, P.~R. 2000, \aap, 353,
  322

\bibitem[{{Ammler-von Eiff} \& {Guenther}(2009)}]{2009A&A...508..677A}
{Ammler-von Eiff}, M. \& {Guenther}, E.~W. 2009, \aap, 508, 677

\bibitem[{{Asplund} {et~al.}(2009){Asplund}, {Grevesse}, {Sauval}, \&
  {Scott}}]{2009ARA&A..47..481A}
{Asplund}, M., {Grevesse}, N., {Sauval}, A.~J., \& {Scott}, P. 2009, \araa, 47,
  481

\bibitem[{{Auman}(1967)}]{1967ApJS...14..171A}
{Auman}, Jr., J. 1967, \apjs, 14, 171

\bibitem[{{Bannister} \& {Jameson}(2007)}]{2007MNRAS.378L..24B}
{Bannister}, N.~P. \& {Jameson}, R.~F. 2007, \mnras, 378, L24

\bibitem[{{Baraffe} {et~al.}(1998){Baraffe}, {Chabrier}, {Allard}, \&
  {Hauschildt}}]{1998A&A...337..403B}
{Baraffe}, I., {Chabrier}, G., {Allard}, F., \& {Hauschildt}, P.~H. 1998, \aap,
  337, 403

\bibitem[{{Baraffe} {et~al.}(2002){Baraffe}, {Chabrier}, {Allard}, \&
  {Hauschildt}}]{2002A&A...382..563B}
{Baraffe}, I., {Chabrier}, G., {Allard}, F., \& {Hauschildt}, P.~H. 2002, \aap,
  382, 563

\bibitem[{{Baraffe} {et~al.}(2009){Baraffe}, {Chabrier}, \&
  {Gallardo}}]{2009ApJ...702L..27B}
{Baraffe}, I., {Chabrier}, G., \& {Gallardo}, J. 2009, \apjl, 702, L27

\bibitem[{{Barenfeld} {et~al.}(2013){Barenfeld}, {Bubar}, {Mamajek}, \&
  {Young}}]{2013arXiv1301.5036B}
{Barenfeld}, S.~A., {Bubar}, E.~J., {Mamajek}, E.~E., \& {Young}, P.~A. 2013,
  ArXiv e-prints

\bibitem[{{Barman} {et~al.}(2011){Barman}, {Macintosh}, {Konopacky}, \&
  {Marois}}]{2011ApJ...733...65B}
{Barman}, T.~S., {Macintosh}, B., {Konopacky}, Q.~M., \& {Marois}, C. 2011,
  \apj, 733, 65

\bibitem[{{Bate}(2012)}]{2012MNRAS.419.3115B}
{Bate}, M.~R. 2012, \mnras, 419, 3115

\bibitem[{{Becklin} \& {Zuckerman}(1988)}]{1988Natur.336..656B}
{Becklin}, E.~E. \& {Zuckerman}, B. 1988, \nat, 336, 656

\bibitem[{{B{\'e}jar} {et~al.}(2008){B{\'e}jar}, {Zapatero Osorio},
  {P{\'e}rez-Garrido}, {{\'A}lvarez}, {Mart{\'{\i}}n}, {Rebolo},
  {Vill{\'o}-P{\'e}rez}, \& {D{\'{\i}}az-S{\'a}nchez}}]{2008ApJ...673L.185B}
{B{\'e}jar}, V.~J.~S., {Zapatero Osorio}, M.~R., {P{\'e}rez-Garrido}, A.,
  {et~al.} 2008, \apjl, 673, L185

\bibitem[{{Biazzo} {et~al.}(2012){Biazzo}, {D'Orazi}, {Desidera}, {Covino},
  {Alcal{\'a}}, \& {Zusi}}]{2012MNRAS.427.2905B}
{Biazzo}, K., {D'Orazi}, V., {Desidera}, S., {et~al.} 2012, \mnras, 427, 2905

\bibitem[{{Bihain} {et~al.}(2010){Bihain}, {Rebolo}, {Zapatero Osorio},
  {B{\'e}jar}, \& {Caballero}}]{2010A&A...519A..93B}
{Bihain}, G., {Rebolo}, R., {Zapatero Osorio}, M.~R., {B{\'e}jar}, V.~J.~S., \&
  {Caballero}, J.~A. 2010, \aap, 519, A93

\bibitem[{{Boccaletti} {et~al.}(2008){Boccaletti}, {Chauvin}, {Baudoz}, \&
  {Beuzit}}]{2008A&A...482..939B}
{Boccaletti}, A., {Chauvin}, G., {Baudoz}, P., \& {Beuzit}, J. 2008, \aap, 482,
  939

\bibitem[{{Boley}(2009)}]{2009ApJ...695L..53B}
{Boley}, A.~C. 2009, \apjl, 695, L53

\bibitem[{{Bonnefoy} {et~al.}(2013){Bonnefoy}, {Boccaletti}, {Lagrange},
  {Allard}, {Mordasini}, {Beust}, {Chauvin}, {Girard}, {Homeier}, {Apai},
  {Lacour}, \& {Rouan}}]{2013arXiv1302.1160B}
{Bonnefoy}, M., {Boccaletti}, A., {Lagrange}, A.-M., {et~al.} 2013, ArXiv
  e-prints

\bibitem[{{Bonnefoy} {et~al.}(2009){Bonnefoy}, {Chauvin}, {Dumas}, {Lagrange},
  {Beust}, {Desort}, {Teixeira}, {Ducourant}, {Beuzit}, \&
  {Song}}]{2009A&A...506..799B}
{Bonnefoy}, M., {Chauvin}, G., {Dumas}, C., {et~al.} 2009, \aap, 506, 799

\bibitem[{{Bonnefoy} {et~al.}(2010){Bonnefoy}, {Chauvin}, {Rojo}, {Allard},
  {Lagrange}, {Homeier}, {Dumas}, \& {Beuzit}}]{2010A&A...512A..52B}
{Bonnefoy}, M., {Chauvin}, G., {Rojo}, P., {et~al.} 2010, \aap, 512, A52+

\bibitem[{{Bonnet} {et~al.}(2004){Bonnet}, {Abuter}, {Baker}, {Bornemann},
  {Brown}, {Castillo}, {Conzelmann}, {Damster}, {Davies}, {Delabre},
  {Donaldson}, {Dumas}, {Eisenhauer}, {Elswijk}, {Fedrigo}, {Finger},
  {Gemperlein}, {Genzel}, {Gilbert}, {Gillet}, {Goldbrunner}, {Horrobin}, {Ter
  Horst}, {Huber}, {Hubin}, {Iserlohe}, {Kaufer}, {Kissler-Patig}, {Kragt},
  {Kroes}, {Lehnert}, {Lieb}, {Liske}, {Lizon}, {Lutz}, {Modigliani}, {Monnet},
  {Nesvadba}, {Patig}, {Pragt}, {Reunanen}, {R{\"o}hrle}, {Rossi}, {Schmutzer},
  {Schoenmaker}, {Schreiber}, {Stroebele}, {Szeifert}, {Tacconi}, {Tecza},
  {Thatte}, {Tordo}, {van der Werf}, \& {Weisz}}]{2004Msngr.117...17B}
{Bonnet}, H., {Abuter}, R., {Baker}, A., {et~al.} 2004, The Messenger, 117, 17

\bibitem[{{Bonnet} {et~al.}(2003){Bonnet}, {Str{\"o}bele}, {Biancat-Marchet},
  {Brynnel}, {Conzelmann}, {Delabre}, {Donaldson}, {Farinato}, {Fedrigo},
  {Hubin}, {Kasper}, \& {Kissler-Patig}}]{2003SPIE.4839..329B}
{Bonnet}, H., {Str{\"o}bele}, S., {Biancat-Marchet}, F., {et~al.} 2003, in
  Presented at the Society of Photo-Optical Instrumentation Engineers (SPIE)
  Conference, Vol. 4839, Society of Photo-Optical Instrumentation Engineers
  (SPIE) Conference Series, ed. {P.~L.~Wizinowich \& D.~Bonaccini}, 329--343

\bibitem[{{Borysow} {et~al.}(1997){Borysow}, {Jorgensen}, \&
  {Zheng}}]{1997A&A...324..185B}
{Borysow}, A., {Jorgensen}, U.~G., \& {Zheng}, C. 1997, \aap, 324, 185

\bibitem[{{Bouy} {et~al.}(2003){Bouy}, {Brandner}, {Mart{\'{\i}}n}, {Delfosse},
  {Allard}, \& {Basri}}]{2003AJ....126.1526B}
{Bouy}, H., {Brandner}, W., {Mart{\'{\i}}n}, E.~L., {et~al.} 2003, \aj, 126,
  1526

\bibitem[{{Bowler} {et~al.}(2011){Bowler}, {Liu}, {Kraus}, {Mann}, \&
  {Ireland}}]{2011ApJ...743..148B}
{Bowler}, B.~P., {Liu}, M.~C., {Kraus}, A.~L., {Mann}, A.~W., \& {Ireland},
  M.~J. 2011, \apj, 743, 148

\bibitem[{{Bowler} {et~al.}(2012){Bowler}, {Liu}, {Shkolnik}, {Dupuy}, {Cieza},
  {Kraus}, \& {Tamura}}]{2012ApJ...753..142B}
{Bowler}, B.~P., {Liu}, M.~C., {Shkolnik}, E.~L., {et~al.} 2012, \apj, 753, 142

\bibitem[{{Brice{\~n}o} {et~al.}(2002){Brice{\~n}o}, {Luhman}, {Hartmann},
  {Stauffer}, \& {Kirkpatrick}}]{2002ApJ...580..317B}
{Brice{\~n}o}, C., {Luhman}, K.~L., {Hartmann}, L., {Stauffer}, J.~R., \&
  {Kirkpatrick}, J.~D. 2002, \apj, 580, 317

\bibitem[{{Brott} \& {Hauschildt}(2005)}]{2005ESASP.576..565B}
{Brott}, I. \& {Hauschildt}, P.~H. 2005, in ESA Special Publication, Vol. 576,
  The Three-Dimensional Universe with Gaia, ed. {C.~Turon, K.~S.~O'Flaherty, \&
  M.~A.~C.~Perryman}, 565--+

\bibitem[{{Burgasser} {et~al.}(2002){Burgasser}, {Kirkpatrick}, {Brown},
  {Reid}, {Burrows}, {Liebert}, {Matthews}, {Gizis}, {Dahn}, {Monet}, {Cutri},
  \& {Skrutskie}}]{2002ApJ...564..421B}
{Burgasser}, A.~J., {Kirkpatrick}, J.~D., {Brown}, M.~E., {et~al.} 2002, \apj,
  564, 421

\bibitem[{{Burrows} {et~al.}(2001){Burrows}, {Hubbard}, {Lunine}, \&
  {Liebert}}]{2001RvMP...73..719B}
{Burrows}, A., {Hubbard}, W.~B., {Lunine}, J.~I., \& {Liebert}, J. 2001,
  Reviews of Modern Physics, 73, 719

\bibitem[{{Caffau} {et~al.}(2011){Caffau}, {Ludwig}, {Steffen}, {Freytag}, \&
  {Bonifacio}}]{2011SoPh..268..255C}
{Caffau}, E., {Ludwig}, H.-G., {Steffen}, M., {Freytag}, B., \& {Bonifacio}, P.
  2011, \solphys, 268, 255

\bibitem[{{Carpenter} {et~al.}(2002){Carpenter}, {Hillenbrand}, {Skrutskie}, \&
  {Meyer}}]{2002AJ....124.1001C}
{Carpenter}, J.~M., {Hillenbrand}, L.~A., {Skrutskie}, M.~F., \& {Meyer}, M.~R.
  2002, \aj, 124, 1001

\bibitem[{{Carson} {et~al.}(2013){Carson}, {Thalmann}, {Janson}, {Kozakis},
  {Bonnefoy}, {Biller}, {Schlieder}, {Currie}, {McElwain}, {Goto}, {Henning},
  {Brandner}, {Feldt}, {Kandori}, {Kuzuhara}, {Stevens}, {Wong}, {Gainey},
  {Fukagawa}, {Kuwada}, {Brandt}, {Kwon}, {Abe}, {Egner}, {Grady}, {Guyon},
  {Hashimoto}, {Hayano}, {Hayashi}, {Hayashi}, {Hodapp}, {Ishii}, {Iye},
  {Knapp}, {Kudo}, {Kusakabe}, {Matsuo}, {Miyama}, {Morino}, {Moro-Martin},
  {Nishimura}, {Pyo}, {Serabyn}, {Suto}, {Suzuki}, {Takami}, {Takato},
  {Terada}, {Tomono}, {Turner}, {Watanabe}, {Wisniewski}, {Yamada}, {Takami},
  {Usuda}, \& {Tamura}}]{2013ApJ...763L..32C}
{Carson}, J., {Thalmann}, C., {Janson}, M., {et~al.} 2013, \apjl, 763, L32

\bibitem[{{Castellani} {et~al.}(2002){Castellani}, {Degl'Innocenti}, {Prada
  Moroni}, \& {Tordiglione}}]{2002MNRAS.334..193C}
{Castellani}, V., {Degl'Innocenti}, S., {Prada Moroni}, P.~G., \&
  {Tordiglione}, V. 2002, \mnras, 334, 193

\bibitem[{{Chabrier} {et~al.}(2000){Chabrier}, {Baraffe}, {Allard}, \&
  {Hauschildt}}]{2000ApJ...542..464C}
{Chabrier}, G., {Baraffe}, I., {Allard}, F., \& {Hauschildt}, P. 2000, \apj,
  542, 464

\bibitem[{{Chauvin} {et~al.}(2010){Chauvin}, {Lagrange}, {Bonavita},
  {Zuckerman}, {Dumas}, {Bessell}, {Beuzit}, {Bonnefoy}, {Desidera}, {Farihi},
  {Lowrance}, {Mouillet}, \& {Song}}]{2010A&A...509A..52C}
{Chauvin}, G., {Lagrange}, A., {Bonavita}, M., {et~al.} 2010, \aap, 509, A52+

\bibitem[{{Chauvin} {et~al.}(2004){Chauvin}, {Lagrange}, {Dumas}, {Zuckerman},
  {Mouillet}, {Song}, {Beuzit}, \& {Lowrance}}]{2004A&A...425L..29C}
{Chauvin}, G., {Lagrange}, A., {Dumas}, C., {et~al.} 2004, \aap, 425, L29

\bibitem[{{Chauvin} {et~al.}(2005{\natexlab{a}}){Chauvin}, {Lagrange},
  {Lacombe}, {Dumas}, {Mouillet}, {Zuckerman}, {Gendron}, {Song}, {Beuzit},
  {Lowrance}, \& {Fusco}}]{2005A&A...430.1027C}
{Chauvin}, G., {Lagrange}, A., {Lacombe}, F., {et~al.} 2005{\natexlab{a}},
  \aap, 430, 1027

\bibitem[{{Chauvin} {et~al.}(2005{\natexlab{b}}){Chauvin}, {Lagrange},
  {Zuckerman}, {Dumas}, {Mouillet}, {Song}, {Beuzit}, {Lowrance}, \&
  {Bessell}}]{2005A&A...438L..29C}
{Chauvin}, G., {Lagrange}, A., {Zuckerman}, B., {et~al.} 2005{\natexlab{b}},
  \aap, 438, L29

\bibitem[{{Chauvin} {et~al.}(2003){Chauvin}, {Thomson}, {Dumas}, {Beuzit},
  {Lowrance}, {Fusco}, {Lagrange}, {Zuckerman}, \&
  {Mouillet}}]{2003A&A...404..157C}
{Chauvin}, G., {Thomson}, M., {Dumas}, C., {et~al.} 2003, \aap, 404, 157

\bibitem[{{Close} {et~al.}(2005){Close}, {Lenzen}, {Guirado}, {Nielsen},
  {Mamajek}, {Brandner}, {Hartung}, {Lidman}, \&
  {Biller}}]{2005Natur.433..286C}
{Close}, L.~M., {Lenzen}, R., {Guirado}, J.~C., {et~al.} 2005, \nat, 433, 286

\bibitem[{{Close} {et~al.}(2007){Close}, {Thatte}, {Nielsen}, {Abuter},
  {Clarke}, \& {Tecza}}]{2007ApJ...665..736C}
{Close}, L.~M., {Thatte}, N., {Nielsen}, E.~L., {et~al.} 2007, \apj, 665, 736

\bibitem[{{Cohen} \& {Kuhi}(1976)}]{1976PASP...88..535C}
{Cohen}, M. \& {Kuhi}, L.~V. 1976, \pasp, 88, 535

\bibitem[{{Cruz} {et~al.}(2009){Cruz}, {Kirkpatrick}, \&
  {Burgasser}}]{2009AJ....137.3345C}
{Cruz}, K.~L., {Kirkpatrick}, J.~D., \& {Burgasser}, A.~J. 2009, \aj, 137, 3345

\bibitem[{{Cushing} {et~al.}(2003){Cushing}, {Rayner}, {Davis}, \&
  {Vacca}}]{2003ApJ...582.1066C}
{Cushing}, M.~C., {Rayner}, J.~T., {Davis}, S.~P., \& {Vacca}, W.~D. 2003,
  \apj, 582, 1066

\bibitem[{{Cushing} {et~al.}(2005){Cushing}, {Rayner}, \&
  {Vacca}}]{2005ApJ...623.1115C}
{Cushing}, M.~C., {Rayner}, J.~T., \& {Vacca}, W.~D. 2005, \apj, 623, 1115

\bibitem[{{Dahn} {et~al.}(2002){Dahn}, {Harris}, {Vrba}, {Guetter}, {Canzian},
  {Henden}, {Levine}, {Luginbuhl}, {Monet}, {Monet}, {Pier}, {Stone}, {Walker},
  {Burgasser}, {Gizis}, {Kirkpatrick}, {Liebert}, \&
  {Reid}}]{2002AJ....124.1170D}
{Dahn}, C.~C., {Harris}, H.~C., {Vrba}, F.~J., {et~al.} 2002, \aj, 124, 1170

\bibitem[{{Delorme} {et~al.}(2013){Delorme}, {Gagn{\'e}}, {Girard}, {Lagrange},
  {Chauvin}, {Naud}, {Lafreni{\`e}re}, {Doyon}, {Riedel}, {Bonnefoy}, \&
  {Malo}}]{2013arXiv1303.4525D}
{Delorme}, P., {Gagn{\'e}}, J., {Girard}, J.~H., {et~al.} 2013, ArXiv e-prints

\bibitem[{{Devillard}(1997)}]{1997Msngr..87...19D}
{Devillard}, N. 1997, The Messenger, 87, 19

\bibitem[{{Draine}(2003{\natexlab{a}})}]{2003ARA&A..41..241D}
{Draine}, B.~T. 2003{\natexlab{a}}, \araa, 41, 241

\bibitem[{{Draine}(2003{\natexlab{b}})}]{2003ApJ...598.1017D}
{Draine}, B.~T. 2003{\natexlab{b}}, \apj, 598, 1017

\bibitem[{{Draine}(2003{\natexlab{c}})}]{2003ApJ...598.1026D}
{Draine}, B.~T. 2003{\natexlab{c}}, \apj, 598, 1026

\bibitem[{{Ducourant} {et~al.}(2008){Ducourant}, {Teixeira}, {Chauvin},
  {Daigne}, {Le Campion}, {Song}, \& {Zuckerman}}]{2008A&A...477L...1D}
{Ducourant}, C., {Teixeira}, R., {Chauvin}, G., {et~al.} 2008, \aap, 477, L1

\bibitem[{{Eisenhauer} {et~al.}(2003){Eisenhauer}, {Abuter}, {Bickert},
  {Biancat-Marchet}, {Bonnet}, {Brynnel}, {Conzelmann}, {Delabre}, {Donaldson},
  {Farinato}, {Fedrigo}, {Genzel}, {Hubin}, {Iserlohe}, {Kasper},
  {Kissler-Patig}, {Monnet}, {Roehrle}, {Schreiber}, {Stroebele}, {Tecza},
  {Thatte}, \& {Weisz}}]{2003SPIE.4841.1548E}
{Eisenhauer}, F., {Abuter}, R., {Bickert}, K., {et~al.} 2003, in Presented at
  the Society of Photo-Optical Instrumentation Engineers (SPIE) Conference,
  Vol. 4841, Society of Photo-Optical Instrumentation Engineers (SPIE)
  Conference Series, ed. {M.~Iye \& A.~F.~M.~Moorwood}, 1548--1561

\bibitem[{{Faherty} {et~al.}(2013){Faherty}, {Rice}, {Cruz}, {Mamajek}, \&
  {N{\'u}{\~n}ez}}]{2013AJ....145....2F}
{Faherty}, J.~K., {Rice}, E.~L., {Cruz}, K.~L., {Mamajek}, E.~E., \&
  {N{\'u}{\~n}ez}, A. 2013, \aj, 145, 2

\bibitem[{{Fern{\'a}ndez} {et~al.}(2008){Fern{\'a}ndez}, {Figueras}, \&
  {Torra}}]{2008A&A...480..735F}
{Fern{\'a}ndez}, D., {Figueras}, F., \& {Torra}, J. 2008, \aap, 480, 735

\bibitem[{{Fortney} {et~al.}(2008){Fortney}, {Marley}, {Saumon}, \&
  {Lodders}}]{2008ApJ...683.1104F}
{Fortney}, J.~J., {Marley}, M.~S., {Saumon}, D., \& {Lodders}, K. 2008, \apj,
  683, 1104

\bibitem[{{Geballe} {et~al.}(2002){Geballe}, {Knapp}, {Leggett}, {Fan},
  {Golimowski}, {Anderson}, {Brinkmann}, {Csabai}, {Gunn}, {Hawley},
  {Hennessy}, {Henry}, {Hill}, {Hindsley}, {Ivezi{\'c}}, {Lupton}, {McDaniel},
  {Munn}, {Narayanan}, {Peng}, {Pier}, {Rockosi}, {Schneider}, {Smith},
  {Strauss}, {Tsvetanov}, {Uomoto}, {York}, \& {Zheng}}]{2002ApJ...564..466G}
{Geballe}, T.~R., {Knapp}, G.~R., {Leggett}, S.~K., {et~al.} 2002, \apj, 564,
  466

\bibitem[{{Gizis}(2002)}]{2002ApJ...575..484G}
{Gizis}, J.~E. 2002, \apj, 575, 484

\bibitem[{{Gizis} {et~al.}(2000){Gizis}, {Monet}, {Reid}, {Kirkpatrick},
  {Liebert}, \& {Williams}}]{2000AJ....120.1085G}
{Gizis}, J.~E., {Monet}, D.~G., {Reid}, I.~N., {et~al.} 2000, \aj, 120, 1085

\bibitem[{{Gorlova} {et~al.}(2003){Gorlova}, {Meyer}, {Rieke}, \&
  {Liebert}}]{2003ApJ...593.1074G}
{Gorlova}, N.~I., {Meyer}, M.~R., {Rieke}, G.~H., \& {Liebert}, J. 2003, \apj,
  593, 1074

\bibitem[{{Goto} {et~al.}(2002){Goto}, {Kobayashi}, {Terada}, {Gaessler},
  {Kanzawa}, {Takami}, {Takato}, {Hayano}, {Kamata}, {Iye}, {Saint-Jacques},
  {Tokunaga}, {Potter}, \& {Cushing}}]{2002ApJ...567L..59G}
{Goto}, M., {Kobayashi}, N., {Terada}, H., {et~al.} 2002, \apjl, 567, L59

\bibitem[{{Grevesse} {et~al.}(1993){Grevesse}, {Noels}, \&
  {Sauval}}]{1993A&A...271..587G}
{Grevesse}, N., {Noels}, A., \& {Sauval}, A.~J. 1993, \aap, 271, 587

\bibitem[{{Guieu} {et~al.}(2007){Guieu}, {Pinte}, {Monin}, {M{\'e}nard},
  {Fukagawa}, {Padgett}, {Noriega-Crespo}, {Carey}, {Rebull}, {Huard}, \&
  {Guedel}}]{2007A&A...465..855G}
{Guieu}, S., {Pinte}, C., {Monin}, J., {et~al.} 2007, \aap, 465, 855

\bibitem[{{Hartmann}(2001)}]{2001AJ....121.1030H}
{Hartmann}, L. 2001, \aj, 121, 1030

\bibitem[{{Hayes}(1985)}]{1985IAUS..111..225H}
{Hayes}, D.~S. 1985, in IAU Symposium, Vol. 111, Calibration of Fundamental
  Stellar Quantities, ed. {D.~S.~Hayes, L.~E.~Pasinetti, \& A.~G.~D.~Philip},
  225--249

\bibitem[{{Helling} {et~al.}(2008{\natexlab{a}}){Helling}, {Ackerman},
  {Allard}, {Dehn}, {Hauschildt}, {Homeier}, {Lodders}, {Marley}, {Rietmeijer},
  {Tsuji}, \& {Woitke}}]{2008MNRAS.391.1854H}
{Helling}, C., {Ackerman}, A., {Allard}, F., {et~al.} 2008{\natexlab{a}},
  \mnras, 391, 1854

\bibitem[{{Helling} {et~al.}(2008{\natexlab{b}}){Helling}, {Dehn}, {Woitke}, \&
  {Hauschildt}}]{2008ApJ...675L.105H}
{Helling}, C., {Dehn}, M., {Woitke}, P., \& {Hauschildt}, P.~H.
  2008{\natexlab{b}}, \apjl, 675, L105

\bibitem[{{Helling} \& {Woitke}(2006)}]{2006A&A...455..325H}
{Helling}, C. \& {Woitke}, P. 2006, \aap, 455, 325

\bibitem[{{Herczeg} {et~al.}(2009){Herczeg}, {Cruz}, \&
  {Hillenbrand}}]{2009ApJ...696.1589H}
{Herczeg}, G.~J., {Cruz}, K.~L., \& {Hillenbrand}, L.~A. 2009, \apj, 696, 1589

\bibitem[{{Ireland} {et~al.}(2011){Ireland}, {Kraus}, {Martinache}, {Law}, \&
  {Hillenbrand}}]{2011ApJ...726..113I}
{Ireland}, M.~J., {Kraus}, A., {Martinache}, F., {Law}, N., \& {Hillenbrand},
  L.~A. 2011, \apj, 726, 113

\bibitem[{{Itoh} {et~al.}(2005){Itoh}, {Hayashi}, {Tamura}, {Tsuji}, {Oasa},
  {Fukagawa}, {Hayashi}, {Naoi}, {Ishii}, {Mayama}, {Morino}, {Yamashita},
  {Pyo}, {Nishikawa}, {Usuda}, {Murakawa}, {Suto}, {Oya}, {Takato}, {Ando},
  {Miyama}, {Kobayashi}, \& {Kaifu}}]{2005ApJ...620..984I}
{Itoh}, Y., {Hayashi}, M., {Tamura}, M., {et~al.} 2005, \apj, 620, 984

\bibitem[{{James} {et~al.}(2006){James}, {Melo}, {Santos}, \&
  {Bouvier}}]{2006A&A...446..971J}
{James}, D.~J., {Melo}, C., {Santos}, N.~C., \& {Bouvier}, J. 2006, \aap, 446,
  971

\bibitem[{{Jones} {et~al.}(1994){Jones}, {Longmore}, {Jameson}, \&
  {Mountain}}]{1994MNRAS.267..413J}
{Jones}, H.~R.~A., {Longmore}, A.~J., {Jameson}, R.~F., \& {Mountain}, C.~M.
  1994, \mnras, 267, 413

\bibitem[{{Kalas} {et~al.}(2008){Kalas}, {Graham}, {Chiang}, {Fitzgerald},
  {Clampin}, {Kite}, {Stapelfeldt}, {Marois}, \& {Krist}}]{2008Sci...322.1345K}
{Kalas}, P., {Graham}, J.~R., {Chiang}, E., {et~al.} 2008, Science, 322, 1345

\bibitem[{{Kenyon} \& {Hartmann}(1995)}]{1995ApJS..101..117K}
{Kenyon}, S.~J. \& {Hartmann}, L. 1995, \apjs, 101, 117

\bibitem[{{King} {et~al.}(2003){King}, {Villarreal}, {Soderblom}, {Gulliver},
  \& {Adelman}}]{2003AJ....125.1980K}
{King}, J.~R., {Villarreal}, A.~R., {Soderblom}, D.~R., {Gulliver}, A.~F., \&
  {Adelman}, S.~J. 2003, \aj, 125, 1980

\bibitem[{{Kirkpatrick}(2005)}]{2005ARA&A..43..195K}
{Kirkpatrick}, J.~D. 2005, \araa, 43, 195

\bibitem[{{Kirkpatrick} {et~al.}(2006){Kirkpatrick}, {Barman}, {Burgasser},
  {McGovern}, {McLean}, {Tinney}, \& {Lowrance}}]{2006ApJ...639.1120K}
{Kirkpatrick}, J.~D., {Barman}, T.~S., {Burgasser}, A.~J., {et~al.} 2006, \apj,
  639, 1120

\bibitem[{{Kirkpatrick} {et~al.}(1997){Kirkpatrick}, {Beichman}, \&
  {Skrutskie}}]{1997ApJ...476..311K}
{Kirkpatrick}, J.~D., {Beichman}, C.~A., \& {Skrutskie}, M.~F. 1997, \apj, 476,
  311

\bibitem[{{Kirkpatrick} {et~al.}(2008){Kirkpatrick}, {Cruz}, {Barman},
  {Burgasser}, {Looper}, {Tinney}, {Gelino}, {Lowrance}, {Liebert},
  {Carpenter}, {Hillenbrand}, \& {Stauffer}}]{2008ApJ...689.1295K}
{Kirkpatrick}, J.~D., {Cruz}, K.~L., {Barman}, T.~S., {et~al.} 2008, \apj, 689,
  1295

\bibitem[{{Kirkpatrick} {et~al.}(2001){Kirkpatrick}, {Dahn}, {Monet}, {Reid},
  {Gizis}, {Liebert}, \& {Burgasser}}]{2001AJ....121.3235K}
{Kirkpatrick}, J.~D., {Dahn}, C.~C., {Monet}, D.~G., {et~al.} 2001, \aj, 121,
  3235

\bibitem[{{Kirkpatrick} {et~al.}(1999){Kirkpatrick}, {Reid}, {Liebert},
  {Cutri}, {Nelson}, {Beichman}, {Dahn}, {Monet}, {Gizis}, \&
  {Skrutskie}}]{1999ApJ...519..802K}
{Kirkpatrick}, J.~D., {Reid}, I.~N., {Liebert}, J., {et~al.} 1999, \apj, 519,
  802

\bibitem[{{Kirkpatrick} {et~al.}(2000){Kirkpatrick}, {Reid}, {Liebert},
  {Gizis}, {Burgasser}, {Monet}, {Dahn}, {Nelson}, \&
  {Williams}}]{2000AJ....120..447K}
{Kirkpatrick}, J.~D., {Reid}, I.~N., {Liebert}, J., {et~al.} 2000, \aj, 120,
  447

\bibitem[{{Konopacky} {et~al.}(2007){Konopacky}, {Ghez}, {Duch{\^e}ne},
  {McCabe}, \& {Macintosh}}]{2007AJ....133.2008K}
{Konopacky}, Q.~M., {Ghez}, A.~M., {Duch{\^e}ne}, G., {McCabe}, C., \&
  {Macintosh}, B.~A. 2007, \aj, 133, 2008

\bibitem[{{Kratter} {et~al.}(2010){Kratter}, {Murray-Clay}, \&
  {Youdin}}]{2010ApJ...710.1375K}
{Kratter}, K.~M., {Murray-Clay}, R.~A., \& {Youdin}, A.~N. 2010, \apj, 710,
  1375

\bibitem[{{Kraus} {et~al.}(2006){Kraus}, {White}, \&
  {Hillenbrand}}]{2006ApJ...649..306K}
{Kraus}, A.~L., {White}, R.~J., \& {Hillenbrand}, L.~A. 2006, \apj, 649, 306

\bibitem[{{Kuzuhara} {et~al.}(2011){Kuzuhara}, {Tamura}, {Ishii}, {Kudo},
  {Nishiyama}, \& {Kandori}}]{2011AJ....141..119K}
{Kuzuhara}, M., {Tamura}, M., {Ishii}, M., {et~al.} 2011, \aj, 141, 119

\bibitem[{{Lagrange} {et~al.}(2010){Lagrange}, {Bonnefoy}, {Chauvin}, {Apai},
  {Ehrenreich}, {Boccaletti}, {Gratadour}, {Rouan}, {Mouillet}, {Lacour}, \&
  {Kasper}}]{2010Sci...329...57L}
{Lagrange}, A.-M., {Bonnefoy}, M., {Chauvin}, G., {et~al.} 2010, Science, 329,
  57

\bibitem[{{Lambrechts} \& {Johansen}(2012)}]{2012A&A...544A..32L}
{Lambrechts}, M. \& {Johansen}, A. 2012, \aap, 544, A32

\bibitem[{{Lan{\c c}on} \& {Mouhcine}(2002)}]{2002A&A...393..167L}
{Lan{\c c}on}, A. \& {Mouhcine}, M. 2002, \aap, 393, 167

\bibitem[{{Lavigne} {et~al.}(2009){Lavigne}, {Doyon}, {Lafreni{\`e}re},
  {Marois}, \& {Barman}}]{2009ApJ...704.1098L}
{Lavigne}, J.-F., {Doyon}, R., {Lafreni{\`e}re}, D., {Marois}, C., \& {Barman},
  T. 2009, \apj, 704, 1098

\bibitem[{{Leggett} {et~al.}(1996){Leggett}, {Allard}, {Berriman}, {Dahn}, \&
  {Hauschildt}}]{1996ApJS..104..117L}
{Leggett}, S.~K., {Allard}, F., {Berriman}, G., {Dahn}, C.~C., \& {Hauschildt},
  P.~H. 1996, \apjs, 104, 117

\bibitem[{{Leggett} {et~al.}(2001){Leggett}, {Allard}, {Geballe}, {Hauschildt},
  \& {Schweitzer}}]{2001ApJ...548..908L}
{Leggett}, S.~K., {Allard}, F., {Geballe}, T.~R., {Hauschildt}, P.~H., \&
  {Schweitzer}, A. 2001, \apj, 548, 908

\bibitem[{{Levine} {et~al.}(2006){Levine}, {Steinhauer}, {Elston}, \&
  {Lada}}]{2006ApJ...646.1215L}
{Levine}, J.~L., {Steinhauer}, A., {Elston}, R.~J., \& {Lada}, E.~A. 2006,
  \apj, 646, 1215

\bibitem[{{Lodieu} {et~al.}(2008){Lodieu}, {Hambly}, {Jameson}, \&
  {Hodgkin}}]{2008MNRAS.383.1385L}
{Lodieu}, N., {Hambly}, N.~C., {Jameson}, R.~F., \& {Hodgkin}, S.~T. 2008,
  \mnras, 383, 1385

\bibitem[{{Lommen} {et~al.}(2008){Lommen}, {J{\o}rgensen}, {van Dishoeck}, \&
  {Crapsi}}]{2008A&A...481..141L}
{Lommen}, D., {J{\o}rgensen}, J.~K., {van Dishoeck}, E.~F., \& {Crapsi}, A.
  2008, \aap, 481, 141

\bibitem[{{Lowrance} {et~al.}(2005){Lowrance}, {Becklin}, {Schneider},
  {Kirkpatrick}, {Weinberger}, {Zuckerman}, {Dumas}, {Beuzit}, {Plait},
  {Malumuth}, {Heap}, {Terrile}, \& {Hines}}]{2005AJ....130.1845L}
{Lowrance}, P.~J., {Becklin}, E.~E., {Schneider}, G., {et~al.} 2005, \aj, 130,
  1845

\bibitem[{{Lowrance} {et~al.}(1999){Lowrance}, {McCarthy}, {Becklin},
  {Zuckerman}, {Schneider}, {Webb}, {Hines}, {Kirkpatrick}, {Koerner}, {Low},
  {Meier}, {Rieke}, {Smith}, {Terrile}, \& {Thompson}}]{1999ApJ...512L..69L}
{Lowrance}, P.~J., {McCarthy}, C., {Becklin}, E.~E., {et~al.} 1999, \apjl, 512,
  L69

\bibitem[{{Lowrance} {et~al.}(2000){Lowrance}, {Schneider}, {Kirkpatrick},
  {Becklin}, {Weinberger}, {Zuckerman}, {Plait}, {Malmuth}, {Heap}, {Schultz},
  {Smith}, {Terrile}, \& {Hines}}]{2000ApJ...541..390L}
{Lowrance}, P.~J., {Schneider}, G., {Kirkpatrick}, J.~D., {et~al.} 2000, \apj,
  541, 390

\bibitem[{{Lucas} \& {Roche}(2000)}]{2000MNRAS.314..858L}
{Lucas}, P.~W. \& {Roche}, P.~F. 2000, \mnras, 314, 858

\bibitem[{{Lucas} {et~al.}(2001){Lucas}, {Roche}, {Allard}, \&
  {Hauschildt}}]{2001MNRAS.326..695L}
{Lucas}, P.~W., {Roche}, P.~F., {Allard}, F., \& {Hauschildt}, P.~H. 2001,
  \mnras, 326, 695

\bibitem[{{Luhman}(1999)}]{1999ApJ...525..466L}
{Luhman}, K.~L. 1999, \apj, 525, 466

\bibitem[{{Luhman}(2004)}]{2004ApJ...617.1216L}
{Luhman}, K.~L. 2004, \apj, 617, 1216

\bibitem[{{Luhman}(2007)}]{2007ApJS..173..104L}
{Luhman}, K.~L. 2007, \apjs, 173, 104

\bibitem[{{Luhman} {et~al.}(2005{\natexlab{a}}){Luhman}, {Adame}, {D'Alessio},
  {Calvet}, {Hartmann}, {Megeath}, \& {Fazio}}]{2005ApJ...635L..93L}
{Luhman}, K.~L., {Adame}, L., {D'Alessio}, P., {et~al.} 2005{\natexlab{a}},
  \apjl, 635, L93

\bibitem[{{Luhman} {et~al.}(2008){Luhman}, {Allen}, {Allen}, {Gutermuth},
  {Hartmann}, {Mamajek}, {Megeath}, {Myers}, \& {Fazio}}]{2008ApJ...675.1375L}
{Luhman}, K.~L., {Allen}, L.~E., {Allen}, P.~R., {et~al.} 2008, \apj, 675, 1375

\bibitem[{{Luhman} {et~al.}(2010){Luhman}, {Allen}, {Espaillat}, {Hartmann}, \&
  {Calvet}}]{2010ApJS..186..111L}
{Luhman}, K.~L., {Allen}, P.~R., {Espaillat}, C., {Hartmann}, L., \& {Calvet},
  N. 2010, \apjs, 186, 111

\bibitem[{{Luhman} {et~al.}(2003{\natexlab{a}}){Luhman}, {Brice{\~n}o},
  {Stauffer}, {Hartmann}, {Barrado y Navascu{\'e}s}, \&
  {Caldwell}}]{2003ApJ...590..348L}
{Luhman}, K.~L., {Brice{\~n}o}, C., {Stauffer}, J.~R., {et~al.}
  2003{\natexlab{a}}, \apj, 590, 348

\bibitem[{{Luhman} {et~al.}(2005{\natexlab{b}}){Luhman}, {D'Alessio}, {Calvet},
  {Allen}, {Hartmann}, {Megeath}, {Myers}, \& {Fazio}}]{2005ApJ...620L..51L}
{Luhman}, K.~L., {D'Alessio}, P., {Calvet}, N., {et~al.} 2005{\natexlab{b}},
  \apjl, 620, L51

\bibitem[{{Luhman} {et~al.}(1997){Luhman}, {Liebert}, \&
  {Rieke}}]{1997ApJ...489L.165L}
{Luhman}, K.~L., {Liebert}, J., \& {Rieke}, G.~H. 1997, \apjl, 489, L165+

\bibitem[{{Luhman} {et~al.}(2009){Luhman}, {Mamajek}, {Allen}, \&
  {Cruz}}]{2009ApJ...703..399L}
{Luhman}, K.~L., {Mamajek}, E.~E., {Allen}, P.~R., \& {Cruz}, K.~L. 2009, \apj,
  703, 399

\bibitem[{{Luhman} {et~al.}(2007){Luhman}, {Patten}, {Marengo}, {Schuster},
  {Hora}, {Ellis}, {Stauffer}, {Sonnett}, {Winston}, {Gutermuth}, {Megeath},
  {Backman}, {Henry}, {Werner}, \& {Fazio}}]{2007ApJ...654..570L}
{Luhman}, K.~L., {Patten}, B.~M., {Marengo}, M., {et~al.} 2007, \apj, 654, 570

\bibitem[{{Luhman} {et~al.}(2004){Luhman}, {Peterson}, \&
  {Megeath}}]{2004ApJ...617..565L}
{Luhman}, K.~L., {Peterson}, D.~E., \& {Megeath}, S.~T. 2004, \apj, 617, 565

\bibitem[{{Luhman} \& {Rieke}(1999)}]{1999ApJ...525..440L}
{Luhman}, K.~L. \& {Rieke}, G.~H. 1999, \apj, 525, 440

\bibitem[{{Luhman} {et~al.}(2005{\natexlab{c}}){Luhman}, {Stauffer}, \&
  {Mamajek}}]{2005ApJ...628L..69L}
{Luhman}, K.~L., {Stauffer}, J.~R., \& {Mamajek}, E.~E. 2005{\natexlab{c}},
  \apjl, 628, L69

\bibitem[{{Luhman} {et~al.}(2003{\natexlab{b}}){Luhman}, {Stauffer}, {Muench},
  {Rieke}, {Lada}, {Bouvier}, \& {Lada}}]{2003ApJ...593.1093L}
{Luhman}, K.~L., {Stauffer}, J.~R., {Muench}, A.~A., {et~al.}
  2003{\natexlab{b}}, \apj, 593, 1093

\bibitem[{{Luhman} {et~al.}(2006){Luhman}, {Wilson}, {Brandner}, {Skrutskie},
  {Nelson}, {Smith}, {Peterson}, {Cushing}, \& {Young}}]{2006ApJ...649..894L}
{Luhman}, K.~L., {Wilson}, J.~C., {Brandner}, W., {et~al.} 2006, \apj, 649, 894

\bibitem[{{Lyubchik} {et~al.}(2007){Lyubchik}, {Jones}, {Pavlenko}, {Martin},
  {McLean}, {Prato}, {Barber}, \& {Tennyson}}]{2007A&A...473..257L}
{Lyubchik}, Y., {Jones}, H.~R.~A., {Pavlenko}, Y.~V., {et~al.} 2007, \aap, 473,
  257

\bibitem[{{Mace} {et~al.}(2012){Mace}, {Prato}, {Torres}, {Wasserman},
  {Mathieu}, \& {McLean}}]{2012AJ....144...55M}
{Mace}, G.~N., {Prato}, L., {Torres}, G., {et~al.} 2012, \aj, 144, 55

\bibitem[{{Malo} {et~al.}(2013){Malo}, {Doyon}, {Lafreni{\`e}re}, {Artigau},
  {Gagn{\'e}}, {Baron}, \& {Riedel}}]{2013ApJ...762...88M}
{Malo}, L., {Doyon}, R., {Lafreni{\`e}re}, D., {et~al.} 2013, \apj, 762, 88

\bibitem[{{Marleau} \& {Cumming}(2013)}]{2013arXiv1302.1517M}
{Marleau}, G.-D. \& {Cumming}, A. 2013, ArXiv e-prints

\bibitem[{{Marois} {et~al.}(2008){Marois}, {Macintosh}, {Barman}, {Zuckerman},
  {Song}, {Patience}, {Lafreni{\`e}re}, \& {Doyon}}]{2008Sci...322.1348M}
{Marois}, C., {Macintosh}, B., {Barman}, T., {et~al.} 2008, Science, 322, 1348

\bibitem[{{Marois} {et~al.}(2010){Marois}, {Zuckerman}, {Konopacky},
  {Macintosh}, \& {Barman}}]{2010Natur.468.1080M}
{Marois}, C., {Zuckerman}, B., {Konopacky}, Q.~M., {Macintosh}, B., \&
  {Barman}, T. 2010, \nat, 468, 1080

\bibitem[{{Mart{\'{\i}}n} {et~al.}(1999){Mart{\'{\i}}n}, {Delfosse}, {Basri},
  {Goldman}, {Forveille}, \& {Zapatero Osorio}}]{1999AJ....118.2466M}
{Mart{\'{\i}}n}, E.~L., {Delfosse}, X., {Basri}, G., {et~al.} 1999, \aj, 118,
  2466

\bibitem[{{Martin} {et~al.}(1996){Martin}, {Rebolo}, \&
  {Zapatero-Osorio}}]{1996ApJ...469..706M}
{Martin}, E.~L., {Rebolo}, R., \& {Zapatero-Osorio}, M.~R. 1996, \apj, 469, 706

\bibitem[{{Mathieu} {et~al.}(2007){Mathieu}, {Baraffe}, {Simon}, {Stassun}, \&
  {White}}]{2007prpl.conf..411M}
{Mathieu}, R.~D., {Baraffe}, I., {Simon}, M., {Stassun}, K.~G., \& {White}, R.
  2007, Protostars and Planets V, 411

\bibitem[{{McElwain} {et~al.}(2007){McElwain}, {Metchev}, {Larkin}, {Barczys},
  {Iserlohe}, {Krabbe}, {Quirrenbach}, {Weiss}, \&
  {Wright}}]{2007ApJ...656..505M}
{McElwain}, M.~W., {Metchev}, S.~A., {Larkin}, J.~E., {et~al.} 2007, \apj, 656,
  505

\bibitem[{{McGovern} {et~al.}(2004){McGovern}, {Kirkpatrick}, {McLean},
  {Burgasser}, {Prato}, \& {Lowrance}}]{2004ApJ...600.1020M}
{McGovern}, M.~R., {Kirkpatrick}, J.~D., {McLean}, I.~S., {et~al.} 2004, \apj,
  600, 1020

\bibitem[{{McLean} {et~al.}(2003){McLean}, {McGovern}, {Burgasser},
  {Kirkpatrick}, {Prato}, \& {Kim}}]{2003ApJ...596..561M}
{McLean}, I.~S., {McGovern}, M.~R., {Burgasser}, A.~J., {et~al.} 2003, \apj,
  596, 561

\bibitem[{{McLean} {et~al.}(2007){McLean}, {Prato}, {McGovern}, {Burgasser},
  {Kirkpatrick}, {Rice}, \& {Kim}}]{2007ApJ...658.1217M}
{McLean}, I.~S., {Prato}, L., {McGovern}, M.~R., {et~al.} 2007, \apj, 658, 1217

\bibitem[{{Mohanty} {et~al.}(2005){Mohanty}, {Jayawardhana}, \&
  {Basri}}]{2005ApJ...626..498M}
{Mohanty}, S., {Jayawardhana}, R., \& {Basri}, G. 2005, \apj, 626, 498

\bibitem[{{Mohanty} {et~al.}(2007){Mohanty}, {Jayawardhana}, {Hu{\'e}lamo}, \&
  {Mamajek}}]{2007ApJ...657.1064M}
{Mohanty}, S., {Jayawardhana}, R., {Hu{\'e}lamo}, N., \& {Mamajek}, E. 2007,
  \apj, 657, 1064

\bibitem[{{Moorwood} {et~al.}(1998){Moorwood}, {Cuby}, \&
  {Lidman}}]{1998Msngr..91....9M}
{Moorwood}, A., {Cuby}, J.-G., \& {Lidman}, C. 1998, The Messenger, 91, 9

\bibitem[{{Mordasini} {et~al.}(2012){Mordasini}, {Alibert}, {Klahr}, \&
  {Henning}}]{2012A&A...547A.111M}
{Mordasini}, C., {Alibert}, Y., {Klahr}, H., \& {Henning}, T. 2012, \aap, 547,
  A111

\bibitem[{{Mountain} {et~al.}(1985){Mountain}, {Selby}, {Leggett}, {Blackwell},
  \& {Petford}}]{1985A&A...151..399M}
{Mountain}, C.~M., {Selby}, M.~J., {Leggett}, S.~K., {Blackwell}, D.~E., \&
  {Petford}, A.~D. 1985, \aap, 151, 399

\bibitem[{{Muench} {et~al.}(2007){Muench}, {Lada}, {Luhman}, {Muzerolle}, \&
  {Young}}]{2007AJ....134..411M}
{Muench}, A.~A., {Lada}, C.~J., {Luhman}, K.~L., {Muzerolle}, J., \& {Young},
  E. 2007, \aj, 134, 411

\bibitem[{{Nakajima} {et~al.}(1995){Nakajima}, {Oppenheimer}, {Kulkarni},
  {Golimowski}, {Matthews}, \& {Durrance}}]{1995Natur.378..463N}
{Nakajima}, T., {Oppenheimer}, B.~R., {Kulkarni}, S.~R., {et~al.} 1995, \nat,
  378, 463

\bibitem[{{Neuh{\"a}user} {et~al.}(2011){Neuh{\"a}user}, {Ginski}, {Schmidt},
  \& {Mugrauer}}]{2011MNRAS.416.1430N}
{Neuh{\"a}user}, R., {Ginski}, C., {Schmidt}, T.~O.~B., \& {Mugrauer}, M. 2011,
  \mnras, 416, 1430

\bibitem[{{Neuh{\"a}user} \& {Guenther}(2004)}]{2004A&A...420..647N}
{Neuh{\"a}user}, R. \& {Guenther}, E.~W. 2004, \aap, 420, 647

\bibitem[{{Neuh{\"a}user} {et~al.}(2000){Neuh{\"a}user}, {Guenther}, {Petr},
  {Brandner}, {Hu{\'e}lamo}, \& {Alves}}]{2000A&A...360L..39N}
{Neuh{\"a}user}, R., {Guenther}, E.~W., {Petr}, M.~G., {et~al.} 2000, \aap,
  360, L39

\bibitem[{{Neuh{\"a}user} {et~al.}(2005){Neuh{\"a}user}, {Guenther},
  {Wuchterl}, {Mugrauer}, {Bedalov}, \& {Hauschildt}}]{2005A&A...435L..13N}
{Neuh{\"a}user}, R., {Guenther}, E.~W., {Wuchterl}, G., {et~al.} 2005, \aap,
  435, L13

\bibitem[{{Nielsen} {et~al.}(2005){Nielsen}, {Close}, {Guirado}, {Biller},
  {Lenzen}, {Brandner}, {Hartung}, \& {Lidman}}]{2005AN....326.1033N}
{Nielsen}, E.~L., {Close}, L.~M., {Guirado}, J.~C., {et~al.} 2005,
  Astronomische Nachrichten, 326, 1033

\bibitem[{{Oasa} {et~al.}(1999){Oasa}, {Tamura}, \&
  {Sugitani}}]{1999ApJ...526..336O}
{Oasa}, Y., {Tamura}, M., \& {Sugitani}, K. 1999, \apj, 526, 336

\bibitem[{{Patience} {et~al.}(2012){Patience}, {King}, {De Rosa}, {Vigan},
  {Witte}, {Rice}, {Helling}, \& {Hauschildt}}]{2012A&A...540A..85P}
{Patience}, J., {King}, R.~R., {De Rosa}, R.~J., {et~al.} 2012, \aap, 540, A85

\bibitem[{{Patten} {et~al.}(2006){Patten}, {Stauffer}, {Burrows}, {Marengo},
  {Hora}, {Luhman}, {Sonnett}, {Henry}, {Raghavan}, {Megeath}, {Liebert}, \&
  {Fazio}}]{2006ApJ...651..502P}
{Patten}, B.~M., {Stauffer}, J.~R., {Burrows}, A., {et~al.} 2006, \apj, 651,
  502

\bibitem[{{Pecaut} {et~al.}(2012){Pecaut}, {Mamajek}, \&
  {Bubar}}]{2012ApJ...746..154P}
{Pecaut}, M.~J., {Mamajek}, E.~E., \& {Bubar}, E.~J. 2012, \apj, 746, 154

\bibitem[{{Phillips} {et~al.}(1987){Phillips}, {Davis}, {Lindgren}, \&
  {Balfour}}]{1987ApJS...65..721P}
{Phillips}, J.~G., {Davis}, S.~P., {Lindgren}, B., \& {Balfour}, W.~J. 1987,
  \apjs, 65, 721

\bibitem[{{Pinte} {et~al.}(2006){Pinte}, {M{\'e}nard}, {Duch{\^e}ne}, \&
  {Bastien}}]{2006A&A...459..797P}
{Pinte}, C., {M{\'e}nard}, F., {Duch{\^e}ne}, G., \& {Bastien}, P. 2006, \aap,
  459, 797

\bibitem[{{Rameau} {et~al.}(2013){Rameau}, {Chauvin}, {Lagrange}, {Klahr},
  {Bonnefoy}, {Mordasini}, {Bonavita}, {Desidera}, {Dumas}, \&
  {Girard}}]{2013arXiv1302.5367R}
{Rameau}, J., {Chauvin}, G., {Lagrange}, A.-M., {et~al.} 2013, ArXiv e-prints

\bibitem[{{Rayner} {et~al.}(2009){Rayner}, {Cushing}, \&
  {Vacca}}]{2009ApJS..185..289R}
{Rayner}, J.~T., {Cushing}, M.~C., \& {Vacca}, W.~D. 2009, \apjs, 185, 289

\bibitem[{{Rayner} {et~al.}(2003){Rayner}, {Toomey}, {Onaka}, {Denault},
  {Stahlberger}, {Vacca}, {Cushing}, \& {Wang}}]{2003PASP..115..362R}
{Rayner}, J.~T., {Toomey}, D.~W., {Onaka}, P.~M., {et~al.} 2003, \pasp, 115,
  362

\bibitem[{{Reid} {et~al.}(2001){Reid}, {Burgasser}, {Cruz}, {Kirkpatrick}, \&
  {Gizis}}]{2001AJ....121.1710R}
{Reid}, I.~N., {Burgasser}, A.~J., {Cruz}, K.~L., {Kirkpatrick}, J.~D., \&
  {Gizis}, J.~E. 2001, \aj, 121, 1710

\bibitem[{{Rice} {et~al.}(2010){Rice}, {Barman}, {Mclean}, {Prato}, \&
  {Kirkpatrick}}]{2010ApJS..186...63R}
{Rice}, E.~L., {Barman}, T., {Mclean}, I.~S., {Prato}, L., \& {Kirkpatrick},
  J.~D. 2010, \apjs, 186, 63

\bibitem[{{Robberto} {et~al.}(2010){Robberto}, {Soderblom}, {Scandariato},
  {Smith}, {Da Rio}, {Pagano}, \& {Spezzi}}]{2010AJ....139..950R}
{Robberto}, M., {Soderblom}, D.~R., {Scandariato}, G., {et~al.} 2010, \aj, 139,
  950

\bibitem[{{Rousselot} {et~al.}(2000){Rousselot}, {Lidman}, {Cuby}, {Moreels},
  \& {Monnet}}]{2000A&A...354.1134R}
{Rousselot}, P., {Lidman}, C., {Cuby}, J., {Moreels}, G., \& {Monnet}, G. 2000,
  \aap, 354, 1134

\bibitem[{{Santos} {et~al.}(2008){Santos}, {Melo}, {James}, {Gameiro},
  {Bouvier}, \& {Gomes}}]{2008A&A...480..889S}
{Santos}, N.~C., {Melo}, C., {James}, D.~J., {et~al.} 2008, \aap, 480, 889

\bibitem[{{Schlegel} {et~al.}(1998){Schlegel}, {Finkbeiner}, \&
  {Davis}}]{1998ApJ...500..525S}
{Schlegel}, D.~J., {Finkbeiner}, D.~P., \& {Davis}, M. 1998, \apj, 500, 525

\bibitem[{{Schmidt} {et~al.}(2008){Schmidt}, {Neuh{\"a}user}, {Seifahrt},
  {Vogt}, {Bedalov}, {Helling}, {Witte}, \& {Hauschildt}}]{2008A&A...491..311S}
{Schmidt}, T.~O.~B., {Neuh{\"a}user}, R., {Seifahrt}, A., {et~al.} 2008, \aap,
  491, 311

\bibitem[{{Schweitzer} {et~al.}(2001){Schweitzer}, {Gizis}, {Hauschildt},
  {Allard}, \& {Reid}}]{2001ApJ...555..368S}
{Schweitzer}, A., {Gizis}, J.~E., {Hauschildt}, P.~H., {Allard}, F., \& {Reid},
  I.~N. 2001, \apj, 555, 368

\bibitem[{{Seifahrt} {et~al.}(2007){Seifahrt}, {Neuh{\"a}user}, \&
  {Hauschildt}}]{2007A&A...463..309S}
{Seifahrt}, A., {Neuh{\"a}user}, R., \& {Hauschildt}, P.~H. 2007, \aap, 463,
  309

\bibitem[{{Sembach} \& {Savage}(1992)}]{1992ApJS...83..147S}
{Sembach}, K.~R. \& {Savage}, B.~D. 1992, \apjs, 83, 147

\bibitem[{{Slesnick} {et~al.}(2004){Slesnick}, {Hillenbrand}, \&
  {Carpenter}}]{2004ApJ...610.1045S}
{Slesnick}, C.~L., {Hillenbrand}, L.~A., \& {Carpenter}, J.~M. 2004, \apj, 610,
  1045

\bibitem[{{Song} {et~al.}(2003){Song}, {Zuckerman}, \&
  {Bessell}}]{2003ApJ...599..342S}
{Song}, I., {Zuckerman}, B., \& {Bessell}, M.~S. 2003, \apj, 599, 342

\bibitem[{{Sparks} \& {Ford}(2002)}]{2002ApJ...578..543S}
{Sparks}, W.~B. \& {Ford}, H.~C. 2002, \apj, 578, 543

\bibitem[{{Spiegel} \& {Burrows}(2012)}]{2012ApJ...745..174S}
{Spiegel}, D.~S. \& {Burrows}, A. 2012, \apj, 745, 174

\bibitem[{{Stephens} {et~al.}(2009){Stephens}, {Leggett}, {Cushing}, {Marley},
  {Saumon}, {Geballe}, {Golimowski}, {Fan}, \& {Noll}}]{2009ApJ...702..154S}
{Stephens}, D.~C., {Leggett}, S.~K., {Cushing}, M.~C., {et~al.} 2009, \apj,
  702, 154

\bibitem[{{Strom} {et~al.}(1989){Strom}, {Strom}, {Edwards}, {Cabrit}, \&
  {Skrutskie}}]{1989AJ.....97.1451S}
{Strom}, K.~M., {Strom}, S.~E., {Edwards}, S., {Cabrit}, S., \& {Skrutskie},
  M.~F. 1989, \aj, 97, 1451

\bibitem[{{Tamura} {et~al.}(1998){Tamura}, {Itoh}, {Oasa}, \&
  {Nakajima}}]{1998Sci...282.1095T}
{Tamura}, M., {Itoh}, Y., {Oasa}, Y., \& {Nakajima}, T. 1998, Science, 282,
  1095

\bibitem[{{Teixeira} {et~al.}(2009){Teixeira}, {Ducourant}, {Chauvin},
  {Krone-Martins}, {Bonnefoy}, \& {Song}}]{2009A&A...503..281T}
{Teixeira}, R., {Ducourant}, C., {Chauvin}, G., {et~al.} 2009, \aap, 503, 281

\bibitem[{{Testi}(2009)}]{2009A&A...503..639T}
{Testi}, L. 2009, \aap, 503, 639

\bibitem[{{Testi} {et~al.}(2001){Testi}, {D'Antona}, {Ghinassi}, {Licandro},
  {Magazz{\`u}}, {Maiolino}, {Mannucci}, {Marconi}, {Nagar}, {Natta}, \&
  {Oliva}}]{2001ApJ...552L.147T}
{Testi}, L., {D'Antona}, F., {Ghinassi}, F., {et~al.} 2001, \apjl, 552, L147

\bibitem[{{Thatte} {et~al.}(2007){Thatte}, {Abuter}, {Tecza}, {Nielsen},
  {Clarke}, \& {Close}}]{2007MNRAS.378.1229T}
{Thatte}, N., {Abuter}, R., {Tecza}, M., {et~al.} 2007, \mnras, 378, 1229

\bibitem[{{Theodossiou} \& {Danezis}(1991)}]{1991Ap&SS.183...91T}
{Theodossiou}, E. \& {Danezis}, E. 1991, \apss, 183, 91

\bibitem[{{Todorov} {et~al.}(2010){Todorov}, {Luhman}, \&
  {McLeod}}]{2010ApJ...714L..84T}
{Todorov}, K., {Luhman}, K.~L., \& {McLeod}, K.~K. 2010, \apjl, 714, L84

\bibitem[{{Torres} {et~al.}(2008){Torres}, {Quast}, {Melo}, \&
  {Sterzik}}]{2008hsf2.book..757T}
{Torres}, C.~A.~O., {Quast}, G.~R., {Melo}, C.~H.~F., \& {Sterzik}, M.~F. 2008,
  {Young Nearby Loose Associations}, ed. {Reipurth, B.}, 757--+

\bibitem[{{Tsuji} {et~al.}(1996){Tsuji}, {Ohnaka}, {Aoki}, \&
  {Nakajima}}]{1996A&A...308L..29T}
{Tsuji}, T., {Ohnaka}, K., {Aoki}, W., \& {Nakajima}, T. 1996, \aap, 308, L29

\bibitem[{{van Belle} {et~al.}(2009){van Belle}, {Creech-Eakman}, \&
  {Hart}}]{2009MNRAS.394.1925V}
{van Belle}, G.~T., {Creech-Eakman}, M.~J., \& {Hart}, A. 2009, \mnras, 394,
  1925

\bibitem[{{van Belle} \& {von Braun}(2009)}]{2009ApJ...694.1085V}
{van Belle}, G.~T. \& {von Braun}, K. 2009, \apj, 694, 1085

\bibitem[{{van der Bliek} {et~al.}(1996){van der Bliek}, {Manfroid}, \&
  {Bouchet}}]{1996A&AS..119..547V}
{van der Bliek}, N.~S., {Manfroid}, J., \& {Bouchet}, P. 1996, \aaps, 119, 547

\bibitem[{{Viana Almeida} {et~al.}(2009){Viana Almeida}, {Santos}, {Melo},
  {Ammler-von Eiff}, {Torres}, {Quast}, {Gameiro}, \&
  {Sterzik}}]{2009A&A...501..965V}
{Viana Almeida}, P., {Santos}, N.~C., {Melo}, C., {et~al.} 2009, \aap, 501, 965

\bibitem[{{Wallace} \& {Hinkle}(2001)}]{2001ApJ...559..424W}
{Wallace}, L. \& {Hinkle}, K. 2001, \apj, 559, 424

\bibitem[{{Weights} {et~al.}(2009){Weights}, {Lucas}, {Roche}, {Pinfield}, \&
  {Riddick}}]{2009MNRAS.392..817W}
{Weights}, D.~J., {Lucas}, P.~W., {Roche}, P.~F., {Pinfield}, D.~J., \&
  {Riddick}, F. 2009, \mnras, 392, 817

\bibitem[{{Weinberger} {et~al.}(2013){Weinberger}, {Anglada-Escud{\'e}}, \&
  {Boss}}]{2013ApJ...762..118W}
{Weinberger}, A.~J., {Anglada-Escud{\'e}}, G., \& {Boss}, A.~P. 2013, \apj,
  762, 118

\bibitem[{{White} \& {Ghez}(2001)}]{2001ApJ...556..265W}
{White}, R.~J. \& {Ghez}, A.~M. 2001, \apj, 556, 265

\bibitem[{{Witte} {et~al.}(2011){Witte}, {Helling}, {Barman}, {Heidrich}, \&
  {Hauschildt}}]{2011A&A...529A..44W}
{Witte}, S., {Helling}, C., {Barman}, T., {Heidrich}, N., \& {Hauschildt},
  P.~H. 2011, \aap, 529, A44+

\bibitem[{{Woitke} \& {Helling}(2003)}]{2003A&A...399..297W}
{Woitke}, P. \& {Helling}, C. 2003, \aap, 399, 297

\bibitem[{{Woitke} \& {Helling}(2004)}]{2004A&A...414..335W}
{Woitke}, P. \& {Helling}, C. 2004, \aap, 414, 335

\bibitem[{{Zapatero Osorio} {et~al.}(2000){Zapatero Osorio}, {B{\'e}jar},
  {Mart{\'{\i}}n}, {Rebolo}, {Barrado y Navascu{\'e}s}, {Bailer-Jones}, \&
  {Mundt}}]{2000Sci...290..103Z}
{Zapatero Osorio}, M.~R., {B{\'e}jar}, V.~J.~S., {Mart{\'{\i}}n}, E.~L.,
  {et~al.} 2000, Science, 290, 103

\bibitem[{{Zapatero Osorio} {et~al.}(2010){Zapatero Osorio}, {Rebolo},
  {Bihain}, {B{\'e}jar}, {Caballero}, \& {{\'A}lvarez}}]{2010arXiv1004.3965Z}
{Zapatero Osorio}, M.~R., {Rebolo}, R., {Bihain}, G., {et~al.} 2010, ArXiv
  e-prints

\end{thebibliography}

\longtab{8}{
\footnotesize
\label{table:2a}              
\begin{longtable}{llllllllllll}
\caption{Observing log. \textbf{sec} \textit{z} is the airmass. EC is encircled energy. DIT is the data integration time. t$_{exp}$ is the total integration time. \textbf{``Tel STD" stands for ``Telluric standard star".}}  \\
\hline\hline       
UT Date				& Target				& Mode		&	Grating 	& R$\mathrm{_{\lambda}}$ 			&	Pre-optic 				& $\mathrm{\langle}$sec \textit{z}$\mathrm{\rangle}$		& $\mathrm{\langle}$FWHM$\mathrm{\rangle}$			 		& $\mathrm{\langle}$EC$\mathrm{\rangle}$\tablefootmark*{} 	& DIT	 	 & $\mathrm{t_{exp}}$	& Note \\
				 		&			 	&		&			 	&		 		&	  (mas/pixel)			&							&   ('')					&   (\%)				&  (s) 	& 	(s)				&		   \\
\hline              
\endfirsthead      
\caption{continued.}  \\  
\hline \hline         
UT Date				& Target				& Mode		&	Grating 	& R$\mathrm{_{\lambda}}$ 			&	Pre-optic 				& $\mathrm{\langle}$sec \textit{z}$\mathrm{\rangle}$		& $\mathrm{\langle}$FWHM$\mathrm{\rangle}$			 		& $\mathrm{\langle}$EC$\mathrm{\rangle}$\tablefootmark*{} 	& DIT	  & $\mathrm{t_{exp}}$	& Note \\
				 		&			 	&		&			 	&		 		&	  (mas/pixel)			&							&   ('')					&   (\%)				&  (s) 		& 	(s)				&		   \\
\hline              
\endhead
\hline              
\endfoot
07/11/2007		&	DH Tau B			&	  AO		&	 $J$   		& 	2000	&  25 $\times$ 12.5	& 1.68		&	1.22			&	13			&	300		&	2700	&	\\	
07/11/2007		&	HIP039776		&	  AO		&	 $J$   		& 	2000	&  25 $\times$ 12.5	& 1.62		&	1.07			&	56			&	5		&	50		&	Tel STD	\\	
16/12/2007		&	DH Tau B			&	  AO		&	 $J$   		& 	2000	&  25 $\times$ 12.5	& 1.75		&	1.43			&	19			&	300		&	2700	&	\\ 
16/12/2007		&	HIP052898		&	  AO		&	 $J$   		& 	2000	&  25 $\times$ 12.5	& 1.63		&	1.61			&	31			&	60		&	120		&	Tel STD	\\	
22/10/2007		&	DH Tau B			&	  AO		&	 $H+K$		& 	1500	&  25 $\times$ 12.5	& 1.69		&	1.15			&	34			&	300		&	2700	&	\\ 
22/10/2007		&	HIP021512		&	  AO		&	 $H+K$		& 	1500	&  25 $\times$ 12.5	& 1.67		&	1.21			&	62			&	5			&	30		&	Tel STD \\ 
\hline                    
05/12/2007 	&	AB Pic b				&	  AO		&	 $J$   		& 	2000	&  25 $\times$ 12.5	&	1.22		&	  0.93		&	36			&	300	 	& 2700		&	   \\	
05/12/2007		&	AB Pic b				&	  AO		&	$ J$   		& 	2000	&  25 $\times$ 12.5 	&	1.20		&	  0.77		&	40			&	300	 	& 	2700		&		   \\  
05/12/2007		&	HIP039640		&	  AO		&	$J$			& 	2000	&  25 $\times$ 12.5	&	1.20		&	  1.27		&	51			&	40	 		&	80			& Tel STD	 \\ 
11/12/2007		&	AB Pic b				&	  AO		&  $J$   		& 	2000	&  25 $\times$ 12.5	&	1.20		&	  0.82		&	35			&	300	 	&	2700		&			 \\ 
11/12/2007		&	HIP023230		&	  AO		&	 $J$   		& 	2000	&  25 $\times$ 12.5	&	1.22		&	  1.10		&	48			&	60	 		&	120			&	Tel STD		 \\	
12/11/2007		&	AB Pic b				&	  AO		&	$H+K$  	&	1500	&  25 $\times$ 12.5	&	1.20		&	  1.70		&	58			&	300		&	2700		&					\\	 
12/11/2007		&	HIP037963		&	  AO		&	$H+K$  	& 	1500	&  25 $\times$ 12.5	&	1.21		&	  2.32		&	49		&	20			&  120			&	Tel STD	 \\	
\hline	
12/12/2007		&	TWA 5B				&	AO		&	 $J$   		& 	2000	&  25 $\times$ 12.5	& 2.36		&	0.69			&	19			&	120		&	840			&	\\	 
12/12:2007		&	HIP031959		&	AO		&	 $J$   		& 	2000	&  25 $\times$ 12.5	& 1.20		&	0.76			&	48			&	30		&	120			&	Tel STD	\\	 
13/12/2007		&	TWA 5B				&	AO		&	 $H+K$		& 	1500	&  25 $\times$ 12.5	& 1.15		&	1.21			&	36			&	60		&	960			&	\\	       
13/12/2007		&	HIP056741		&	AO		&	 $H+K$		& 	1500	&  25 $\times$ 12.5	& 1.17		&	1.43			&	72			&	10		&	20			&	Tel STD	\\	 
\hline
07/01/2008		&	GSC08047  B		&	  AO		&	 $J$   		& 	2000	&  25 $\times$ 12.5	&	1.25		&	  0.89		&	12			&	300	&	2700		&	\\ 
07/01/2008		&	HIP034339		&	  AO		&	 $J$   		& 	2000	&  25 $\times$ 12.5	&	1.23		&	  1.03		&	47			&	4			&	24			&	Tel STD	\\ 
19/01/2008		&	GSC08047  B		&	  AO		&	 $J$   		& 	2000	&  25 $\times$ 12.5	&	1.27		&	  0.88		&	10			&	300			&	2700		&	\\ 
19/01/2008		&	HIP035267		&	  AO		&	 $J$   		& 	2000	&  25 $\times$ 12.5	&	1.21		&	  0.77		&	49			&	15		&	150			&	Tel STD	\\	
20/01/2008		&	GSC08047  B		&	  AO		&	 $J$   		& 	2000	&  25 $\times$ 12.5	&	1.46		&	  0.56		&	18			&	300	&	2100		&	\\ 
20/01/2008		&	HIP011337		&	  AO		&	 $J$   		& 	2000	&  25 $\times$ 12.5	&	1.44		&	  0.77		&	53			&	5			&	50			&	Tel STD \\
06/01/2008		&	GSC08047  B		&	  AO		&	$H+K$		&	1500	&  25 $\times$ 12.5	&	1.24		&	  1.13		&	54			&	300		&	1500		&	 \\ 
06/01/2008		&	HIP036096		&	  AO		&	$H+K$		&	1500	&  25 $\times$ 12.5	&	1.23		&	  1.78		&	63			&	60			&	120			&	Tel STD \\ 
10/01/2008		&	GSC08047  B		&	  AO		&	$H+K$		&	1500	&  25 $\times$ 12.5	&	1.20		&	  1.01		&	46			&	300		&	2700		&	 \\ 
10/01/2008		&	HIP031959		&	  AO		&	$H+K$		&	1500	&  25 $\times$ 12.5	&	1.22		&	  0.97		&	77			&	10			&	100			&	Tel STD \\ 
\hline	       
19/10/2006		&	2M0141			&	NOAO	&	 $J$   		& 	2000	&  250 $\times$ 125	&	1.08		&	  0.65		&	38			&	300		&	1200		&	\\
19/10/2006		&	HIP014898		&	NOAO	&	 $J$   		& 	2000	&  250 $\times$ 125	&	1.05		&	  0.61		&	38			&	2			&	20			&	Tel STD	\\
19/10/2006		&	2M0141			&	NOAO	&	 $H+K$		&	1500	&  250 $\times$ 125	&  1.08		&	  0.70		&	48			&	60			&	240			&	\\
19/10/2006		&	HIP014898		&	NOAO	&	 $H+K$		&	1500	&  250 $\times$ 125	&  1.04		&	  0.67		&	51			&	0.83			&	8.3			&	Tel STD	\\
\hline	       
21/10/2007		&	KPNO Tau 4		&	NOAO	&	 $J$   		& 	2000	&  250 $\times$ 125	&	1.66		&	  1.02		&	31			&	300		&	2400		&	\\	
21/10/2007		&	HIP037044		&	NOAO	&	 $J$   		& 	2000	&  250 $\times$ 125	&	1.66		&	  1.05		&	33			&	10		&	20			&	Tel STD \\ 
21/10/2007		&	KPNO Tau 4		&	NOAO	&	$H+K$		&	1500	&  250 $\times$ 125	&  1.58		&	  0.91		&	51			&	300			&	2400		&	\\ 
21/10/2007		&	HIP021333		&	NOAO	&	$H+K$		&	1500	&  250 $\times$ 125	&	1.63		&	  0.99		&	47			&	2				&	20			&	Tel STD \\	
\hline	       
25/10/2007		&	2M0345			&	NOAO	&	 $J$   		& 	2000	&  250 $\times$ 125	&	1.57		&	  1.66		&	21			&	300		&	1800		&	\\
25/10/2007		&	HIP033468		&	NOAO	&	 $J$   		& 	2000	&  250 $\times$ 125	&	1.48		&	  2.15		&	25			&	10			&	20			&	Tel STD \\
25/10/2007		&	2M0345			&	NOAO	&	 $H+K$  	& 	1500	&  250 $\times$ 125	&	1.63		&	  1.89		&	24			&	100		&	1400		&	\\
25/10/2007		&	HIP043374		&	NOAO	&	 $H+K$  	& 	1500	&  250 $\times$ 125	&	1.60		&	  1.65		&	39			&	2			&	20			&	Tel STD \\
\hline
07/12/2007		&	Cha1109			&	NOAO	&	 $J$   		& 	2000	&  250 $\times$ 125	& 1.83		&	 1.11		&	25			&	300		&	2400		&	\\
07/12/2007		&	HIP061829		&	NOAO	&	 $J$   		& 	2000	&  250 $\times$ 125	& 1.83		&	 1.63		&	12			&	30		&	120			&	Tel STD \\
11/12/2007		&	Cha1109			&	NOAO	&	 $J$   		& 	2000	&  250 $\times$ 125	& 1.80		&	 1.01		&	18			&	300			&	2400		&	\\
11/12/2007		&	Cha1109			&	NOAO	&	 $J$   		& 	2000	&  250 $\times$ 125	& 1.89		&	 1.08		&	17			&	300		&	2400		&	\\
11/12/2007		&	HIP024611		&	NOAO	&	 $J$   		& 	2000	&  250 $\times$ 125	& 1.89		&	 0.92		&	20			&	10			&	60			&	Tel STD \\
05/12/2007		&	Cha1109			&	NOAO	&	 $H+K$		& 	1500	&  250 $\times$ 125	& 1.90		&	 1.06		&	31			&	300		&	1500		&	\\		
05/12/2007		&	Cha1109			&	NOAO	&	 $H+K$		& 	1500	&  250 $\times$ 125	& 1.83		&	 0.93		&	34			&	300		&	1500		&	\\		
05/12/2007		&	HIP063258		&	NOAO	&	 $H+K$		& 	1500	&  250 $\times$ 125	& 1.86		&	 0.91		&	26			&	5				&	40			&	Tel STD \\
\hline                    
14/12/2007		&	OTS 44				&	NOAO	&	 $J$   		& 	2000	&  250 $\times$ 125	&	1.88		&	  1.71		&	21			&	300			&	2400		&	\\
14/12/2007		&	HIP025888		&	NOAO	&	 $J$   		& 	2000	&  250 $\times$ 125	&	1.67		&	  1.05		&	31			&	5				&	20			&	Tel STD	\\
21/12/2007		&	OTS 44				&	NOAO	&	 $J$   		& 	2000	&  250 $\times$ 125	&  1.67		&	  0.98		&	31			&	300		&	2400		&	\\
21/12/2007		&	HIP059899		&	NOAO	&	 $J$   		& 	2000	&  250 $\times$ 125	&	1.76		&	  0.92		&	36			&	8			&	80			&	Tel STD	\\
18/12/2007		&	OTS 44				&	NOAO	&	$H+K$		&	1500	&  250 $\times$ 125	&	1.82		&	  1.01		&	33			&	300	&	1800		&	\\			   
18/12/2007		&	HIP059043		&	NOAO	&	$H+K$		&	1500	&  250 $\times$ 125	&	1.76		&	  1.03		&	37			&	1			&	20			&	Tel STD \\               
21/12/2007		&	OTS 44				&	NOAO	&	$H+K$		&	1500	&  250 $\times$ 125	&	1.80		&	  1.07		&	43			&	300			&	1800		&	\\			   
21/12/2007		&	HIP059899		&	NOAO	&	$H+K$		&	1500	&  250 $\times$ 125	&	1.80		&	  1.02		&	50			&	5				&	50			&	Tel STD	\\
\hline	       
20/01/2008		&	Gl 417 B			&	NOAO	&	 $J$   		& 	2000	&  250 $\times$ 125	&	2.08		&	  1.17		&	22			&	300			&	2100		&	\\
20/01/2008		&	HIP050374		&	NOAO	&	 $J$   		& 	2000	&  250 $\times$ 125	&	2.12		&	  1.02		&	20			&	7			&	28			&	Tel STD \\
25/12/2007		&	Gl	417	B			&	NOAO	&	 $H+K$  	& 	1500	&  250 $\times$ 125	&	2.82		&	  0.81		&	37			&	100			&	800			&	\\
25/12/2007		&	HIP026334		&	NOAO	&	 $H+K$  	& 	1500	&  250 $\times$ 125	&	2.39		&	  0.97		&	18			&	5				&	20			&	Tel STD \\
\hline	
09/04/2009		&	UScoCTIO108B   &  AO  &  $J$  &  2000 &  100 $\times$ 50  &  1.03  &  0.95  &  6  &  130    &  6240  &  \\
09/04/2009	&  HIP088201  &  AO  &    $J$  &  2000 &  100 $\times$ 50  & 1.03  &  0.97  &  41   &  20		&  20  &  Tel STD \\
09/04/2009	&	UScoCTIO108B   &  AO  &  $H+K$  &  1500 &  100 $\times$ 50  &  1.15  &  1.28  & 12   &  133  &  1596  &  \\
09/04/2009   &  HIP073881   &   AO   &    $J$  &  2000 &  100 $\times$ 50  & 1.09  &  0.97  &  43   &  20		 &  20  &  Tel STD \\
\hline	
19/05/2009 	&   HR7329B   &  AO   &  $J$   &  2000     &  100 $\times$ 50  &  1.18  &  0.60  &  27   &  300  &   	1200  &   \\
19/05/2009 	&   HIP094378  &  AO  &   $J$   &  2000     &  100 $\times$ 50  &  1.11   &  0.56  &   60  &  2   &  10  &  Tel STD \\
19/05/2009 &   HR7329B   &  AO   &  $H+K$   &  1500     &  100 $\times$ 50  &  1.20   &  0.59  &  42  &  80  &  320   &   \\
19/05/2009	&   HIP086951   &  AO   &  $H+K$   &  1500     &  100 $\times$ 50  & 1.05  &  0.75  &  46  &  2  &  10  & Tel STD \\
\hline	
\end{longtable}
\tablefoot{ \tablefoottext*{}{At 1.25 $\mu$m in the J band and at 2.2 $\mu$m in the H+K band over appertures of 300 mas in AO mode and 600 mas in NOAO mode.}}
}

\end{document}